\documentclass[twocolumn]{openjournal}
\makeatletter
\def\frontmatter@title@below{\vspace*{0.25in}}
\def\frontmatter@above@affilgroup{\vspace*{0.05in}}
\AtBeginDocument{%
  \long\def\@makecaption#1#2{%
   \noindent\begin{minipage}{0.9999\linewidth}
     \if\csname ftype@\@captype\endcsname 2
     \vskip 2ex\noindent \@table@type@size{\@eapj@cap@font  #1.}~%
     #2\par\medskip
     \else
     \vspace*{\abovecaptionskip}\noindent\footnotesize #1 #2\par\vskip \belowcaptionskip
     \fi
   \end{minipage}\par
   }%
}
\makeatother

\usepackage{newtxtext,newtxmath}

\usepackage[T1]{fontenc}

\DeclareRobustCommand{\VAN}[3]{#2}
\let\VANthebibliography\thebibliography
\def\thebibliography{\DeclareRobustCommand{\VAN}[3]{##3}\VANthebibliography}
\makeatletter
\def\fnum@table{{\@eapj@cap@font Table~\thetable}}
\long\def\@makecaption#1#2{%
 \noindent\begin{minipage}{0.9999\linewidth}
   \if\csname ftype@\@captype\endcsname 2 
   \vskip 2ex\noindent \centering\footnotesize{#1}~#2\par\medskip
   \else
   \vspace*{\abovecaptionskip}\noindent\footnotesize #1 #2\par\vskip \belowcaptionskip
   \fi
 \end{minipage}\par
}
\makeatother

\usepackage{graphicx} 
\usepackage{amsmath}
\usepackage[dvipsnames]{xcolor}
\usepackage{hyperref}
\usepackage{orcidlink}
\usepackage{lineno}

\begin{document}
\title{Intrinsic tidal shear at the largest scales: galaxy multiplet alignment in DESI DR2}

\author{
\mbox{Claire Lamman\orcidlink{0000-0002-6731-9329}$^{1,2,3}$,}
\mbox{Jessica Nicole Aguilar$^{4}$,}
\mbox{Steven Ahlen\orcidlink{0000-0001-6098-7247}$^{5}$,}
\mbox{Alejandro Aviles\orcidlink{0000-0001-5998-3986}$^{6,7}$,}
\mbox{Davide Bianchi\orcidlink{0000-0001-9712-0006}$^{8,9}$,}
\mbox{David Brooks$^{10}$,}
\mbox{Aurelio Carnero Rosell\orcidlink{0000-0003-3044-5150}$^{11,12}$,}
\mbox{Francisco Javier Castander\orcidlink{0000-0001-7316-4573}$^{13,14}$,}
\mbox{Todd Claybaugh$^{4}$,}
\mbox{Axel de la Macorra\orcidlink{0000-0002-1769-1640}$^{15}$,}
\mbox{Daniel Eisenstein$^{16}$,}
\mbox{Andreu Font-Ribera\orcidlink{0000-0002-3033-7312}$^{17,18}$,}
\mbox{Jaime E.~Forero-Romero\orcidlink{0000-0002-2890-3725}$^{19,20}$,}
\mbox{Cristhian Garcia-Quintero\orcidlink{0000-0003-1481-4294}$^{16}$,}
\mbox{Enrique Gazta\~{n}aga\orcidlink{0000-0001-9632-0815}$^{13,21,14}$,}
\mbox{Gaston Gutierrez$^{22}$,}
\mbox{Klaus Honscheid\orcidlink{0000-0002-6550-2023}$^{1,3,23}$,}
\mbox{Mustapha Ishak\orcidlink{0000-0002-6024-466X}$^{24}$,}
\mbox{Stephanie Juneau\orcidlink{0000-0002-0000-2394}$^{25}$,}
\mbox{Tanveer Karim\orcidlink{0000-0002-5652-8870}$^{26}$,}
\mbox{David Kirkby\orcidlink{0000-0002-8828-5463}$^{27}$,}
\mbox{Ofer Lahav\orcidlink{0000-0002-1134-9035}$^{10}$,}
\mbox{Martin Landriau\orcidlink{0000-0003-1838-8528}$^{4}$,}
\mbox{Michael E.~Levi\orcidlink{0000-0003-1887-1018}$^{4}$,}
\mbox{Marc Manera\orcidlink{0000-0003-4962-8934}$^{28,18}$,}
\mbox{Aaron Meisner\orcidlink{0000-0002-1125-7384}$^{25}$,}
\mbox{Ramon Miquel$^{17,18}$,}
\mbox{Seshadri Nadathur\orcidlink{0000-0001-9070-3102}$^{21}$,}
\mbox{Hernan Enrique Noriega\orcidlink{0000-0002-3397-3998}$^{7,15}$,}
\mbox{Nathalie Palanque-Delabrouille\orcidlink{0000-0003-3188-784X}$^{29,4}$,}
\mbox{Will Percival\orcidlink{0000-0002-0644-5727}$^{30,31,32}$,}
\mbox{Miguel Perdomo$^{19}$,}
\mbox{Francisco Prada\orcidlink{0000-0001-7145-8674}$^{33}$,}
\mbox{Ignasi P\'erez-R\`afols\orcidlink{0000-0001-6979-0125}$^{34}$,}
\mbox{Corentin Ravoux\orcidlink{0000-0002-3500-6635}$^{35}$,}
\mbox{Ashley J.~Ross\orcidlink{0000-0002-7522-9083}$^{1,2,23}$,}
\mbox{Graziano Rossi$^{36}$,}
\mbox{Rossana Ruggeri\orcidlink{0000-0002-0394-0896}$^{37}$,}
\mbox{Lado Samushia\orcidlink{0000-0002-1609-5687}$^{38,39}$,}
\mbox{Eusebio Sanchez\orcidlink{0000-0002-9646-8198}$^{40}$,}
\mbox{Christoph Saulder\orcidlink{0000-0002-0408-5633}$^{41}$,}
\mbox{David Schlegel$^{4}$,}
\mbox{Michael Schubnell$^{42,43}$,}
\mbox{Joseph Harry Silber\orcidlink{0000-0002-3461-0320}$^{4}$,}
\mbox{Ma\l{}gorzata Siudek\orcidlink{0000-0002-2949-2155}$^{14,12}$,}
\mbox{Gregory Tarl\'{e}\orcidlink{0000-0003-1704-0781}$^{43}$,}
\mbox{Rajeev Vaisakh\orcidlink{0009-0001-2732-8431}$^{44}$,}
\mbox{Benjamin Alan Weaver$^{25}$}
\\[6pt] {\normalfont\itshape Affiliations are listed at the end of the paper}}

\begin{abstract}
We measure the intrinsic alignment (IA) of galaxy multiplets with the large-scale tidal field traced by galaxy positions in DESI Data Release 2. Using the BGS, LRG, and ELG samples spanning $0.01 < z < 1.6$, we optimize the scales which define multiplets in each sample to minimize measurement noise. We measure multiplet tidal alignment $\mathcal{E}_+$ on scales beyond 100 $h^{-1}$Mpc in all samples, and detect alignment over 200 $h^{-1}$Mpc, up to 4.5$\sigma$ in the highest LRG redshift bin. We additionally present the multiplet shape–shape autocorrelation, $\mathcal{E}_{++}$, which is independent of galaxy bias and provides a consistency check of our NLA (nonlinear alignment) model. We do not detect a baryon acoustic feature in the projected estimator. We also find the alignment strength of multiplets is correlated with redshift and galaxy morphology. Multiplet alignment offers direct access to the tidal shear field in all galaxy samples, even where individual galaxy alignment cannot, and extends IA detection to the largest scales yet probed.
\end{abstract}

\section{Introduction}\label{sec:intro}

While often traced directly with galaxy positions, the large-scale structure (LSS) of the universe also generates a measurable gravitational field. This tidal field encodes information that is complementary to, and in some cases inaccessible from, two-point clustering. Such information is particularly valuable when measured on large scales, where the field is close to linear and retains a clean connection to primordial physics and the initial conditions of the universe. The large-scale tidal field can be used to constrain primordial non-Gaussianity (PNG), cosmic B-modes, and parity violation, particularly around scales of 100-1000 $h^{-1}$Mpc, where these signatures are concentrated \citep{akitsuImprintAnisotropicPrimordial2021, kuritaConstraintsAnisotropicPrimordial2023, achucarroInflationTheoryObservations2022, paulApparentParityViolation2024}. However, largest current direct detections of the tidal field only extend to around $100$ $h^{-1}$Mpc (Figure \ref{fig:snr-comparison}).

Cosmic shear surveys probe the tidal field of LSS through the integrated effect that gravitational lensing has on background galaxies \citep{thelsstdarkenergysciencecollaborationLSSTDarkEnergy2018}. More directly, the non-projected tidal shear field can be traced by objects that are intrinsically aligned with it, such as galaxies \citep{joachimiGalaxyAlignmentsOverview2015, troxelIntrinsicAlignmentGalaxies2015, lammanIAGuideBreakdown2024, chisariRisingTideIntrinsic2025}. Measuring the intrinsic alignment (IA) of galaxy shapes to the underlying density is essentially correlating galaxy positions with a spin-2 object and is comparable to the collapsed form of the 3-point correlation function. This collapsed form is expected to have unique sensitivity to several of the cosmological effects listed above \citep{chisariCosmologicalInformationIntrinsic2013, kuritaParityViolationGalaxy2025}. 
However, measuring IA is challenging in practice; it is significantly susceptible to imaging systematics and restricted to the populations which show measurable alignment -- large, red galaxies \citep{samuroffDarkEnergySurvey2023, siegelIntrinsicAlignmentDemographics2025, hervaspetersUNIONSDirectMeasurement2025}. With current surveys, this also functionally limits detection to below redshift $z=1$.

\cite{lammanDetectionLargescaleTidal2024} presented an alternative that circumvents these issues: multiplet intrinsic alignment (MIA). This correlates the projected orientations of small sets of galaxies with the surrounding large-scale matter, as opposed to the orientations of individual ones. Galaxy multiplets typically consist of 2-5 galaxies within a few $h^{-1}$Mpc of each other. They are not necessarily clusters, groups, or a proxy for halos. We show in this work that they often contain fewer galaxies spread over a larger volume, and may be better understood as tracing filaments. Like IA, their shapes are reduced to an orientation at a single point. It is a form of a projected, collapsed 3-point correlation function, where the collapsed side is weighted over the positions of the multiplet's member galaxies. This configuration is motivated by our understanding of galaxy formation, where matter falls in along tidally-aligned filaments, and is considerably simpler to measure and interpret than the full bispectrum. Above $\sim20 h^{-1}$Mpc, it can be modeled with existing IA frameworks such as the nonlinear alignment model, or NLA \citep{bridleDarkEnergyConstraints2007}. 

Using spectroscopic data from the Dark Energy Spectroscopic Instrument (DESI)'s first data release, multiplet alignment has been detected out to projected separations of $100$ $h^{-1}$Mpc, in populations of blue galaxies, and beyond redshift $z=1$ \citep{lammanDetectionLargescaleTidal2024}. It also shows promise for cross-survey analysis: multiplets identified in imaging surveys around spectroscopic tracers display similar signals with improved signal-to-noise ratios \citep{2025RNAAS...9..305K}.

Here we explore different definitions of multiplets, identifying the scales which contain the most information and how best to weight the relative positions of member galaxies. Combined with updated data from DESI's second data release, we extend the scales at which intrinsic tidal shear can be detected and take a preliminary look at a potential BAO feature in the projected tidal shear correlation - which has yet to be detected. We also, for the first time, measure the multiplet-multiplet autocorrelation. This is a form of the collapsed 4-point correlation function and of interest for several reasons. The autocorrelation is independent of galaxy bias and provides a valuable check for systematics and the internal consistency of our model. It is also the lowest-order estimator sensitive to cosmic B-modes and certain classes of parity violation. These measurements also provide the parameters necessary to extend existing IA forecasts to multiplet alignment, including characteristic number densities, alignment strengths, and redshift dependence.

Section \ref{sec:catalogs} describes the DESI DR2 catalogs. Section \ref{sec:multiplets} introduces multiplets, how they are defined, and their demographics across the DESI catalogs. Section \ref{sec:formalism} summarizes the formalism and modeling of multiplet alignment. Section \ref{sec:measurement} documents our methods of measurement and error estimation, which are also performed on mock data, described in Section \ref{sec:mocks}. Our main results are presented in Section \ref{sec:results}, including large-scale measurements across all samples, autocorrelation functions, correlations of alignment strength with redshift, and investigations of a potential BAO feature.

Throughout this paper, we assume the cosmological parameters of $H_0=69.6$ km s$^{-1}$ Mpc$^{-1}$, $\Omega_{m,0}=0.286$, $\Omega_{\Lambda,0}=0.714$.  

\vspace{.1in}\section{DESI Catalogs}
\label{sec:catalogs}

DESI \citep{DESI2016b.Instr, desicollaborationOverviewInstrumentationDark2022} is a robotic, fiber-fed spectroscopic surveyor mounted on the Mayall 4-meter telescope at Kitt Peak National Observatory. It obtains simultaneous spectra of nearly 5000 targets across a $\sim3^\circ$-diameter field of view via a robotic focal plane \citep{silberRoboticMultiobjectFocal2023}, a corrected wide-field optical assembly \citep{millerOpticalCorrectorDark2024}, and a fiber system feeding ten three-arm spectrographs \citep{poppettOverviewFiberSystem2024}. Targets are selected from DR9 of the Legacy Imaging Surveys \citep{deyOverviewDESILegacy2019, myersTargetselectionPipelineDark2023}, and observations are planned and executed by the survey operations pipeline described in \citet{schlaflySurveyOperationsDark2023}. Spectra are reduced and redshifts extracted with the DESI spectroscopic pipeline \citep{guySpectroscopicDataProcessing2023}. The final survey will consist of over 63 million spectra of galaxies and quasars taken over eight years.

In this work, we use 19 million spectroscopic redshifts from DESI's first three years of observations, Data Release 2 (DR2). The DR1 dataset and its data model are described in \citet{desicollaborationDataRelease12026}, and the DR2 samples and their cosmological analysis are presented in \citet{abdulkarimDESIDR2Results2025}. We use the large-scale structure (LSS) catalogs \citep{rossConstructionLargescaleStructure2025a} for all analysis. The catalogs contain weights to account for redshift failures and targeting incompleteness, which are used to weight the tracer sample in each cross-correlation measurement, and associated random catalogs matched to the survey geometry, which are also used in Section~\ref{sec:measurement}. 

We use three LSS catalogs. The first is the Bright Galaxy Survey (BGS), which is a relatively dense sample containing galaxies with a variety of morphologies and nearly 100\% completeness at low redshifts \citep{hahnDESIBrightGalaxy2023}. We split this catalog into five redshift samples between $0.01 < z < 0.5$. We use the full BGS catalog (``BGS BRIGHT'') rather than the absolute-magnitude limited subsample adopted for the BAO analysis \citep{abdulkarimDESIDR2Results2025}, which includes more galaxies at the cost of a density highly correlated with redshift. The next sample consists of Luminous Red Galaxies (LRG), which are large elliptical galaxies that are strong tracers of the underlying matter \citep{zhouTargetSelectionValidation2023}. We split the LRGs into three redshift bins between $0.4<z<1.1$. The final sample contains Emission Line Galaxies (ELG), which are largely blue, spiral galaxies \citep{raichoorTargetSelectionValidation2023}. We split these into two redshift bins between $0.8<z<1.6$. The LRG and ELG samples overlap in $0.8<z<1.1$, allowing for cross-correlations. All subsamples are labeled as their galaxy type and a number which increases with redshift, e.g. the lowest-redshift BGS sample is BGS1. The properties of these subsamples are displayed in Table~\ref{tab:samples}. 

\vspace{.1in}
\section{Defining Multiplets}
\label{sec:multiplets} 

\subsection{Identifying Multiplets}
Multiplets are sets of galaxies separated from each other by a maximum linking length $\ell$ and identified similarly to the friends-of-friends (FOF) algorithm \citep{moreOverdensityMassesFriendsoffriends2011}. We begin by finding all pairs of galaxies separated by $\ell$ or less. We then identify multiplets as the connected components of the resulting pairs using a Union Find algorithm, setting no limit on the physical size or number of members of a group. To account for the differing scales of line-of-sight (LOS) and transverse uncertainty, largely from redshift-space distortions, we split the linking length into a transverse component $\ell_{p}$ and a LOS component $\ell_{\|}$.

Our goal is to measure tidal shear with the highest signal-to-noise ratio (SNR), rather than to identify a consistent set of physical objects. The scales at which multiplets are defined therefore vary between samples, with denser samples and higher clustering favoring smaller linking scales. We do not expect variations within the $\ell{\sim}1\,h^{-1}{\rm Mpc}$ regime to affect the scale dependence of the signal, and expect only a small impact on its overall amplitude.

To identify the optimal linking lengths, we measured multiplet-galaxy alignment across a range of scales in each sample (described later in Section \ref{sec:measurement}). The signal was averaged in six logarithmic bins between projected separations of $10$ and $50\,h^{-1}{\rm Mpc}$ and the errors are the standard error from measuring in different sky regions; we found no improvement from a full jackknife. 

We sampled linking scales of both  $\ell_{p}$ and $\ell_{\|}$ in steps of $0.5\,h^{-1}{\rm Mpc}$, up to $\ell=8\,h^{-1}{\rm Mpc}$ and selected the smallest $\ell$ which produced a measurement within $1\%$ of the peak SNR, on the basis that adding a large number of additional multiplets is not warranted for a marginal gain in signal. We perform this test on a subset of the area in all samples, using the minimum area necessary to obtain smooth results with respect to $\ell$. This is more resource efficient and ensures we are not sensitive to a SNR spike in the overall measurements.

This SNR scales as $A/\sigma$, where $\sigma$ is the per-multiplet scatter, the same quantity that governs the final measurement significance, making it a well-motivated cut. The scale that maximizes this SNR need not coincide with the one that maximizes the signal amplitude $A$, since a larger linking scale alters the multiplet population and can also increase $\sigma$; in practice, however, the two peaks fall at the same scale for the samples considered here. Additionally, since the signal appears to be constant over a range of $\ell$, our chosen scales are the ones which minimize noise. 

For LRG, we restricted the measurement to a patch defined by $130 < {\rm RA} < 150$ and $0 < {\rm Dec} < 20$, in degrees. For the sparser ELG sample, we used the full catalog and increase the measured scale to $100\,h^{-1}{\rm Mpc}$. We found $\ell_{p}=\ell_{\|}=5h^{-1}$Mpc to be optimal for all redshift bins in the LRG sample, and $\ell_{p}=7 h^{-1}$Mpc, $\ell_{\|}=3h^{-1}$Mpc for ELGs. This agreement across redshift bins within the samples is expected, given their near-constant comoving number density.

\begin{figure}
\centering
\includegraphics[width=.48\textwidth]{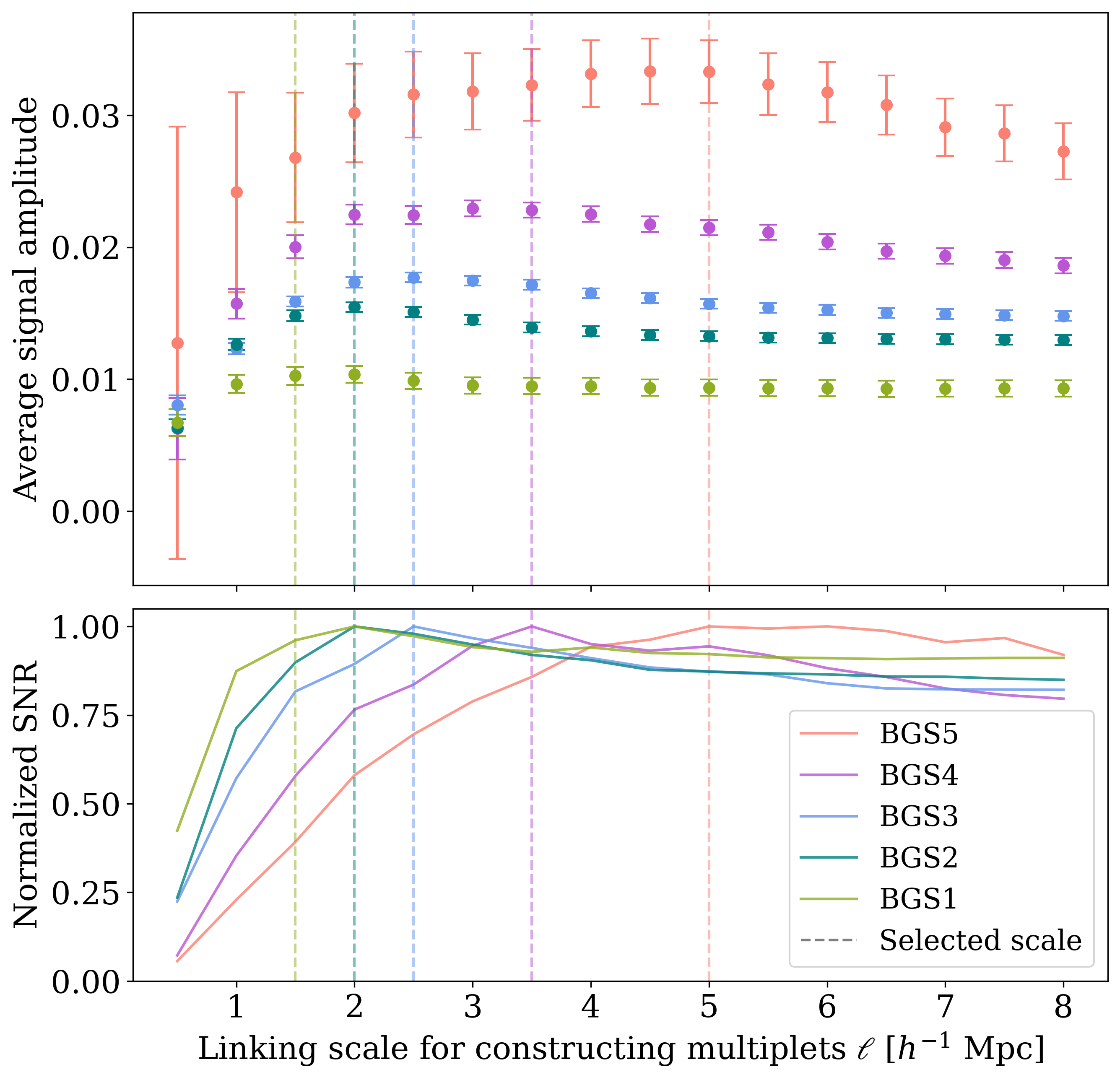}
\caption[]{The results of measuring multiplet alignment in a subset of the BGS catalog (defined in Table~\ref{tab:samples}) as a function of linking scale, the maximum separation between two galaxies for them to be considered part of the same multiplet. The choice of linking scale has a minimal effect on the signal amplitude (top panel) above $\ell{\sim}1.5\,h^{-1}{\rm Mpc}$. We  select the scale based on the total SNR (bottom panel), which captures where most of the alignment information lies. This is especially important for BGS, whose number density varies strongly with redshift. The chosen linking scale for each redshift bin is marked by a vertical dashed line and, as expected, increases as the number density decreases.
}
\label{fig:bgs-snr}
\vspace{.2in}
\end{figure}

Conversely, the density of the magnitude-limited BGS sample drops substantially toward higher redshifts. We tested a range of linking scales in a subset of the footprint for each redshift bin, using $800$, $800$, $1500$, $1500$, and $3000\,{\rm deg}^2$ for BGS1 through BGS5 respectively, with the largest areas reserved for the sparser high-redshift bins and the smallest for the two densest, lowest-redshift bins. Across all BGS redshift bins, we found no significant benefit to adopting different scales for $\ell_{p}$ and $\ell_{\|}$ and set the two equal. The results are shown in Figure~\ref{fig:bgs-snr}. Especially for lower redshifts, the scale choice does not have a dramatic impact on the signal amplitude above 1.5 $h^{-1}$Mpc.

The existence of a peak in SNR across varying linking scales reflects a competition between two effects. Increasing the linking scale preferentially builds larger objects, which trace the tidal field more strongly (a known effect for galaxies and halos), and galaxies with larger separations are less likely to influence each other's positions, which washes out external tidal information. Beyond a certain scale, however, the linking becomes loose enough to associate galaxies with progressively weaker physical connection, diluting the coherent orientation signal. The balance of these effects for a given galaxy sample sets the optimal scale. Notably, the linking lengths we find are considerably larger than those adopted in \cite{lammanDetectionLargescaleTidal2024} of $\ell_p = 1\,h^{-1}{\rm Mpc}$ and $\ell_{\|}=6\,h^{-1}{\rm Mpc}$, suggesting that large-scale structure imprints a coherent orientation across more loosely separated galaxies than previously utilized.

It is also somewhat surprising that we recover $\ell_{p}=\ell_{\|}$ for every sample except ELG. We attribute this to the fact that the optimal scales are large enough that redshift-space distortions are negligible, removing the primary motivation for an elongated LOS linking length. The exception is the ELG sample, where we find a transverse scale more than twice the LOS scale ($7$ versus $3\,h^{-1}{\rm Mpc}$). We interpret this difference as a consequence of sample variance; ELGs are the sparsest and most diffuse of our samples. We nonetheless retain the recovered values, as our aim here is to obtain the highest-SNR measurement possible rather than a physically uniform multiplet definition across samples.

Ideally, future analyses will similarly identify the optimal scales for each sample. We note that, with these samples, the optimal linking scale roughly scales with galaxy number density as $\ell\simeq \ell_0 - \tfrac{1}{4}\log n$. 

This selection is robust to cosmology: because we quote all separations in $h^{-1}$Mpc, $\ell$ does not depend on $H_0$, and only negligibly on the shape of the distance-redshift relation.

\vspace{.1in}
\subsection{Multiplet Shapes}\label{subsec:shapes}
The orientation of each multiplet is determined from the projected, plane-of-sky positions of its member galaxies. Using projected positions mitigates the effect of small-scale RSD on the inferred orientation. Following \cite{lammanDetectionLargescaleTidal2024}, the multiplet orientation $\Theta$ is the angle of the $N$-member-averaged complex position relative to the multiplet centroid,
\begin{equation}\label{eq:orientation}
    \epsilon_{\rm multiplet} = \frac{1}{N}\sum_{i=1}^{N} r_i\, e^{2i\theta_i} = a + bi,
    \qquad
    \Theta = \frac{1}{2}\arctan\frac{b}{a},
\end{equation} where $r_i$ and $\theta_i$ are the projected distance and position angle of member $i$ relative to the centroid. This is a ``stick'' model: each multiplet is characterized by orientation alone, with no axis-ratio information. 

The error of a shape alignment measurement scales with the `shape noise' of the shape sample, conventionally taken as the variance of one component of the complex ellipticity. For multiplets modeled as sticks, the orientation is a unit spin-2 phasor $e^{2i\theta}$ with $\theta$ uniformly distributed under the null hypothesis, so $\langle \cos^2 2\theta\rangle = \langle\sin^2 2\theta\rangle = 1/2$ and the per-component shape noise is $\sigma_s = 1/\sqrt{2}$, about three times higher than many galaxy samples.

To potentially bring this down, we try using the full shape information of multiplets. The full ellipticity $(E_1, E_2)$ projected perpendicular to the $\hat{z}$ line of sight can be computed from the second moments $Q$ of member positions about the multiplet center:
\begin{equation}
    E_1 + iE_2 = \frac{Q_{xx} - Q_{yy} + 2iQ_{xy}}{Q_{xx} + Q_{yy}}.
\end{equation}
This is the standard second-moment definition used to compute ellipticities from particle distributions \citep{bartelmannWeakGravitationalLensing2001}, and reduces to the orientation-only `stick' ($|E|=1$) when a multiplet has only two members. For a multiplet of $N$ members drawn from the same orientation distribution, the ellipticity is the mean of $N$ such phasors, so its per-component variance scales as $1/N$ and the shape noise as $\sigma_s \propto 1/\sqrt{N}$.

However, this also decreases the measured amplitude of multiplet alignment. The total SNR of a shape-tracer measurement scales with alignment amplitude $A$ and shape noise $\sigma_s$ as $A/\sigma_s$. When measuring alignment with the full ellipticity, we find the amplitude decreases faster than the shape noise: averaging member orientations within a multiplet washes out the alignment signal more than it suppresses the noise. The SNR is not improved by using the full ellipticity, so we retain the stick model. 

\begin{figure}
\centering
\includegraphics[width=.48\textwidth]{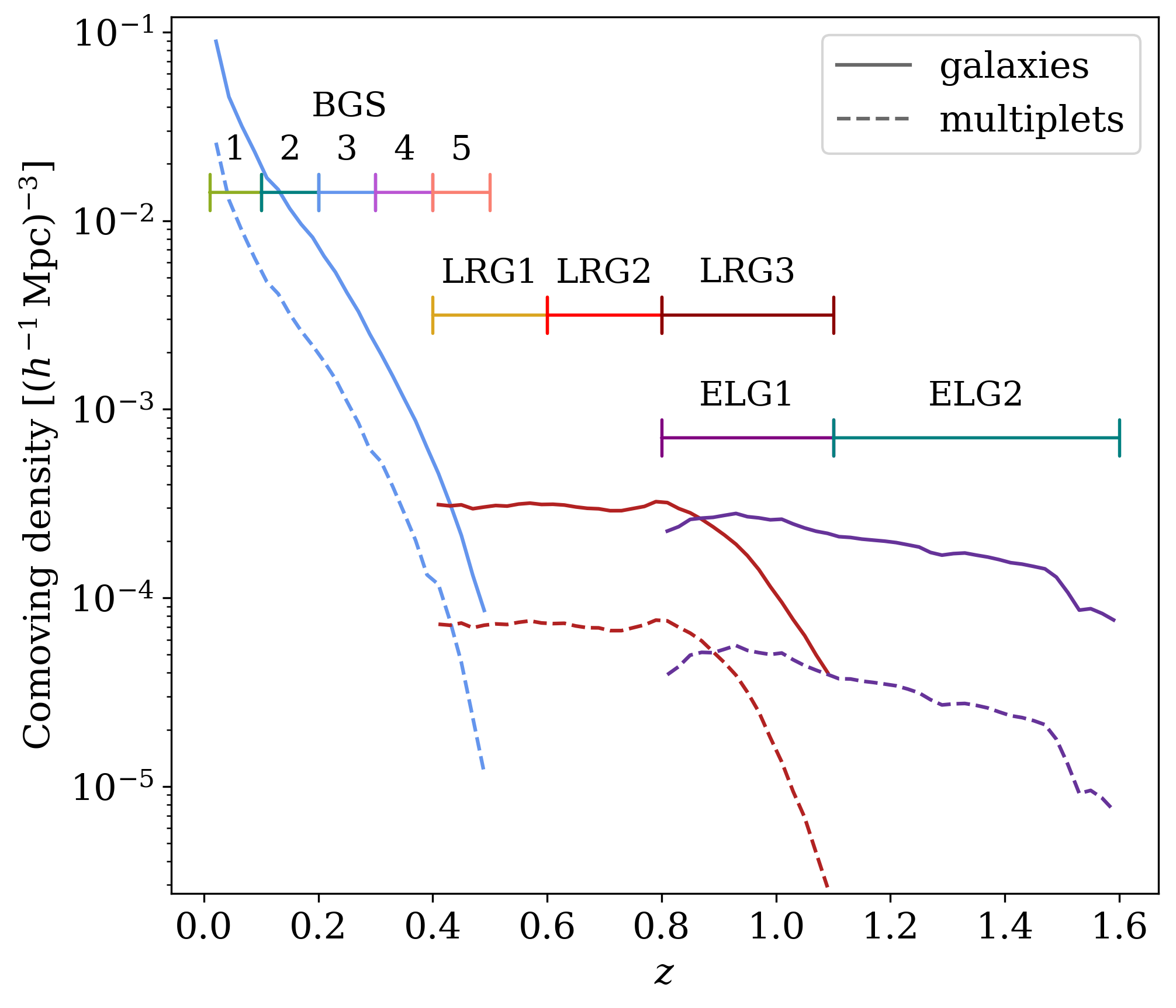}
\caption[]{Comoving number densities of galaxies (solid) and the multiplets identified within them (dashed) as a function of redshift, for the BGS, LRG, and ELG samples. The horizontal bars mark the redshift extent of each sample bin. The multiplet density tracks the galaxy density approximately linearly across all samples, remaining near $1/4$ of the galaxy density for BGS and LRG, $1/6$ for ELG.}
\label{fig:sample-densities}
\vspace{.2in}
\end{figure}

\begin{figure*}
\centering
\includegraphics[width=1\textwidth]{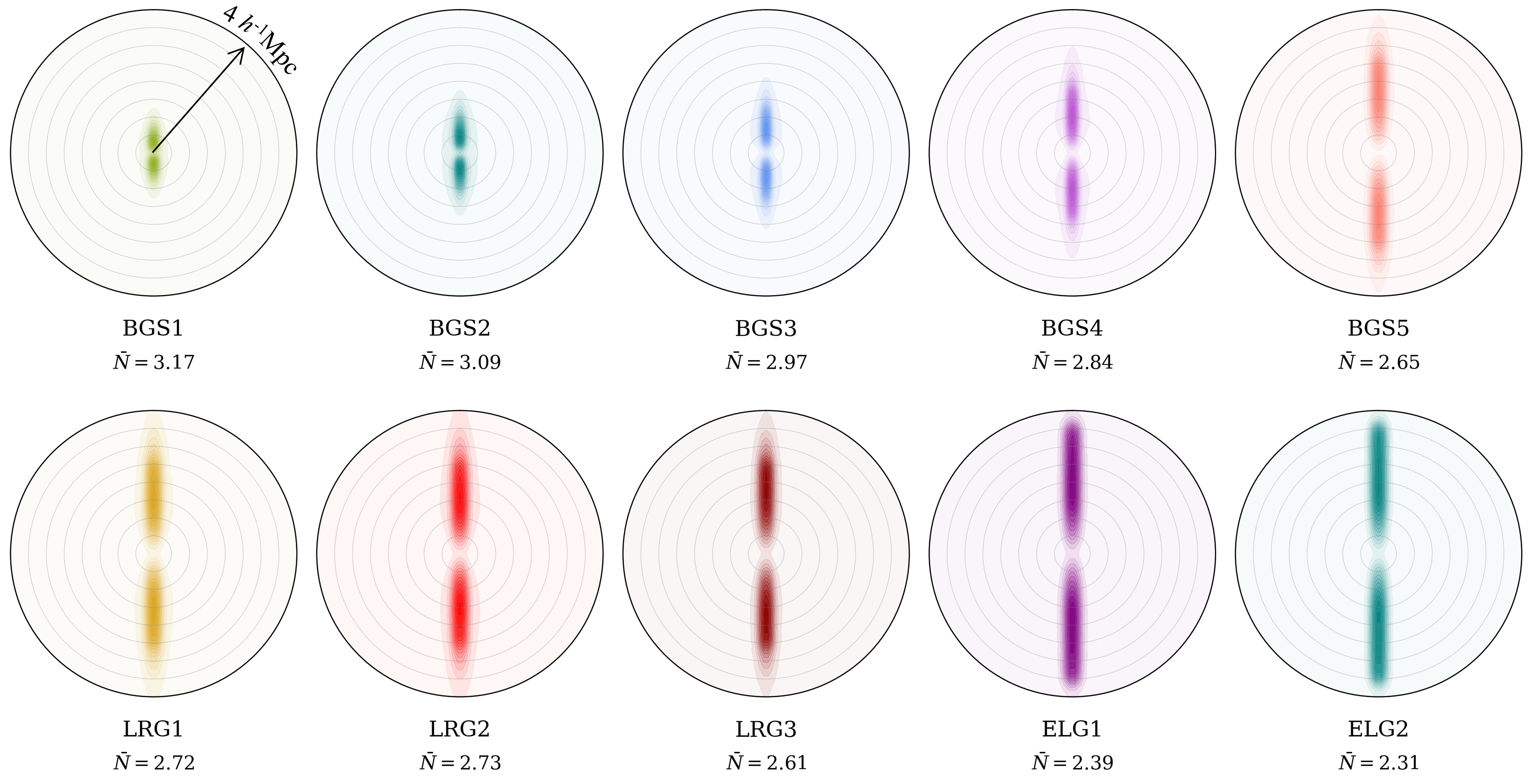}
\caption[]{Stacked visualization of the average multiplet shape in each galaxy sample and redshift bin, projected onto the plane of the sky. Member galaxies are rotated into each multiplet's own orientation frame before stacking, so the common elongation is preserved; density is mapped to opacity. All panels are shown out to a fixed comoving radius of $4\,h^{-1}\mathrm{Mpc}$ from the multiplet center, and the average number of members per multiplet $\bar{N}$ is shown below each panel. Because most multiplets are pairs, the stacks are dominated by elongated, stick-like shapes. However, even galaxies in $N>2$ multiplets fall closely along the multiplet's primary axis. The central gap reflects the minimum resolved pair separation, which increases with redshift. The overall size grows toward sparser, higher-redshift samples with larger linking lengths.}
\label{fig:shapes}
\vspace{.2in}
\end{figure*}

\subsection{Multiplet Demographics}\label{sec:demographics}
 Here we examine the number densities of multiplets, along with their sizes, shapes, and number of members. Although the physical nature of a multiplet is not directly relevant to the large-scale alignment measurement, their characterization is helpful in building physical intuition and astrophysical interpretations. 

To estimate number densities, the comoving volume of each sample was computed by treating its galaxies as tracers of the survey footprint. The RA/Dec positions were pixelized on a HEALPix grid ($\texttt{nside}=64$), and the number of occupied pixels multiplied by the pixel area gives the effective solid angle $\Omega$. Combined with the comoving distances at the bin's redshift edges, this yields the shell volume. The number density $n(z)$ for both galaxies and multiplets was then computed on a finer redshift grid within each bin using this same footprint estimate, ensuring the summary volumes and densities share consistent convention. 

The number densities of multiplets and the galaxy samples in which they were identified are shown in Table~\ref{tab:samples} and Figure~\ref{fig:sample-densities}. In BGS, the multiplet number density is consistently close to $1/4$ of the galaxy density, despite the linking length varying between the redshift subsamples. For LRG and ELG, the multiplet fraction is somewhat lower, roughly $1/5$ to $1/4$. Across all samples, the multiplet number density scales only slightly more steeply than linearly with galaxy number density. This is notable given that the number of candidate neighbors available to form multiplets scales as $n^2$, which would naively suggest a much steeper relation. The near-linear scaling instead reflects the clustering of galaxies: the close pairs that seed multiplets reside in overdense regions, which are already populated at low mean density, so increasing the overall density predominantly adds galaxies in underdense regions that do not form new links. This effect is compounded by our Union Find definition, which places no cap on membership. In the densest regions, added galaxies tend to grow existing multiplets rather than nucleate new ones, further flattening the relation.

We also visualize typical multiplet shapes in Figure~\ref{fig:shapes}. For each sample and redshift bin, we constructed a stacked image of multiplet shapes as follows. Each multiplet's member galaxies were first rotated into a frame aligned with that multiplet's own orientation axis, defined in Section \ref{subsec:shapes}. The rotated members of every multiplet in a given subsample were then pooled into a single point cloud of comoving separations from the multiplet centroid. The two-dimensional density of this point cloud was estimated via a Gaussian-smoothed histogram in Cartesian coordinates and resampled onto a polar grid for display. Density was mapped to opacity at fixed hue, shown out to a fixed comoving radius of $4\,h^{-1}\mathrm{Mpc}$. The average number of galaxies per multiplet for each sample is shown below its corresponding panel. Histograms of the number of multiplet members for the DESI DR1 sample are shown in \cite{lammanDetectionLargescaleTidal2024}, and are qualitatively unchanged besides a larger linking length resulting in a larger average number of members.

The majority of multiplets are pairs, which dominate the stacked shapes and give them their characteristic elongated, stick-like appearance. However, galaxies in multiplets with more than two members also tend to align closely with the multiplet's primary axis. The gap in the center of each panel is set by the minimum physical separation of pairs resolved in the survey, which grows with redshift. Many multiplets extend well beyond the scale of a typical halo or cluster ($R_{200}\sim1$–$2\,h^{-1}\mathrm{Mpc}$ for massive clusters). This, combined with their distinct elongation, suggests that the typical multiplet more closely resembles a filamentary structure, though groups and clusters are also captured in the multiplet catalogs. We also emphasize that the physical meaning of a multiplet varies somewhat across galaxy samples and redshift bins, depending on the number density of the parent sample and chosen linking length.

\section{Multiplet Alignment}\label{sec:formalism}
\subsection{Alignment Estimators}
\label{sec:estimators}
We measure the alignment of multiplet orientations with respect to a density tracer ($\mathcal{E}_+$), and the autocorrelation of multiplet orientations with each other ($\mathcal{E}_{++}$). The former follows \cite{lammanDetectionLargescaleTidal2024}; we summarize the relevant points here.

The cross-correlation estimator averages the projected ellipticity $\epsilon_+(j|i)$ of each shape $i\in S$ relative to each tracer $j\in D$,
\begin{equation}\label{eq:rele}
    \mathcal{E}_+(r) = \frac{1}{N_{\rm pairs}}\sum_{i \in S,\, j \in D} \epsilon_+(j|i)
    = \frac{S_+D(r)}{DD(r)},
\end{equation}
where $\epsilon_+$ quantifies the multiplet orientation relative to the axis connecting the multiplet and the tracer, $\theta$, as 
\begin{equation}
\epsilon_+=\cos2\theta_i.
\end{equation}
A positive value indicates tidal alignment. $S_+D$ represents the number of alignment-weighted shape-tracer pairs and $DD$ the number of tracer-tracer pairs. This is related to the shape-density correlation function $\xi_{\rm g+}$, a generalized form of the Landy-Szalay estimator \citep{mandelbaumDetectionLargescaleIntrinsic2006},
\begin{equation}\label{eq:xi_gp}
    \xi_{\rm g+} = \frac{S_+D - S_+R_D}{R_SR_D} \approx \frac{S_+D}{R_SR_D}.
\end{equation}
The key difference is that $\xi_{\rm g+}$ is not averaged on a per-pair basis. The two estimators are related by
\begin{equation}\label{eq:rele_to_xi}
    \xi_{\rm g+}(s) = \frac{DD}{R_SR_D}\,\mathcal{E}_+(s)
    = \big(1+\xi(s)\big)\,\mathcal{E}_+(s).
\end{equation}
The signal is projected along the line of sight up to a maximum separation $\Pi_{\rm max}$. Following \cite{lammanOptimalIntrinsicAlignment2025}, we adopt a variable $\Pi_{\rm max}$ that depends on the bin of transverse separation: $\Pi_{\rm max}(R_{\rm bin}) = [8 + \frac{2}{3}R_{\rm bin}]$ $h^{-1}$Mpc. This optimizes signal-to-noise and captures comparable information to 3D estimators, while remaining simpler to measure and model. We do not apply the full Gaussian LOS weighting introduced in that work, as the marginal gain in signal-to-noise does not justify the additional computational cost for this analysis.

In this work we additionally measure the multiplet orientation autocorrelation, $\mathcal{E}_{++}$. For a pair of multiplets $i$ and $j$, we project each orientation onto their connecting separation vector and correlate the two projected components as a function of transverse separation $R$,
\begin{equation}\label{eq:rele_auto}
    \mathcal{E}_{++}(R) = \frac{1}{N_{\rm pairs}}
    \sum_{i,\,j \in S} \epsilon_+(j|i)\,\epsilon_+(i|j)
    = \big\langle \cos 2\theta_i^{(R)}\, \cos 2\theta_j^{(R)} \big\rangle,
\end{equation}
where $\theta_i^{(R)}$ and $\theta_j^{(R)}$ are the orientation angles of the two multiplets relative to the vector connecting them. This is the orientation-only analog of the projected shape--shape correlation $w_{++}$ commonly measured for galaxy ellipticities \citep{singhIntrinsicAlignmentsBOSS2016, troxelIntrinsicAlignmentGalaxies2015}, with the multiplet ellipticity modulus folded into the amplitude parameter introduced in Section~\ref{sec:modeling}. As with $\mathcal{E}_+$, the signal is projected along the LOS using the same variable $\Pi_{\rm max}$.

Unlike the cross-correlation between multiplet orientations and the density field, the orientation autocorrelation does not require a random catalog. In the cross-correlation, randoms serve to normalize the summed shears by the available pair counts and to subtract the spurious alignment induced by the survey footprint. Neither role applies here. Both quantities entering the autocorrelation are shape orientations, which have zero mean and carry no monopole sourced by the mask. 

\begin{table*}
\centering
\begin{tabular}{lcccccccc}
\hline
Sample & $z$ range & Volume & $\ell_p$ & $\ell_\|$ & $N_{\rm galaxies}$ & $N_{\rm multiplets}$ & $-\tau$ & $A_{\rm IA}$ \\
 & & [Gpc$^3 h^{-3}$] & [$h^{-1}$Mpc] & [$h^{-1}$Mpc] & [$10^{6}$] & [$10^{6}$] & [$10^{-2}$]& \\
\hline

BGS1 & 0.01 - 0.1 & 0.04 & 1.5 & 1.5 & 1.19 & 0.33 & 4.05 $\pm$ 0.45 & 8.64 $\pm$ 0.96 \\
BGS2 & 0.1 - 0.2 & 0.25 & 2 & 2     & 2.74 & 0.75 & 6.07 $\pm$ 0.39 & 10.67 $\pm$ 0.69 \\
BGS3 & 0.2 - 0.3 & 0.61 & 2.5 & 2.5 & 2.49 & 0.65 & 7.60 $\pm$ 0.17 & 11.09 $\pm$ 0.25 \\
BGS4 & 0.3 - 0.4 & 1.07 & 3.5 & 3.5 & 1.24 & 0.30 & 9.65 $\pm$ 0.49 & 11.92 $\pm$ 0.61\\
BGS5 & 0.4 - 0.5 & 1.58 & 5 & 5 & 0.36 & 0.08 & 11.42 $\pm$ 1.76 & 12.19 $\pm$ 1.88 \\
\hline

LRG1 & 0.4 - 0.6 & 3.44 & 5 & 5 & 1.07 & 0.24 & 11.80 $\pm$ 0.09 & 11.37 $\pm$ 0.09 \\
LRG2 & 0.6 - 0.8 & 5.41 & 5 & 5 & 1.64 & 0.37 & 13.62 $\pm$ 0.27 & 10.41 $\pm$ 0.21 \\
LRG3 & 0.8 - 1.1 & 11.3 & 5 & 5 & 1.83 & 0.35 & 14.66 $\pm$ 0.38 & 9.04 $\pm$ 0.23 \\
\hline

ELG1 & 0.8 - 1.1 & 11.3 & 7 & 3 & 2.85 & 0.50 & 5.51 $\pm$ 0.24 & 3.34 $\pm$ 0.15 \\
ELG2 & 1.1 - 1.6 & 24.8 & 7 & 3 & 3.84 & 0.56 & 7.86 $\pm$ 0.17 & 3.68 $\pm$ 0.08 \\
\hline
\end{tabular}
\caption{Properties of our multiplet samples and the galaxy catalogs they were identified in. The last two columns are our reported values for the alignment strength of the multiplets, as measured by the multiplet-galaxy cross-correlation. Shown is both our parametrization of the tidal response, $\tau$, and the commonly-used alignment amplitude for the intrinsic alignments of galaxies, $A_{\rm IA}$. Errors on both of these scale-independent amplitudes come from the variance in estimating them in 12 logarithmically spaced $R$ bins between 20-100 $h^{-1}$Mpc.}
\label{tab:samples}
\end{table*}

\subsection{Modeling}\label{sec:modeling}

We model the multiplet alignment signal by assuming a linear relationship between orientations and the large-scale tidal field, following the nonlinear alignment (NLA) framework \citep{bridleDarkEnergyConstraints2007} as applied to multiplets in \cite{lammanDetectionLargescaleTidal2024}. The projected ellipticity is related to the traceless tidal tensor $T_{\alpha\beta}$ by a single amplitude parameter $\tau$,
\begin{equation}\label{eq:eps_tau}
    \epsilon_{\alpha\beta} = \tau\left(T_{\alpha\beta} + \tfrac{1}{2}T_{zz}\right),
\end{equation}
where $\tau$ encodes the response of multiplet orientations to the tidal field, including any contribution from the full-shape information and any systematic misalignment, assuming neither introduces scale dependence at large separations. For the cross-correlation, this yields the model prediction derived in \citep{lammanDetectionLargescaleTidal2024}
\begin{equation}\label{eq:model_gI}
    \mathcal{E}_+(R) = \frac{-\tau}{2\Pi_{\rm max} + \bar{w}_p}
    \int K\,dK\, \mathcal{J}_2(K,R)\,\mathcal{P}_\Pi(K),
\end{equation}
where $\mathcal{J}_2$ is the bin-averaged second-order Bessel function of the first kind, $\mathcal{P}_\Pi(K)$ is the LOS-projected matter power spectrum weighted by the tidal factor $K^2/k^2$, and $\bar{w}_p$ is the bin-averaged projected cross-correlation between the multiplet and tracer catalogs. The power spectrum enters as $b_g P_{mm}(k)$, with $b_g$ the galaxy bias of the tracer sample.

We characterize the alignment amplitude with $\tau$ rather than the more common $A_{\rm IA}$ used in most weak-lensing-related works \citep{bridleDarkEnergyConstraints2007, joachimiConstraintsIntrinsicAlignment2011, johnstonKiDSGAMAIntrinsic2019}, as $\tau$ is a more natural characterization of the pure tidal response, although we report both here. The two are related by
\begin{equation}\label{eq:tau_to_AIA}
    A_{\rm IA}(z) = -\frac{\tau}{C_1}\,\frac{D(z)}{\rho_{m}(z)},
\end{equation}
where $C_1 = 5\times10^{-14}\,M_\odot^{-1}h^{-2}{\rm Mpc}^3$ \citep{brownMeasurementIntrinsicAlignments2002}, $\rho_{m}$ is the matter density, and $D(z)$ is the growth factor normalized such that $\bar{D}(z) = (1+z)D(z)$ is unity at $z=0$. Note that here we have changed the $D(z)$ normalization from \cite{lammanDetectionLargescaleTidal2024}, which normalized at matter domination, to match the majority of recent works. Since $A_{\rm IA}$ contains the growth factor while $\tau$ does not, the assumed cosmology enters directly into $A_{\rm IA}$ and is another reason to adopt $\tau$ as our primary parameterization. 

The autocorrelation $\mathcal{E}_{++}$ follows from the same framework with two modifications. First, because both fields being correlated are shapes rather than one shape and one density tracer, the amplitude enters quadratically: $\mathcal{E}_{++} \propto\tau^2$, and the galaxy bias does not appear, since neither field is a density tracer. The relevant power spectrum is therefore $P_{mm}(k)$, with the tidal factor entering squared, $(K^2/k^2)^2$. Second, the projection geometry changes. Correlating two spin-2 (shape) fields, rather than a shape and a scalar, replaces the $\mathcal{J}_2$ kernel of the cross-correlation with the sum $\mathcal{J}_0 + \mathcal{J}_4$. This is the standard projection for shape--shape correlations \citep{singhIntrinsicAlignmentsBOSS2016, hervaspetersUNIONSDirectMeasurement2025}. The bin-averaged Bessel kernels are defined as in \cite{lammanDetectionLargescaleTidal2024} for $\mathcal{J}_2$, with $\mathcal{J}_0$ and $\mathcal{J}_4$ following the same convention. The $\mathcal{J}_0$ term makes $\mathcal{E}_{++}$ more isotropic in the $(r_{\rm p}, \Pi)$ plane than $\mathcal{E}_+$, while the $\mathcal{J}_4$ term retains the transverse dependence. We confirm that these bin-averaged transforms converge with respect to the upper integration limit. The model prediction is then
\begin{multline}\label{eq:model_II}
    \mathcal{E}_{++}(R) = \\
    \frac{\tau^2}{2\Pi_{\rm max} + \bar{w}_p}
    \int K\,dK\, 2\big[\mathcal{J}_0(K,R) + \mathcal{J}_4(K,R)\big]\,
    \mathcal{P}_\Pi^{(2)}(K),
\end{multline}
where $\mathcal{P}_\Pi^{(2)}$ carries the squared tidal factor $(K^2/k^2)^2$ and $\bar{w}_p$ is the bin-averaged multiplet--multiplet projected correlation function. We do not include a factor of $\frac{1}{2}$ in this expression (for a per-multiplet quantity) in order to be consistent with the $\mathcal{E}_+$ estimator, which is defined as an average over pairs. 

Crucially, if $\tau$ is determined from the better-constrained cross-correlation, Equation~\ref{eq:model_II} is a parameter-free prediction of the autocorrelation amplitude: there is no free normalization, and the galaxy bias does not enter. If, instead, $\tau$ is measured from the autocorrelation, the cross-correlation provides an independent measurement of the galaxy bias. In the following sections, we adopt the $\tau$ value from the $\mathcal{E}_+$ fit and overlay the resulting $\tau^2$ prediction on the measured $\mathcal{E}_{++}$ as a consistency check of the alignment framework. We additionally fit $\tau$ directly to $\mathcal{E}_{++}$, compare the two amplitude values, and check consistency with the predicted galaxy bias.

\vspace{.1in}
\section{Measurements}\label{sec:measurement}

\subsection{Measurement Method}
We measure both alignment estimators of Section~\ref{sec:estimators} with the pipeline introduced in \cite{lammanDetectionLargescaleTidal2024}, updated for the larger DR2 samples. The code is publicly available in the \href{https://github.com/cmlamman/spec-IA}{\texttt{spec-IA}}\footnote{\href{https://github.com/cmlamman/spec-IA}{github.com/cmlamman/spec-IA}} repository. We summarize the procedure here and describe the changes made for this work.

We compute each projected estimator in bins of transverse separation, where larger bins have larger $\Pi_{\rm max}$, as defined in Section~\ref{sec:estimators}. The measurement is carried out independently in $N_{\rm reg}$ sky regions of equal multiplet count, divided in right ascension and declination. Within each region, the multiplet orientations are measured relative to the full tracer sample, or full multiplet sample for autocorrelations. The average signal and errors are then estimated from these regions. We increase the number of regions relative to \cite{lammanDetectionLargescaleTidal2024}, using $N_{\rm reg}=2500$ for BGS2 (which contains the largest number of multiplets), $N_{\rm reg}=2000$ for the other BGS samples, and $N_{\rm reg}=1000$ for LRG and ELG. The larger count for BGS is driven by its much higher density, both to keep the per-region pair count tractable in memory and because the denser sample supports a finer regional division at fixed statistical precision. Alongside this, we made minor optimizations to the pair-counting code; the measured signal is unchanged from the original implementation.

Errors are estimated by bootstrap resampling of the regions: regions are drawn with replacement $N_{\rm boot}=10^5$ times and the per-bin uncertainty is the standard deviation of the resampled mean. The same resampling defines the covariance,
\begin{equation}
\mathsf{C} = \mathrm{Cov}\!\left[\bar{\mathbf{d}}^{(b)}\right]_{b=1}^{N_{\rm boot}},
\label{eq:cov}
\end{equation}
where $\bar{\mathbf{d}}^{(b)}$ is the mean over the $b$-th resampling, so error bars and covariance follow from a single procedure.

For some analysis, we combine the measurements for each tracer across their redshift subsamples. For this, we use an amplitude-based weighting: each subsample is divided by its best-fit alignment amplitude $\tau$ before the inverse-variance combination. This weighting has little effect for the LRG and ELG samples, whose subsamples have similar amplitudes, but modestly improves the combined BGS signal-to-noise. When combining, the bins are disjoint and treated as independent. They are combined at the covariance level: the combined inverse covariance is $\mathsf{C}_{\rm comb}^{-1} = \sum_z \mathsf{C}_z^{-1}$ and the combined signal is $\bar{\mathbf{d}}_{\rm comb} = \mathsf{C}_{\rm comb}\sum_z \mathsf{C}_z^{-1}\bar{\mathbf{d}}_z$. Inverse covariances are debiased with the factor $(N_{\rm reg}-N_{\rm bin}-2)/(N_{\rm reg}-1)$ before combination \citep{hartlapWhyYourModel2007}.

We validated the pipeline against an independent implementation built on \textsc{TreeCorr} \citep{jarvisSkewnessApertureMass2004}, finding agreement in the measured signal. We nonetheless retain our own code for the production measurements, as \textsc{TreeCorr} does not natively support the variable $\Pi_{\rm max}(R_{\rm bin})$ binning or the three-dimensional pair selection used here, and offers no sufficient speed advantage once these are imposed.

\begin{figure*}
\centering
\includegraphics[width=.33\textwidth]{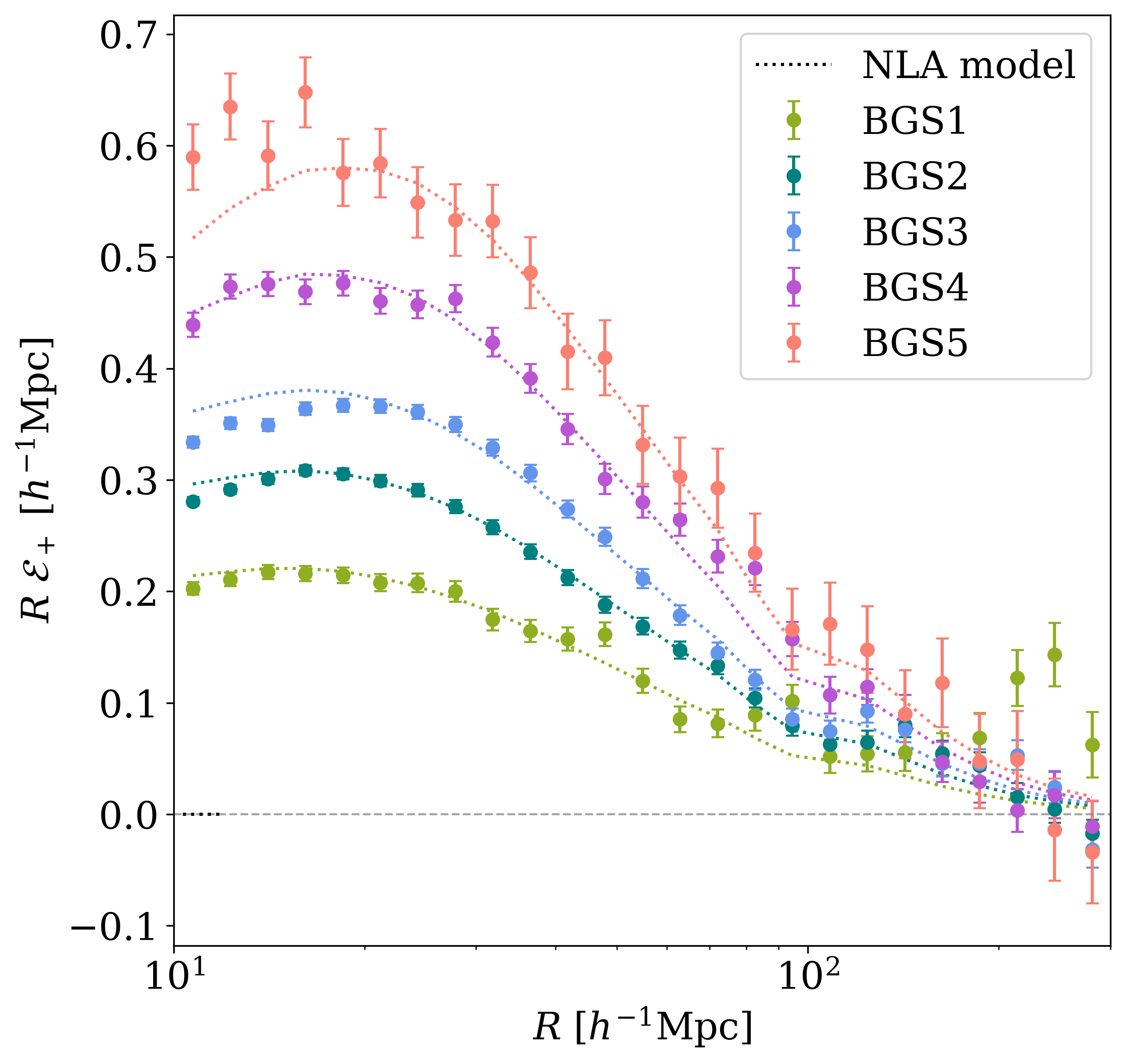}
\hfill
\includegraphics[width=.33\textwidth]{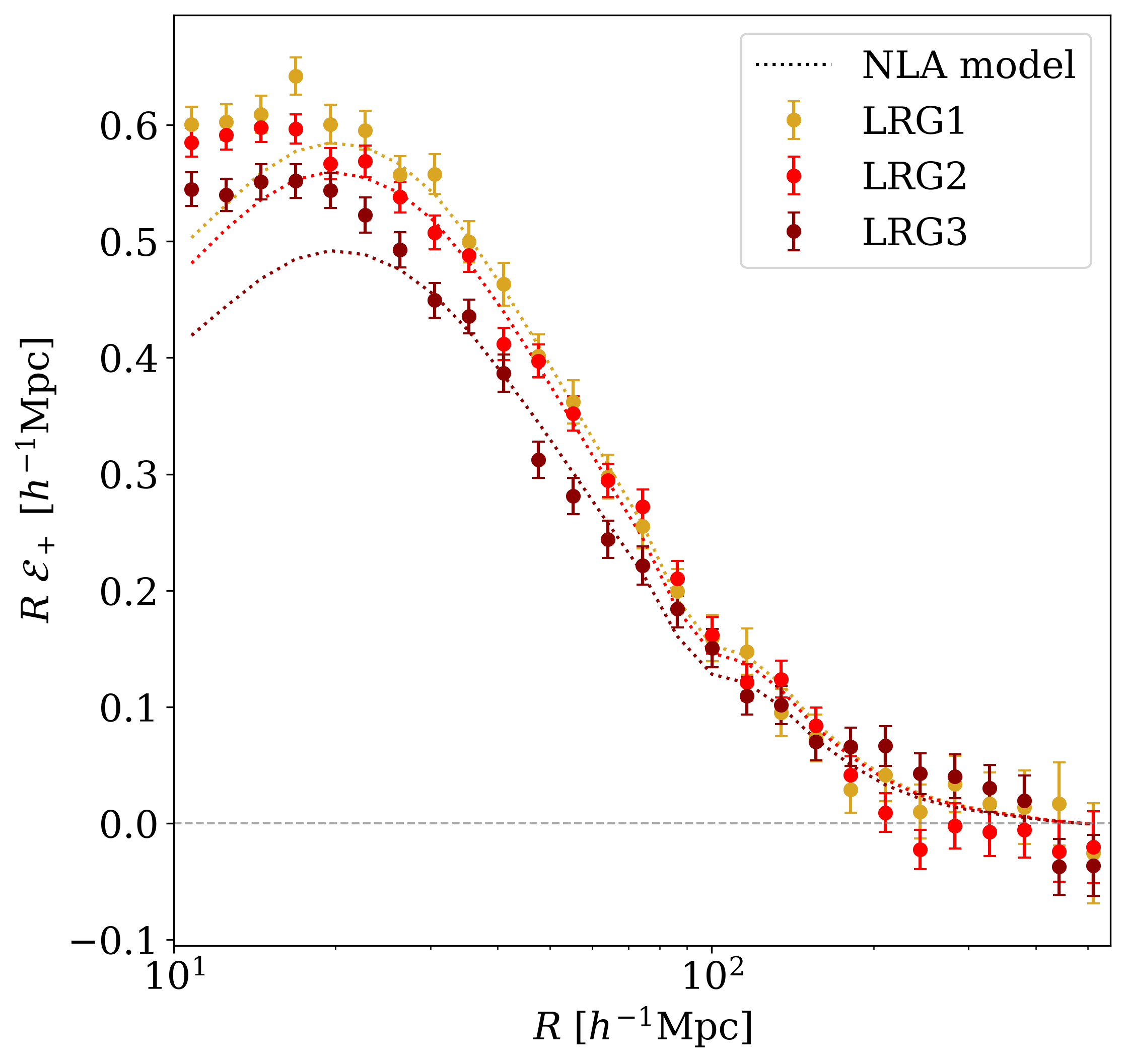}
\hfill
\includegraphics[width=.33\textwidth]{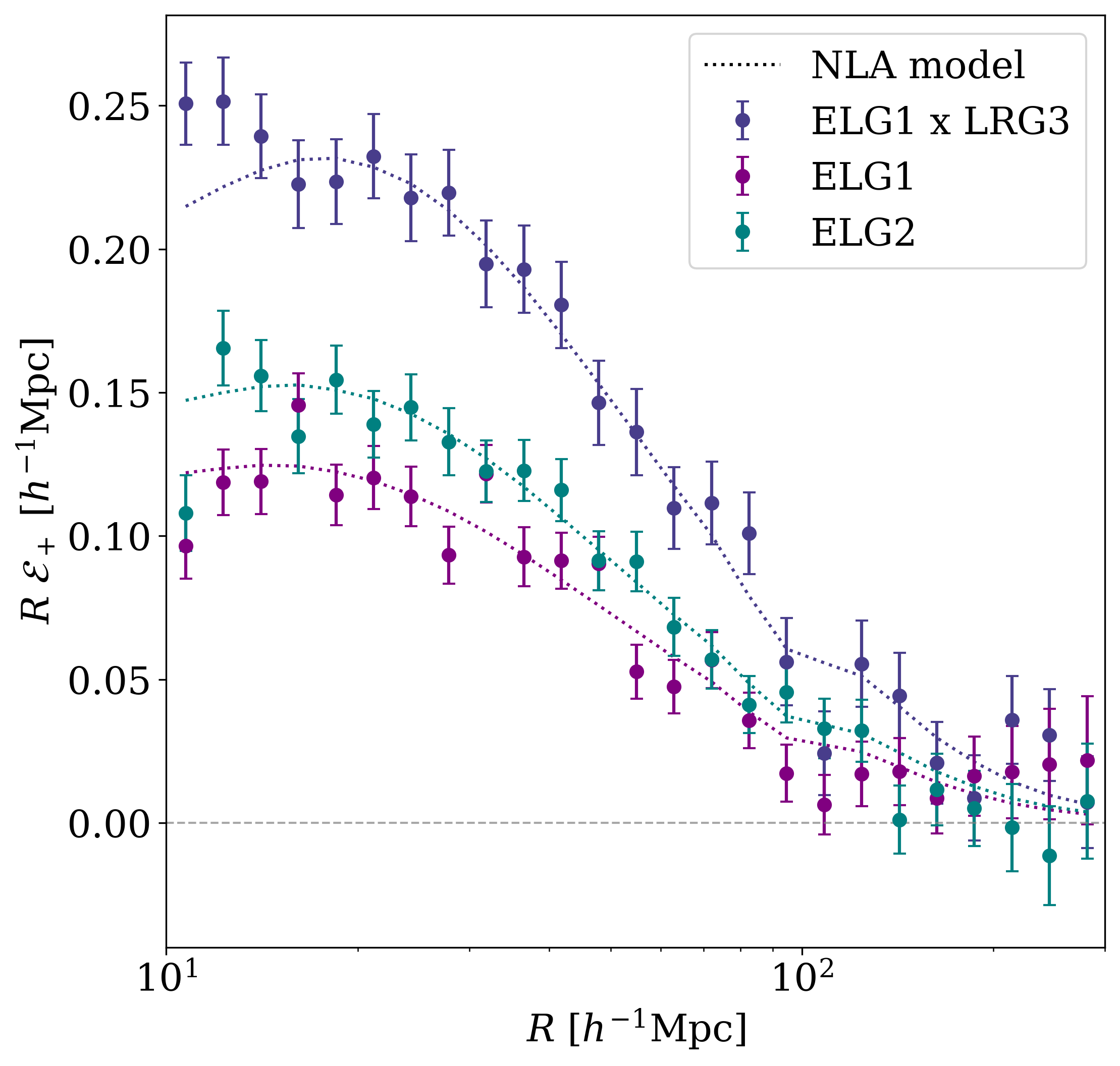}
\caption[]{Correlations between the projected shape of multiplets and the surrounding matter field, as traced by the same galaxy samples they were identified in. This is measured as function of projected separation between multiplets and tracers, $R$, and with an $R$-dependent LOS distance, described in Section \ref{sec:estimators}. Within a given $R$ bin, a larger value corresponds to stronger tidal alignment. The signals are compared to the data-normalized prediction of the nonlinear alignment model (NLA), shown in dotted lines. The BGS measurements (left) have a high correlation with galaxy clustering, which increases with redshift across the BGS sample. Measurements from the LRG sample (center) are more consistent across redshift bins as the sample is designed to have a constant co-moving number density. Measurements made with ELGs (right) show the most striking improvement over DR1, and are especially notable given that the measured alignment of individual ELGs is consistent with 0. The right panel also displays the cross-correlation of ELG multiplets with LRG tracers. The increased signal demonstrates the dependence of signal amplitude on galaxy clustering, however it is still half of the amplitude of LRG multiplets measured relative to the same tracer sample. This demonstrates the tendency of ELG multiplets to fall in less dense regions and is discussed more in Section \ref{sec:amp-redshift}.}
\label{fig:crosscorrs}
\vspace{.2in}
\end{figure*}

\begin{figure*}
\centering
\includegraphics[width=.33\textwidth]{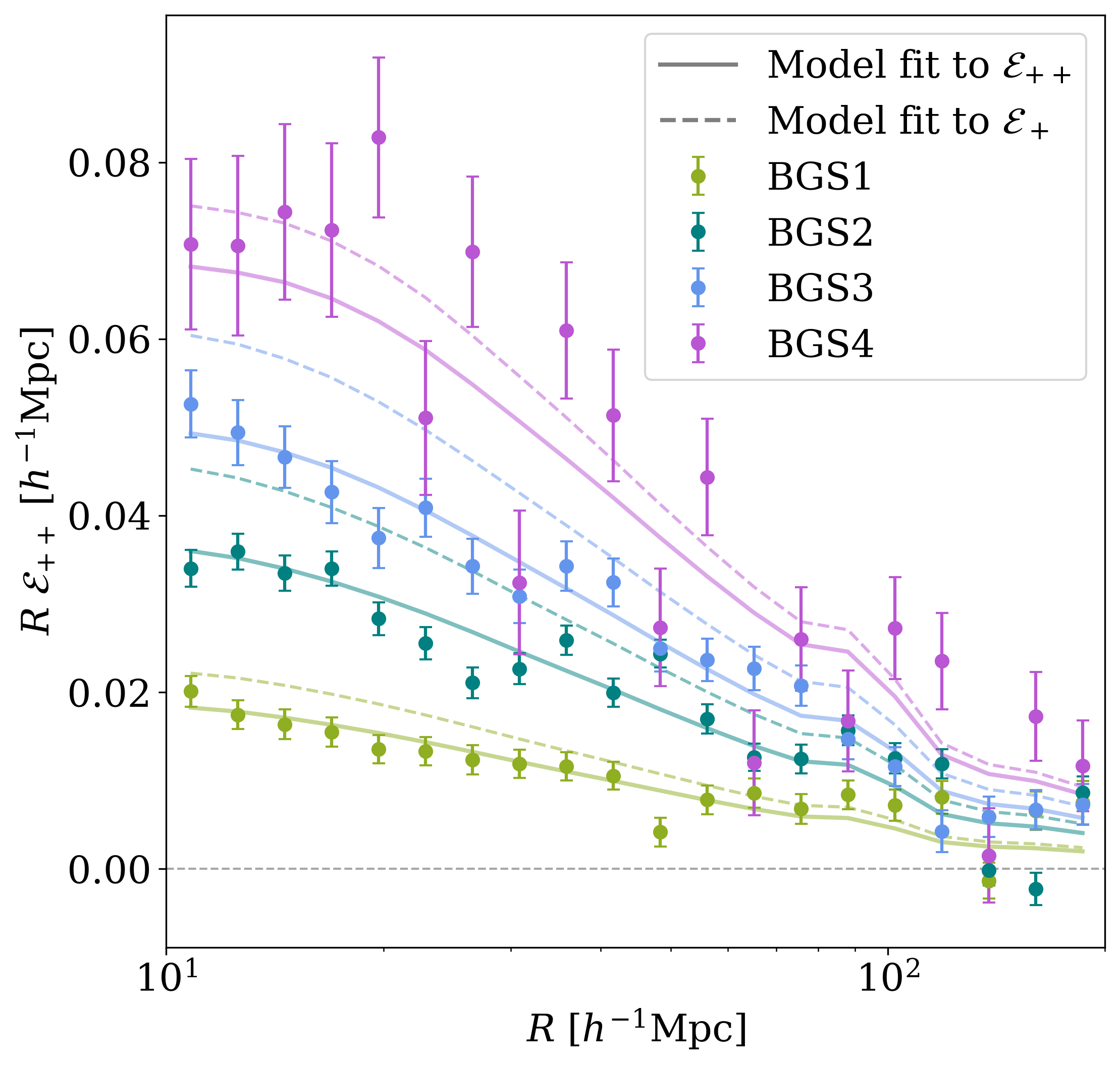}
\hfill
\includegraphics[width=.33\textwidth]{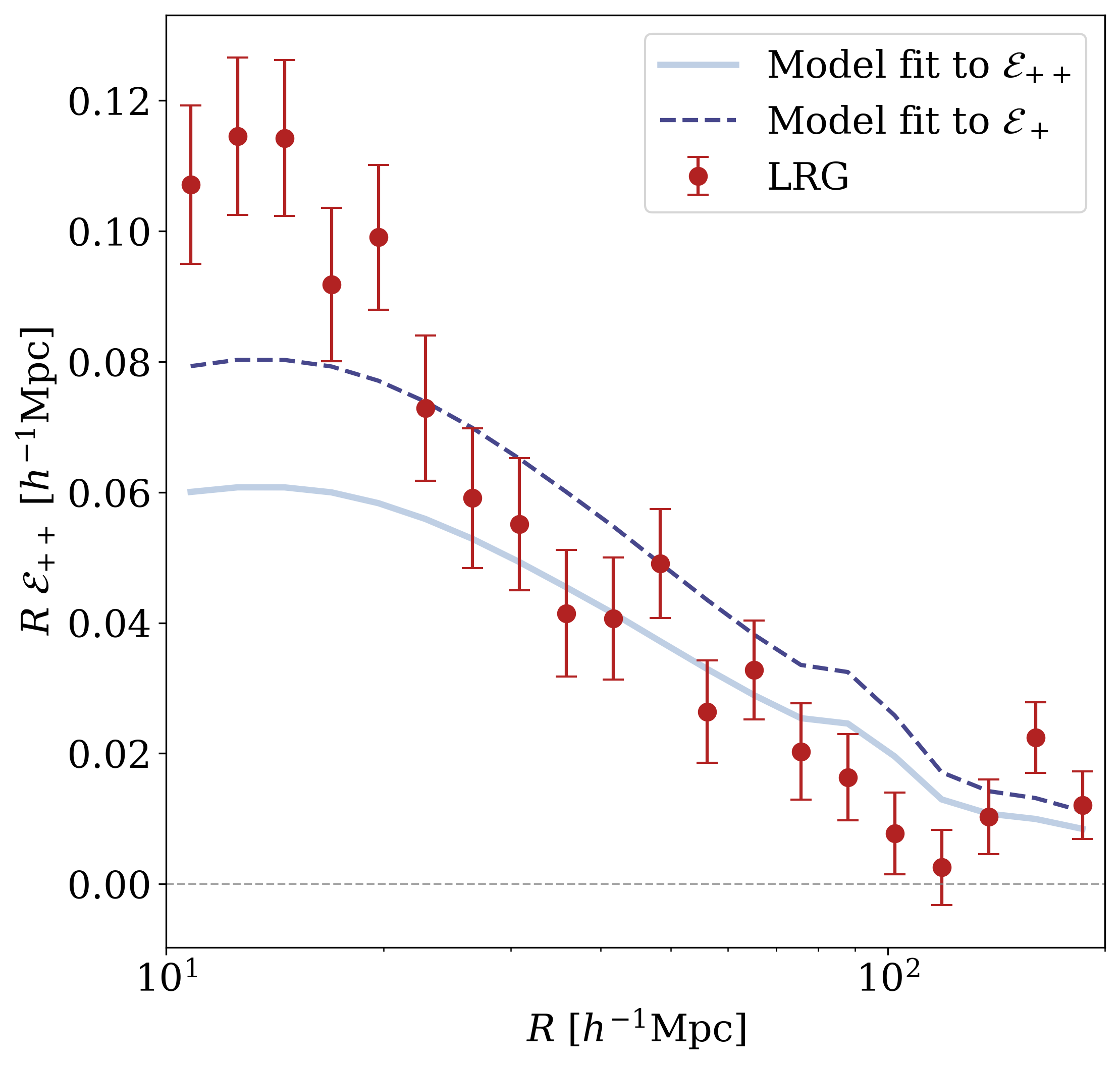}
\hfill
\includegraphics[width=.33\textwidth]{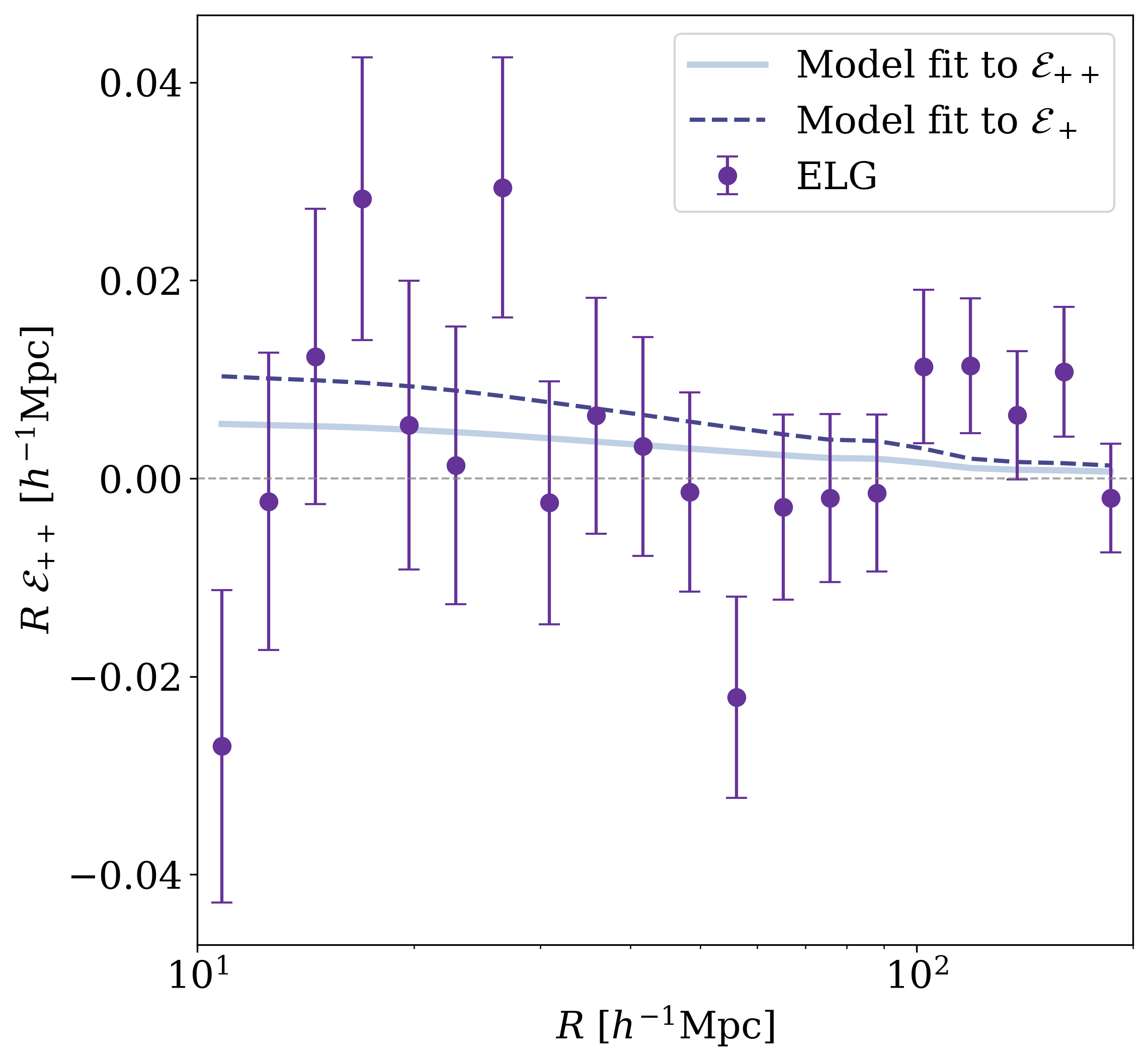}
\caption[]{The autocorrelation of multiplet shapes in the BGS, LRG, and ELG samples, measured as a function of projected separation $R$ and the same variable LOS distance used in Figure \ref{fig:crosscorrs}. The multiplet autocorrelation is a weaker signal than the cross-correlation because shape-shape correlations are intrinsically weaker, and there are far fewer multiplets than individual galaxies, which are used in the cross-correlation. This is especially apparent in less dense regions, such as ELGs, where the signal is marginally detectable. The highest redshift BGS bin is similarly noisy and has been removed from this figure for clarity.
The measurements of LRG and ELG are averaged over the redshift bins within each sample, and all measurements are compared to the NLA model prediction made with two amplitudes. The first is directly fit to the measured autocorrelation, $\mathcal{E}_{++}$, shown in a faint solid line. The second is the amplitude fit to the cross-correlation, $\mathcal{E}_+$, shown in a dashed line. The model fit to the autocorrelation amplitude is systematically smaller, likely due to an underestimated galaxy bias (Section \ref{subsec:autocorr_measurement}).}
\label{fig:autocorrs}
\vspace{.2in}
\end{figure*}

\subsection{Cross-correlations}

For the cross-correlation $\mathcal{E}_+$, we compute the projected orientation of each multiplet relative to each tracer as described in Section~\ref{sec:estimators}. Unless otherwise stated, the tracer catalog is the same catalog used to identify the multiplets; the exception is the overlapping LRG and ELG region at $0.8 < z < 1.1$, where we cross-correlate the two samples.

 For every measurement, we also measure the multiplet orientations relative to DESI's random catalogs, which are constructed to match the survey geometry, and subtract this from the signal. The randoms subtraction differs from \cite{lammanDetectionLargescaleTidal2024}. Previously, the mean random signal was subtracted from the mean measured signal after averaging over regions. Here we instead subtract the randoms on a per-region basis before averaging. The two procedures yield the same combined signal, but the per-region subtraction removes the coherent, geometry-sourced scatter between regions before the variance is estimated. The resulting errors are more accurate and smaller at large separations, where the geometric term contributes most to the region-to-region variance. 

 The resulting cross-correlation measurements and NLA model fit are shown in Figure \ref{fig:crosscorrs}. Across all samples, we see a significant improvement from the DR1 measurements presented in \cite{lammanDetectionLargescaleTidal2024}. This is partially due to improved measurement methods, but mostly due to more data and adopting a larger linking length when identifying multiplets. This especially improved the ELG signal, which had the largest increase in galaxy density from DR1 and, as the most diffuse sample, had the most to gain from a larger maximum linking length. Compared to DR1, the number of BGS and LRG galaxies roughly doubled, while ELG increased by a factor of about 2.5. The number of multiplets in each sample, largely due to the new definition, increased by factors of 3 for BGS, 9 for LRG, and 25 for ELG. 

Measurements from BGS and ELG are consistent with the NLA model in the large-scale linear regime, generally agreeing over the scales it is expected to be valid, $R>10h^{-1}$Mpc \citep{chisariRisingTideIntrinsic2025}. However, \cite{lammanDetectionLargescaleTidal2024} found NLA to break down for LRGs starting below $30h^{-1}$Mpc, which is similarly reflected in DR2 LRGs. This could be because density is correlated with alignment strength in ways that NLA does not capture \citep{ichikawaEnvironmentalDependenceHalo2026}. In particular, the observed shape field carries a density weighting factor of $(1+\delta)$, which our $1/(1+\xi)$ normalization (Equation \ref{eq:rele_to_xi}) can only partially account for. This increases the minimum valid NLA scales for samples with higher clustering (Shi et al., in prep), and would be especially pronounced for highly biased tracers like multiplets. 

 Within the BGS sample, signal amplitude is highly correlated with redshift. This is largely due to the clustering differences between redshift bins, as the galaxy bias increases significantly when fainter galaxies drop out of the more distant bins. Because of this, the maximum linking length set for identifying multiplets also changes across the bins (Table \ref{tab:samples}). This difference in clustering and multiplet definition is accounted for in the alignment amplitudes discussed in Section \ref{sec:amp-redshift}; this plot should not be used to infer any redshift dependence. Conversely, the LRG and ELG catalogs have a more constant number density (Figure \ref{fig:sample-densities}) and identical multiplet definitions within each sample, so the amplitude of the correlations are similar across redshift bins. 

 The overlap of LRGs and ELGs in the redshift range $0.8<z<1.1$ provides an interesting opportunity to directly compare the alignment of multiplets made with galaxies of different morphologies, so the cross-correlation of ELG multiplets with LRG tracers is also shown in Figure \ref{fig:crosscorrs}. This is discussed further in Section \ref{sec:amp-redshift}.

\begin{figure*}
\centering
\includegraphics[width=1\textwidth]{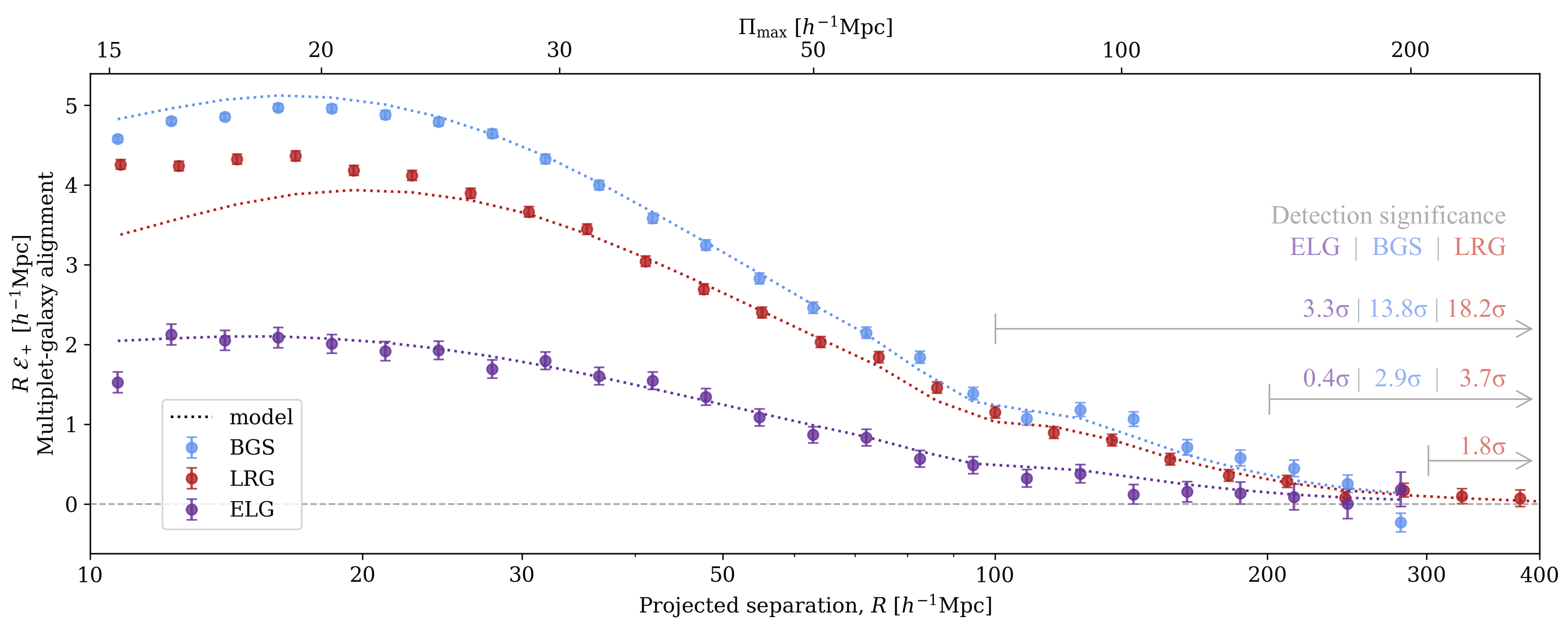}
\caption[]{Intrinsic tidal shear measured with galaxy multiplets in DESI's DR2 BGS, LRG, and ELG samples. This is the largest-scale detection of its kind. The measurements are made as a function of projected separation $R$, with a $\Pi_{\rm max}$ value that increases with $R$ (shown on the top axis). They are the signals in Figure \ref{fig:crosscorrs}, combined across redshift bins, and compared to NLA model predictions. The significance of each signal being detected above 100, 200, and 300 $h^{-1}$Mpc is shown on the plot. These significances are computed with a matched filter based on the model shape, downweighting measurements where the correlation is known to be weaker. The high density of the BGS sample and the strong bias of the LRG sample result in high-significance detections, even when the correlation strength drops at large separations. We note that the BGS significance is an upper limit as our covariance may underestimate sample variance (Section \ref{sec:significance}). Although a weaker detection, ELG multiplet alignment is especially notable given that the equivalent signal for individual ELGs is consistent with 0; this represents our best approach for detecting intrinsic shear above redshift $z=1$.}
\label{fig:large-scale-averaged}
\vspace{.2in}
\end{figure*}

\begin{figure}
\centering
\includegraphics[width=.48\textwidth]{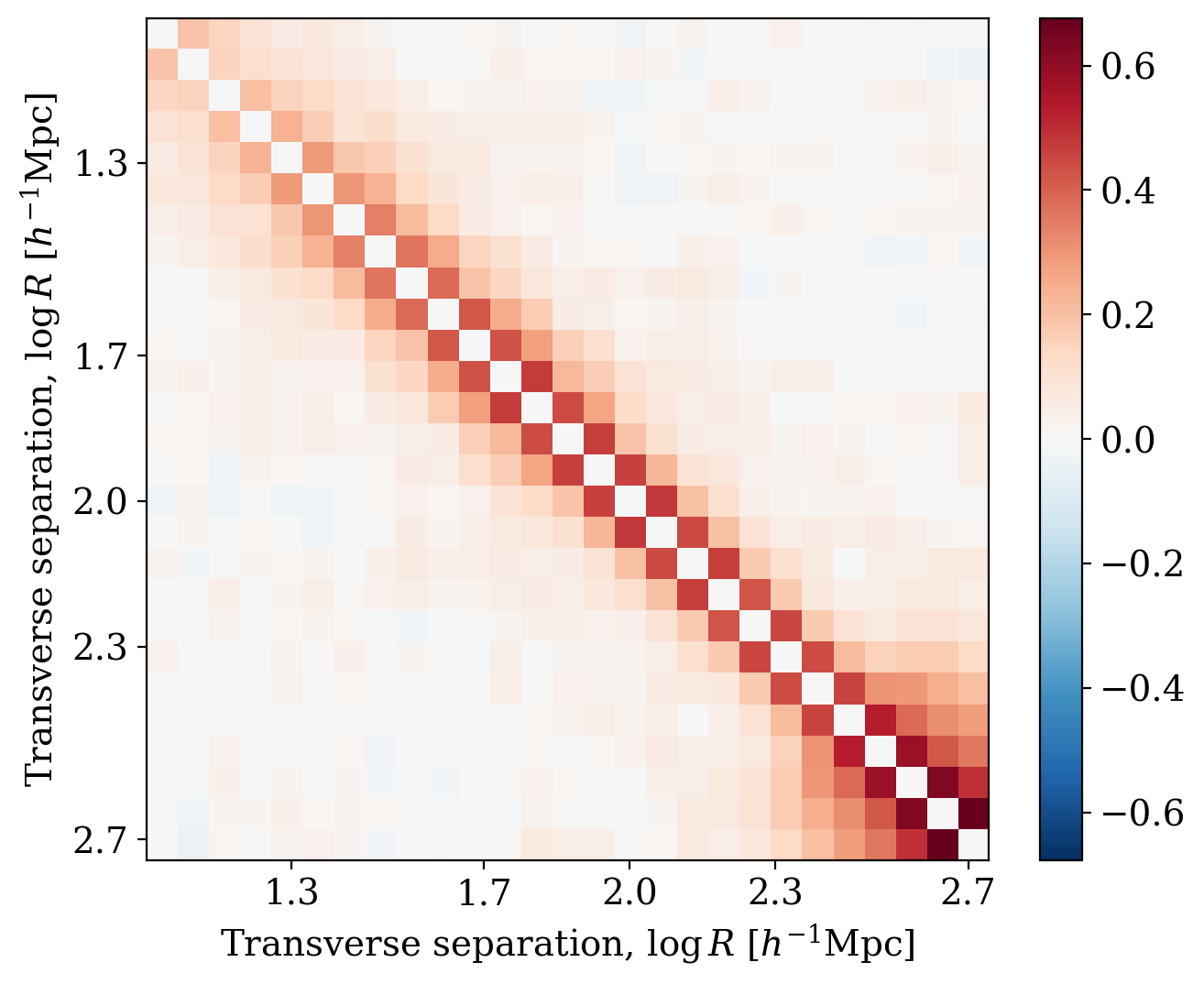}
\caption[]{The correlation matrix of the combined signal from the three LRG redshift bins, with the
identity subtracted to highlight off-diagonal structure. Our quoted significances consider this positive correlation between adjacent bins, which is particularly pronounced at large separations.}
\label{fig:cov-LRG}
\vspace{.2in}
\end{figure}

Figure \ref{fig:large-scale-averaged} shows the average multiplet alignment made with BGS, LRG, and ELG. They are a combination of the measurements in all redshift bins for each tracer, with errors taking into account the difference in amplitude across bins, as described above. Differences in amplitudes on this plot are from both clustering variations between the samples and differences in the alignment strength of multiplets. These represent our cleanest measurements of large-scale tidal shear with each tracer. The correlation matrix of the combined LRG signal is characteristic of all measurements and is shown in Figure \ref{fig:cov-LRG}. The diagonal of the underlying covariance reproduces the standard errors in Figure \ref{fig:large-scale-averaged}. Adjacent $R$ bins are positively correlated, especially at large separations. More details of this covariance estimation, the impact of region size, and detection significance are discussed in Section \ref{sec:significance}. 

\subsection{Autocorrelations}\label{subsec:autocorr_measurement}
 
The multiplet orientation autocorrelation $\mathcal{E}_{++}$ (Section~\ref{sec:estimators}) is measured with the same infrastructure as $\mathcal{E}_+$, differing only in the quantity correlated and in the absence of a random catalog. Rather than correlating a multiplet orientation with a tracer position, we take pairs of multiplets, project each multiplet's orientation onto the separation vector connecting the pair, and average the product of the two projected components. The line-of-sight pair selection $\Pi_{\rm max}(R_{\rm bin})$, the division into sky regions of equal multiplet count, and the estimation of errors and covariance from the region-to-region scatter are all unchanged from the cross-correlation.

The resulting autocorrelation measurements are shown in Figure \ref{fig:autocorrs}. Shape-shape correlations are generally weaker than shape-density cross-correlations, and this effect is more pronounced with multiplets than individual intrinsic alignments since there are fewer shape tracers (multiplets) than density tracers (galaxies). The ELG measurements especially suffer from this, where the density of multiplets are low, and we find only a marginal detection of the ELG autocorrelation. The signal for all samples is shown in Figure \ref{fig:autocorrs}, where the signal for LRGs and ELGs has been averaged over the redshift bins. 

They are compared to the model predictions normalized with two amplitudes: one directly fit to the measured autocorrelation, $\mathcal{E}_{++}$, and one fit to the cross-correlation, $\mathcal{E}_+$. Across all samples, the amplitude fit to $\mathcal{E}_+$ is larger and shows evidence of a true tension in LRG and BGS. For LRGs, the amplitudes are consistent for LRG1, but the combined disagreement across the three redshift bins reaches $\sim5\sigma$. This could be from the assumption of one galaxy bias value across the redshift range, as opposed to an issue with the NLA model. For BGS, we use the independent, magnitude-calibrated bias values from the U\textsc{chuu}-BGS mocks (see below), but still see disagreement at the 4-8$\sigma$ level depending on the adopted magnitude percentile. There is also a large fractional offset in ELG, although this is consistent with its noisy detection. We find no evidence that this is a result of the model itself. The Bessel-kernel projections and tidal-factor treatment used in $\mathcal{E}_+$ and $\mathcal{E}_{++}$ are unchanged between samples, so the redshift-dependence of the tension is more consistent with an underestimated galaxy bias. Where the discrepancy is worst, LRG3, it would be reconciled with an increase in the galaxy bias of 25\%.

The high density of multiplets in BGS allows for significantly cleaner measurements of the multiplet autocorrelation. In part due to this, and in part due to the difficulty of estimating galaxy bias for a sample that varies so much across redshift bins, we decided to use these correlations to independently estimate the galaxy bias directly from the samples, as described below. 

\subsection{Galaxy Bias}\label{sec:validation_bias}

The inferred alignment amplitude $\tau$ is inversely proportional to the linear galaxy bias, $b_g$, so accurate bias estimates are essential, particularly when comparing measurements across redshift bins with different sample properties. For the LRG and ELG samples we adopt the linear bias values from DESI's fiducial cosmological analyses of $b_{\rm LRG}=1.99$ and $b_{\rm ELG1}=1.18$,  $b_{\rm ELG2}=1.40$ \citep{mena-fernandezHODdependentSystematicsLuminous2025, vaisakhDESIDR2Reference2026}. For BGS, the situation is more complicated, and we describe our approach in this section. We compare to reference mocks but ultimately take advantage of the fact that $\mathcal{E}_+$ depends on $b$ but $\mathcal{E}_{++}$ doesn't, and estimate it from our samples directly. 

The BGS BRIGHT sample used in this work spans $0.01 < z < 0.5$, extending well beyond the redshift range $0.1 < z < 0.4$ used in DESI's standard cosmological BGS analyses. Unlike the LRG and ELG samples, which are designed to maintain a roughly constant comoving number density with redshift, BGS is flux-limited: the $r < 19.5$ magnitude cut preferentially removes fainter, less-biased galaxies at higher redshifts. As a result, both the effective luminosity and the linear bias of the BGS sample evolve strongly across our five redshift bins.

We first estimate $b_g$ using the fit to the U\textsc{chuu}-BGS mock catalogs from \citet{fernandez-garciaDESIDR2Reference2026}, who measure the linear bias of BGS BRIGHT as a function of absolute magnitude threshold. They parameterize the bias of a volume-limited sample of galaxies with $M_r$ brighter than a given threshold as:
\begin{equation}
    b(<M_r) = B_0 + B_1 \times 10^{B_2(M^*_r - M_r)/2.5},
    \label{eq:fernandez_bias}
\end{equation}
with best-fit parameters $B_0 = 1.087$, $B_1 = 0.196$, $B_2 = 1.12$, and $M^*_r = -20.44$ for the DR2 BGS BRIGHT sample. This fit is calibrated over $-22 \leq M_r \leq -20$, but the functional form is monotonic and well-behaved outside this range.

\begin{figure}
\centering
\includegraphics[width=.5\textwidth]{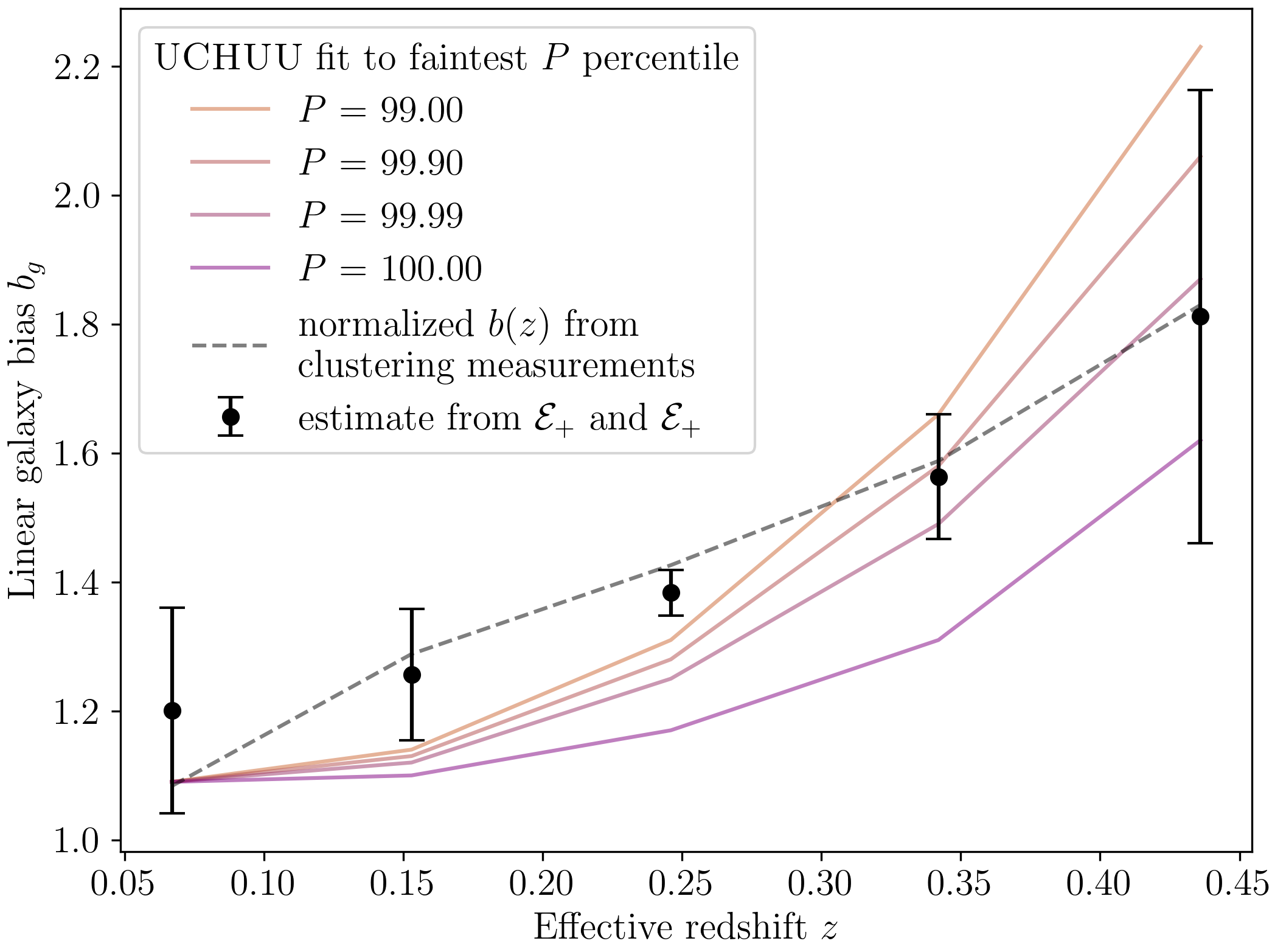}
\caption[]{Estimates of galaxy bias in our BGS redshift samples. The solid lines show predictions from the U\textsc{chuu} simulations using the faintest $P$ percentiles of galaxies in each bin. The dashed line shows the relative difference in bias between each bin, as predicted from comparing their 2-point clustering, normalized to our estimates from multiplet correlations. These estimates come from comparing the measured amplitude of the cross and auto correlations. They are shown in black, with errors from the variance of determining the amplitudes in independent $R$ bins. The bias values estimated directly from our sample are consistent with the trend expected from 2-point clustering and are the ones we adopt for the rest of this work.}
\label{fig:bias}
\vspace{.2in}
\end{figure}

Because our BGS bins are flux-limited rather than volume-limited, we cannot directly apply Equation \ref{eq:fernandez_bias} at a single threshold. Instead, we approximate the effective bias in each bin by identifying $M_r^{\rm eff}$ as a high percentile of the observed $M_r$ distribution in that bin. 
This approximation becomes less accurate at low redshift, where the flux limit permits a wider range of intrinsic luminosities. We compute the bias using 99.9th, 99.99th, and 100th (brightest galaxy) percentiles; results are shown in Figure \ref{fig:bias}. The estimates are relatively stable at low $z$ and diverge modestly at high $z$, reflecting the extrapolation into the bright tail.

The multiplet autocorrelation, $\mathcal{E}_{++}$, provides an alternative estimate of $b_g$ that is independent of the mock calibration. As described in Section \ref{sec:formalism}, the cross-correlation amplitude scales as $\tau b_g$ while the autocorrelation amplitude scales as $\tau^2$. Taking the ratio for each redshift bin $i$,
\begin{equation}
    b_g^{(i)} = \frac{|\tau_{+}^{(i)}|}{\sqrt{|\tau_{++}^{(i)}|}}.
    \label{eq:b_from_auto}
\end{equation}
These estimates and their uncertainties are also shown in Figure \ref{fig:bias}.

As an independent cross-check, we predict the relative bias between bins using our own projected two-point correlation function measurements, $w_p$. In the linear regime, $b_g \propto \sqrt{w_p}/D(z)$, where $D(z)$ is the linear growth factor.
We use measurements of the average of our projected correlation function, $w_p^{(i)}$, over $R$ bins of $10-50$ $h^{-1}$Mpc. In Figure \ref{fig:bias}, this relation $b(z)$ is shown, scaled to the average multiplet estimate. The shape of $b(z)$ here depends only on our own $w_p$ measurements, and requires no assumption about the alignment amplitude $\tau$ or its evolution.

Figure \ref{fig:bias} compares all three methods. At low redshift, the multiplet autocorrelation and the U\textsc{chuu}-BGS fit agree to within $\sim 5\%$. At higher redshift, the U\textsc{chuu}-BGS estimates rise more steeply than the autocorrelation-derived values, with the discrepancy sensitive to the extrapolation of Equation \ref{eq:fernandez_bias} into the bright end of the luminosity distribution. $b(z)$, normalized to the mean of the autocorrelation estimates, provides a smooth interpolation that lies within the uncertainties of the autocorrelation measurements. 

We adopt these per-bin autocorrelation-derived values as our fiducial bias estimates. This choice has three advantages: (1) it is measured from the same data used for the alignment signal, avoiding any mismatch in sample selection or cosmology relative to a mock calibration; (2) each per-bin estimate is independent of $\tau$ and its potential redshift evolution, which is a particular concern for BGS given the strong sample-composition changes across our redshift range; and (3) the estimates are consistent with the independent clustering-based shape from $w_p$ and $D(z)$, providing internal validation.


\section{Mock Catalogs}\label{sec:mocks}

We reproduce our measurements with mock catalogs built from the \textsc{AbacusSummit} $N$-body simulations \citep{maksimovaAbacusSummitMassiveSet2021}, the same suite used to validate DESI's clustering analysis. In \cite{lammanDetectionLargescaleTidal2024}, multiplet alignment measured in these mocks was found to reproduce the scale dependence of the real signal, as expected if the alignment traces the tidal field in the manner described by the NLA framework, while under-predicting its amplitude by about 30\% for LRGs. The amplitude offset is a reflection of the simulation's limited ability to capture the small-scale, higher-order clustering that sources the multiplet orientation. Because the multiplets identified in this work are built with larger linking lengths than in \cite{lammanDetectionLargescaleTidal2024} (Section~\ref{sec:multiplets}), it is not obvious that this behavior should persist.

We measure multiplet alignment in the LRG and ELG mocks; no analogous BGS LSS mocks were available. We averaged over all $25$ realizations and compared with the real measurement by taking the ratio of the two averaged over $20$ logarithmic bins in projected separation between $10$ and $200\,h^{-1}{\rm Mpc}$. For the three LRG redshift bins the ratio of the measured signal to the mock is $1.20 \pm 0.01$, $1.28 \pm 0.01$, and $1.33 \pm 0.01$ for LRG1, LRG2, and LRG3 respectively. These are marginally closer to unity than the ratio of $1.33$ reported for LRGs in \cite{lammanDetectionLargescaleTidal2024}, consistent with the expectation that the larger multiplets used here are less diluted by the nonlinear effects the $N$-body simulation does not capture. For the two ELG bins, the ratios are $1.97 \pm 0.05$ and $1.80 \pm 0.05$. The larger offset for ELGs follows the same trend: the ELG multiplets are on average smaller and sparser than the LRG multiplets, and so retain a larger contribution from the small-scale dynamics that the mocks under-represent.

In all cases, the scale dependence of the mock signal remains consistent with the measurement, as in \cite{lammanDetectionLargescaleTidal2024}, even though the amplitude does not match. We use this agreement in shape to validate the analytic model underlying our BAO analysis in Section~\ref{sec:bao}, where it is the scale dependence of the template rather than its overall amplitude that determines the result.

\vspace{.2in}
\section{Results}\label{sec:results}

\subsection{Detection Significance}\label{sec:significance}
To compute the significance of our detections above a transverse scale $R_{\rm min}$, we use the bins with $R > R_{\rm min}$ and the $\chi^2$ against the null of zero signal,
\begin{equation}
\chi^2 = \bar{\mathbf{d}}^{\mathsf{T}} \mathsf{C}^{-1} \bar{\mathbf{d}}.
\label{eq:chi2}
\end{equation}
Assuming a prior of a positive tidal alignment, this is converted to a one-sided significance.
The $\chi^2\!\to\!\sigma$ conversion assumes Gaussian per-region errors. We
validate this with an empirical null distribution of the statistic, built two
ways from the same regions: (i) recentering each bootstrap resampling
(Eq.~\ref{eq:cov}) to zero signal; (ii) a sign-flip null, randomly flipping the
sign of each $\mathbf{d}_i$. Significance is read from the fraction of null
realizations exceeding the observed $\chi^2$. Both nulls agree with the
analytic value, confirming Gaussian errors on these scales. With this method, we find a detection significance for the combined LRG signal over $R>200h^{-1}$Mpc at 3.25$\sigma$.

To avoid the dilution of our significance by measurements made where the signal is expected to be indistinguishable from 0, we apply a matched filter using the NLA model shape $\mathbf{t}$. This weights bins by their expected signal:
\begin{equation}
{\rm S/N} = \frac{\mathbf{t}^{\mathsf{T}} \mathsf{C}^{-1}
\bar{\mathbf{d}}}{\sqrt{\mathbf{t}^{\mathsf{T}} \mathsf{C}^{-1} \mathbf{t}}} .
\label{eq:matchedfilter}
\end{equation}
The statistic is independent of the empirical model amplitude and does not depend on the assumed cosmology. Equation~\ref{eq:chi2} tests the data against a null of zero signal, and the matched filter of Equation~\ref{eq:matchedfilter} uses only the shape of the model, not its amplitude. To determine the significance over a given scale, we apply this only to the bins larger than $R_{\rm min}$. Combined over the three LRG redshift bins above $R>200\,h^{-1}{\rm Mpc}$, the matched filter gives $3.7\sigma$.

Our covariances are estimated from the scatter between sky regions, which in some cases are smaller than the measurement scales. Therefore, the very large-scale modes we measure span multiple regions, and we may underestimate large-scale covariance. However, we also expect the measurements to be close to shape-noise dominated, especially with the $\sigma_s=\frac{1}{\sqrt{2}}$ of multiplets, so this sample variance issue may not dramatically impact our detection significance. 

To test this, we merge spatially adjacent regions into larger ones spanning the scales of interest and re-estimate the covariance. The origin of the merging grid is arbitrary and the resulting significance is moderately sensitive to it, so we enumerate several offsets and quote the median. We also test a range of region sizes. Across all samples, the number of merged regions ranges from the originally used 2500 in BGS2, down to tens of merged regions, corresponding to physical sizes of around 10 $h^{-1}$Mpc for the smallest BGS regions and up to 600 $h^{-1}$Mpc for ELG. Following the original region construction, each is built with the same number of multiplets and randoms are subtracted on a per-region basis. To see if any change in covariance reflects the inclusion of larger-scale modes rather than a lower number of regions, we repeat this but by randomly merging non-adjacent regions. These randomly-merged regions produce consistent covariance at all merging levels, as would be expected if the regions are independent, while adjacently-merged regions will depart from it at different merging levels. The difference between them isolates the contribution of modes larger than a single region. 

For LRG3 and both ELG bins, whose regions are already comparable to the separations of interest, the adjacent and randomly merged results agree. Therefore, we keep the full number of regions that the measurements were originally made in for these samples. For the other samples, the effect appears once the regions are smaller than the separation being tested: for region sizes larger than $\sim140h^{-1}$Mpc, the merged and randomized results agree and below that the difference grows as region size shrinks. Since the combined LRG signal above $200h^{-1}$Mpc is dominated by LRG3, our large-scale detection significance for all LRGs is largely unaffected. For BGS, the significance above $200h^{-1}$Mpc falls from 4.1$\sigma$ to $2.9\sigma$. The randomized control does not reproduce this drop, indicating modes which supersede region scales. We also find that this significance does not converge as the number of regions becomes too small to estimate covariance. The lowest redshift BGS bins do not contain multiple independent regions which span more than $200h^{-1}$Mpc. Therefore, we use the larger-region estimation and quote the BGS significances above $200h^{-1}$Mpc as an upper bound. 

We note that these cannot be validated with the \textsc{AbacusSummit} mocks, as the realizations involve replications of the box volume and are not suitable for characterizing the measurement scatter. A full analytic treatment, combining the Gaussian and super-sample covariance terms calibrated against the region-based estimate, would resolve this properly. However, this would require a density-response function for the multiplet alignment statistic that has not been derived; we leave this to future work.

Detection significances from the matched filter and estimated with the merged regions are shown on Figure \ref{fig:large-scale-averaged}, which displays the signals used to compute them.  
For LRGs, over scales of 100, 200, and 300 $h^{-1}$Mpc, we find detection significances of $18.2\sigma$, $3.7\sigma$, and $1.8\sigma$ respectively. As mentioned above, the large-scale LRG signal is concentrated in the highest-redshift bin, $4.5\sigma$ above $200h^{-1}$Mpc when determined from LRG3 alone, while LRG1 and LRG2 are individually consistent with no large-scale signal ($1.9\sigma$, $0.3\sigma$). We adopt the three-bin combination as the headline result, since the per-bin variation is consistent with cosmic variance.

Measurements for BGS and ELG were not made over 300$h^{-1}$Mpc because we expect it would be an impractical use of computing resources. At these scales, the angular limits of the survey footprint for BGS and the low density of ELG multiplets will dilute significance well below that of LRGs. However, the high density of BGS (resulting in many more multiplets) produces more significant detections than LRGs below this. We detect tidal shear from BGS multiplets over 100 and 200 $h^{-1}$Mpc at $\le 13.8\sigma$ and $\le2.9\sigma$. The equivalent for ELGs is 3.3$\sigma$ and $0.4\sigma$.

\vspace{.1in}
\subsection{Comparison to Previous Measurements}

\begin{figure}
\centering
\includegraphics[width=.48\textwidth]{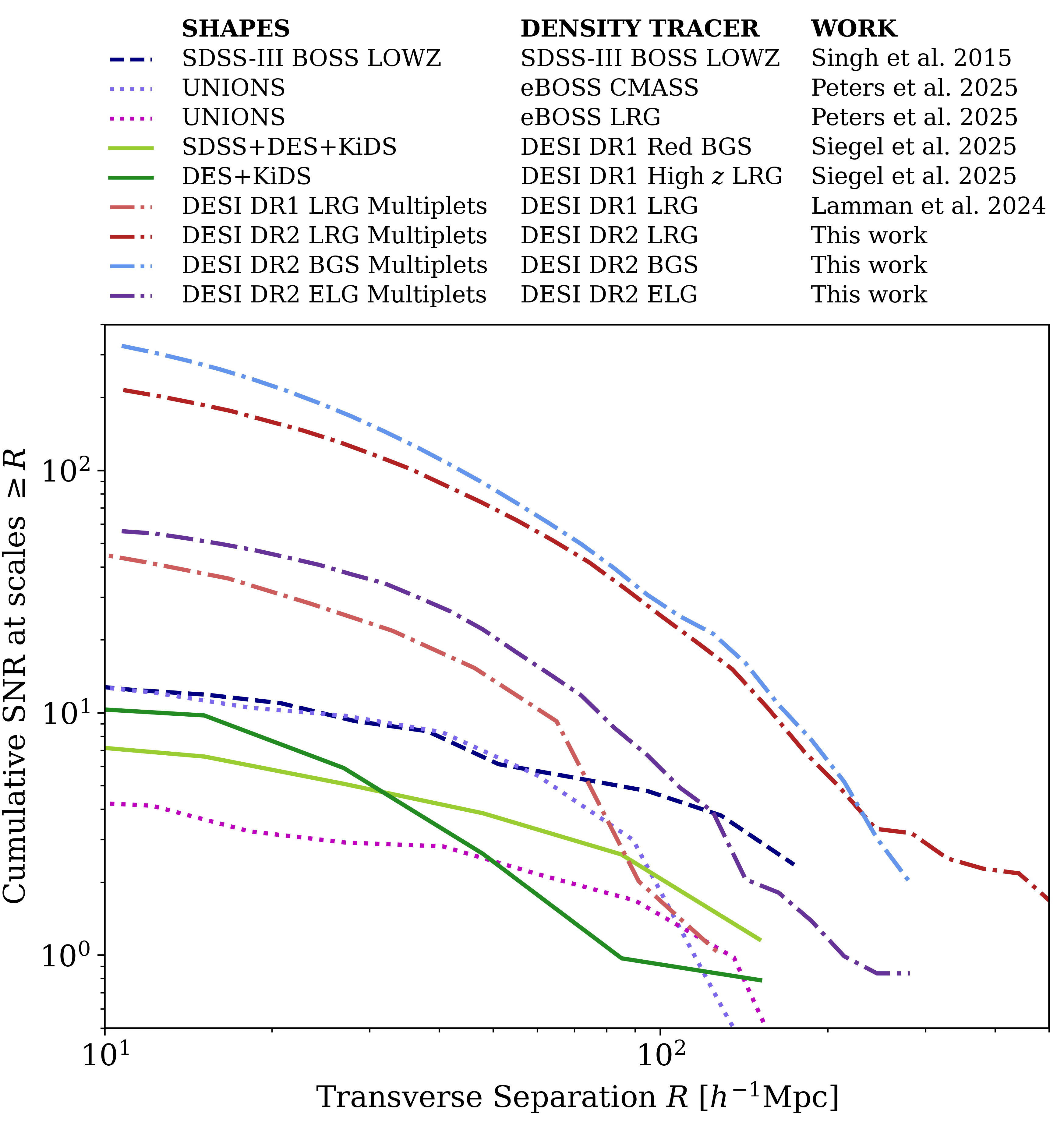}
\caption[]{The landscape of large-scale intrinsic tidal alignment detections. For each measurement we plot the cumulative signal-to-noise ratio accumulated from the largest measured scale inward. Solid, dotted, and dashed lines denote galaxy-shape intrinsic alignment measurements ($w_{g+}$); dash-dotted lines denote multiplet alignment ($\mathcal{E}_+$). All curves use the binning of the original published analyses. The cumulative statistic assumes diagonal covariance and therefore represents an upper bound on the true combined significance.}
\label{fig:snr-comparison}
\vspace{.2in}
\end{figure}

Reliable measurements of intrinsic alignment on large scales are a crucial ingredient for many of the applications discussed in Section~\ref{sec:intro}. Here we place our multiplet measurements in the context of the existing landscape of large-scale alignment detections.

Figure~\ref{fig:snr-comparison} compares the cumulative signal-to-noise of the strongest published alignment measurements that extend beyond projected separations of $100\,h^{-1}\,\mathrm{Mpc}$. These include the galaxy-shape intrinsic alignment of BOSS LOWZ \citep[][$w_{g+}$]{singhIntrinsicAlignmentsSDSSIII2015}, the CMASS-UNIONS and eBOSS~LRG-UNIONS measurements of \citet{hervaspetersUNIONSDirectMeasurement2025}, and the DESI DR1 red BGS and high-redshift LRG measurements of \citet{siegelIntrinsicAlignmentDemographics2025}. Against these we plot our DESI DR1 LRG multiplet result \citep{lammanDetectionLargescaleTidal2024} and the DESI DR2 LRG, BGS, and ELG multiplet measurements presented in this work. We use the reported alignment signal and its uncertainty, in the published binning, for each measurement. 

Several relevant measurements are not shown. Fourier-space analyses such as \citet{kuritaConstraintsAnisotropicPrimordial2023} report a high total signal-to-noise for the E-mode intrinsic-alignment power spectrum, but that significance is accumulated across a broad range in wavenumber dominated by quasi-nonlinear scales, and a bandpower at a given $k$ does not correspond to a single physical separation, so it cannot be placed on the horizontal axis of Figure~\ref{fig:snr-comparison}. 
The BAO-scale intrinsic alignment of \citet{xuEvidenceBaryonAcoustic2023} does reach large scales, but is measured in linear three-dimensional bins around the BAO feature rather than in projected separation, so it is not directly comparable; we note that its signal-to-noise is comparable to the CMASS-UNIONS measurement.

Although the galaxy-shape measurements report the projected correlation $w_{g+}$ and the multiplet measurements report $\mathcal{E}_+$, the two can be compared on a signal-to-noise basis. The multiplet estimator averages a relative-orientation signal over multiplet-tracer pairs, while $w_{g+}$ averages a shear over shape-density pairs; in both cases the uncertainty on the associated two-point clustering term is negligible relative to that on the alignment signal, so the per-bin signal-to-noise values are directly comparable even though the underlying estimators differ. The most significant difference between all measurements is the choice of $\Pi_{\rm max}$. IA measurements are typically made with a flat $\Pi_{\rm max}$ across all bins of 60-100$h^{-1}$Mpc. The multiplet measurements are the only ones using the suggested variable $\Pi_{\rm max}$ from \cite{lammanOptimalIntrinsicAlignment2025}, and therefore also represent an advancement in the alignment estimator.

Two caveats apply to this comparison. First, the per-bin signal-to-noise depends on the choice of binning; finer binning divides a fixed signal among more bins. The cumulative statistic in Figure~\ref{fig:snr-comparison} largely removes this sensitivity, since quadrature-summed signal-to-noise is approximately conserved under rebinning of independent bins. Second, and less tractably, neighboring bins are correlated, and the cumulative statistic neglects this off-diagonal covariance; it therefore overstates the true combined significance. Both effects should be borne in mind, but neither qualitatively changes the comparison.

The multiplet measurements reach a cumulative large-scale signal-to-noise more than an order of magnitude above the strongest galaxy-shape measurements over the same range of separations. We emphasize, however, that the galaxy-shape measurements shown here are drawn from earlier data sets; a like-for-like comparison against galaxy intrinsic alignment measured on DESI DR2 would be more equitable. We also note that an ELG-like galaxy population has never yielded a successful intrinsic-alignment detection, so the ELG multiplet measurement has no galaxy-shape counterpart to compare against.

Despite the challenges of comparing measurements made with different methods and data, Figure \ref{fig:snr-comparison} demonstrates the remarkable growth of this field over the past decade. Improvements are driven by new estimators, the use of multiplet shapes, and the growing spectroscopic data available from large cosmological surveys.

\begin{figure*}
\centering
\includegraphics[width=1\textwidth]{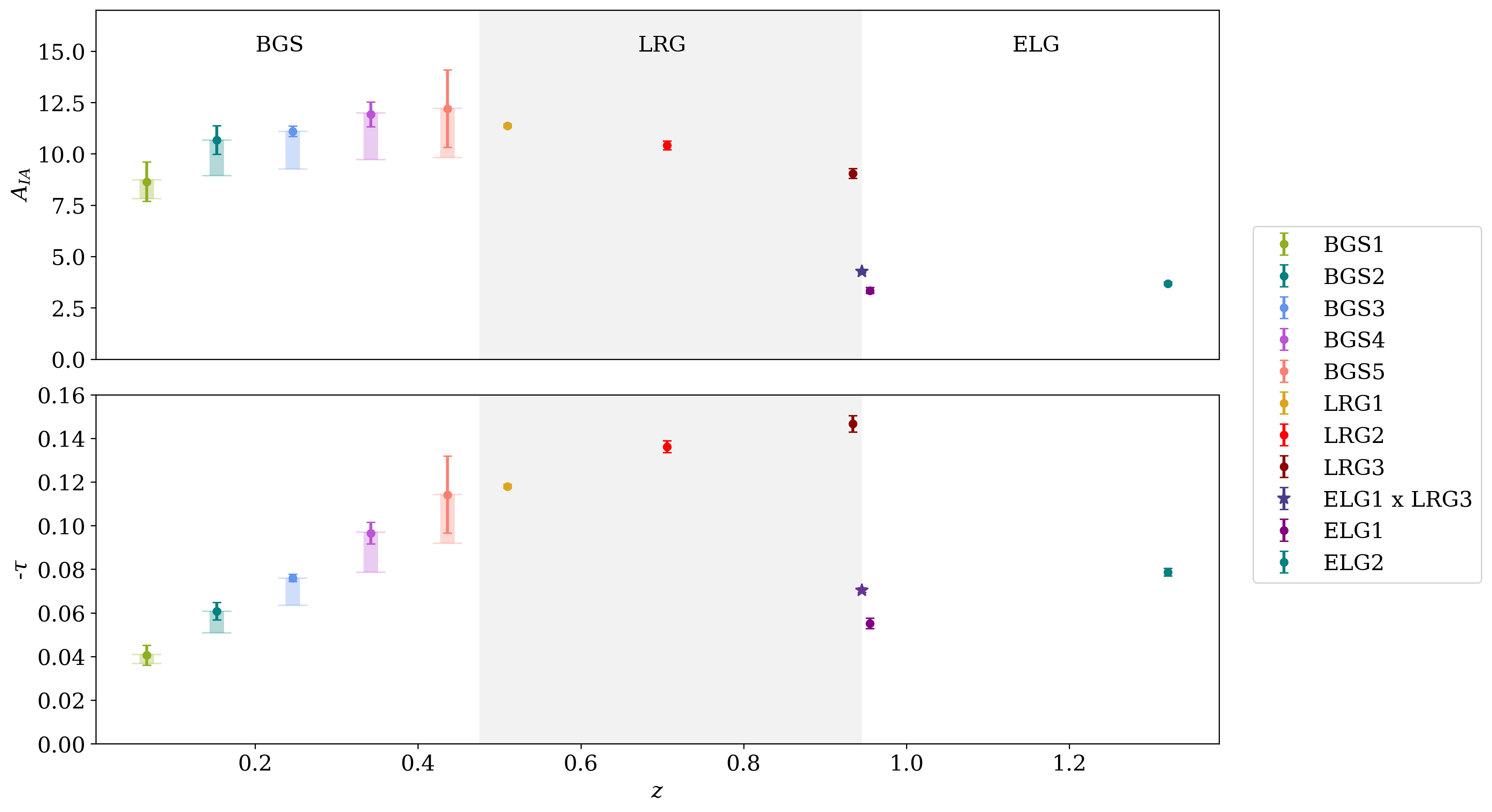}
\caption[]{Redshift dependence of the multiplet alignment amplitude for the BGS, LRG, and ELG samples, shown as both the shear-response parameter $\tau$ and the conventional intrinsic-alignment amplitude $A_{\rm IA}$. The two are related by a factor of the growth factor, $A_{\rm IA}\propto\tau\,D(z)$ (Equation~\ref{eq:tau_to_AIA}), so a sample whose $\tau$ is roughly constant with redshift can show a mild trend in $A_{\rm IA}$. For BGS, the galaxy bias used to set the amplitude is measured directly from the data (Figure~\ref{fig:bias}), and its uncertainty is included in the error bars. Because the BGS density varies strongly with redshift, the multiplet definition is also changed between BGS redshift bins; the wider shaded bars show the spread in amplitude obtained when the linking length is varied between $1.5$ and $8\,h^{-1}\mathrm{Mpc}$ (Figure~\ref{fig:bgs-snr}). Even so, the multiplets in each BGS bin differ intrinsically because of the changing density, so the BGS points should not be read as a clean evolution at fixed sample. The trend is more safely compared within the LRG and ELG samples, which hold a more nearly constant comoving density and a single linking length across their redshift bins. The star point marks the cross-correlation of ELG multiplets with LRG tracers over the overlapping range $0.8<z<1.1$, cleanly demonstrating that ELG multiplets are weaker tracers of intrinsic shear than LRGs.}
\label{fig:amp-evolution}
\vspace{.2in}
\end{figure*}

\vspace{.1in}
\subsection{Alignment Strength and Redshift}\label{sec:amp-redshift}

There is evidence, from both hydrodynamic simulations and observations, that the intrinsic alignment amplitude $A_{\rm IA}$ of individual galaxies increases with redshift over $0 < z < 1$ in LRG-like samples \citep{samuroffAdvancesConstrainingIntrinsic2021, herleAssemblyBiasRedshift2026, siegelIntrinsicAlignmentDemographics2025}. We might expect multiplet alignment to follow a similar trend, though it is not obvious that it should track the individual-galaxy result. Here we take an initial look at the redshift dependence of multiplet alignment, although a pure comparison across redshifts is limited by variations across and within samples, particularly the fact that our multiplet definitions vary between redshift bins in BGS. Figure~\ref{fig:amp-evolution} shows the measured amplitude of multiplet alignment as a function of redshift for all three tracers, reported both as the shear-response parameter $\tau$ and as the more common $A_{\rm IA}$. For this comparison, the important distinction between the two is that $A_{\rm IA}$ includes the growth factor, $A_{\rm IA}\propto D(z)$.

For BGS the interpretation is complicated by the sample itself. As described in Section~\ref{sec:catalogs}, the flux-limited BGS sample changes substantially in both density and intrinsic luminosity across our five redshift bins, and we adopt a different multiplet definition in each bin to track the changing density (Section~\ref{sec:multiplets}). The bias used to convert the measured signal to an amplitude is determined directly from the data (Section~\ref{sec:validation_bias}, Figure~\ref{fig:bias}), and its uncertainty is propagated into the amplitude errors shown here. To assess the sensitivity of the amplitude to the multiplet definition, we additionally recompute it in each bin over a range of linking lengths between $1.5$ and $8\,h^{-1}{\rm Mpc}$ (Figure~\ref{fig:bgs-snr}); the resulting spread is shown as the wider shaded error bars in Figure~\ref{fig:amp-evolution}. Even with this accounted for, the multiplets in each BGS bin remain physically distinct because of the changing density, so the trend should not be read as a clean evolution at fixed sample. With that caveat, we do see the multiplet alignment amplitude increase with redshift across the BGS range, consistent with the trend reported for individual galaxies.

The redshift evolution is more safely compared within the LRG and ELG samples, which maintain a more nearly constant comoving density and use a single linking length across their redshift bins. Within LRGs, the direction of the apparent trend depends on which amplitude is considered: the multiplet $A_{\rm IA}$ decreases modestly with redshift while $\tau$ increases, reflecting the opposite growth-factor scaling of the two parameters. The overall picture is that the LRG multiplet alignment amplitude is fairly constant across this redshift range. 

The ELG multiplets are similarly constant with redshift, but with a substantially lower amplitude than either BGS or LRG. This is notable. Individual ELGs show no measurable intrinsic alignment at all, but it was not a given that ELG \emph{multiplets} would have a much lower amplitude than LRG multiplets, since the multiplet orientation does not depend on resolving individual galaxy shapes. We find that they do. A cleaner comparison is provided by the cross-correlation of ELG multiplets with LRG tracers over the overlapping redshift range $0.8 < z < 1.1$. The ELG multiplet amplitude remains much lower, demonstrating that the difference is not simply a matter of the tracer clustering, although the two multiplet definitions still differ somewhat. This indicates that ELGs are found in less dense environments and are, perhaps, weaker tracers of the filamentary structure that sources multiplet alignment, consistent with expectations from morphology--environment correlations. This is a clear motivation to explore these trends in hydrodynamic simulations, where the astrophysical causes can be examined directly.

\begin{figure}
\centering
\includegraphics[width=.48\textwidth]{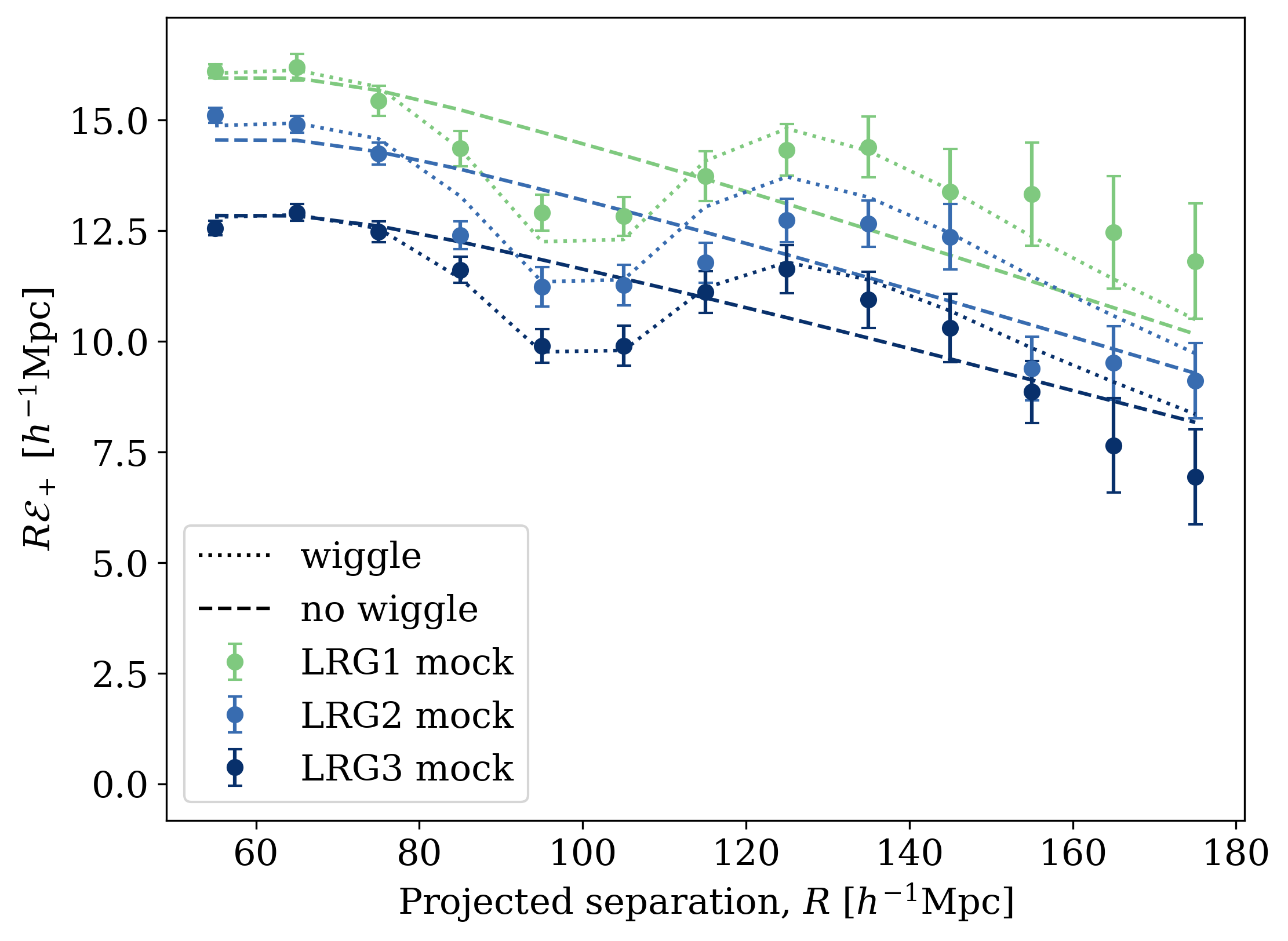}
\caption[]{The average of 25 realizations of a mock DESI DR2 LRG catalog, compared to the NLA model predictions with and without an acoustic feature. Each model's amplitude is fit to the measurements. The measurements are consistent with the BAO model, displaying a characteristic dip at the acoustic scale and peak beyond it.}
\label{fig:bao-lrg-measurement-mock}
\vspace{.2in}
\end{figure}

\vspace{.1in}
\subsection{The BAO Feature}\label{sec:bao}

The signature of Baryon Acoustic Oscillations (BAO) in the tidal alignment signal can be understood through the NLA framework \citep{okumuraTestingTidalAlignment2020, vandompselerAlignmentGalaxiesBaryon2023b}.
A tracer at the center of the acoustic ``ring'' sees an excess of matter at the BAO scale distributed isotropically around it. This symmetric shell exerts no net tidal stretching in any preferred transverse direction, \textit{diluting} radial alignment at separations near the sound horizon. At separations just inside and outside the shell, by contrast, the overdensity is anisotropic with respect to the tracer and \textit{enhances} alignment. The net effect is a localized dip in alignment at the BAO scale relative to the smooth-model expectation. This has been measured in the galaxy density-ellipticity correlation by \citet{xuEvidenceBaryonAcoustic2023}, who reported $2-3\sigma$ evidence of the BAO feature using BOSS CMASS galaxies with DESI Legacy Imaging shape measurements. Here we assess whether an analogous feature is detectable in multiplet alignment.

We restrict this analysis to the BGS and LRG samples, in linear bins around the expected BAO feature, $55 < R < 180\,h^{-1}\mathrm{Mpc}$. The acoustic feature occupies a fixed comoving scale with an approximately constant width in $R$, and linear binning samples it more evenly than the scale-invariant logarithmic binning used earlier. To test for the presence of the feature, we compare two model templates for $\mathcal{E}_+(R)$, computed as in Section~\ref{sec:modeling} but differing only in the input matter power spectrum: one using the full spectrum with acoustic oscillations, and one using a smooth ``no-wiggle'' spectrum with the oscillations removed. These are the same models used in DESI's BAO main analysis \citep{andradeValidationDESIDR22025, briedenModelagnosticInterpretation102022}. The amplitude of each model, $\tau$, is fit to the measured signal independently and in each redshift bin. 

We validated this model using the 25 mock realizations of the LRG sample, as described in Section \ref{sec:mocks}. Their measurements, averaged over all realizations but kept in the three redshift bins, are compared to models in Figure \ref{fig:bao-lrg-measurement-mock}. As discussed there, the amplitude of these mock measurements disagree with the true signal, but, as predicted by NLA, is consistent its scale dependence. Here we clearly see the acoustic feature in all redshift bins, consistent with the model prediction. 

For each redshift bin in the LRG and BGS sample, we form $\Delta\chi^2 = \chi^2_{\mathrm{nw}} - \chi^2_{\mathrm{w}}$. A positive value indicates a preference for the acoustic template, zero indicates no acoustic information, and a negative value corresponds to a non-physical negative acoustic amplitude -- which indicates a residual which is not described well by either template. We also fit the no-wiggle amplitude and apply a matched filter to the residual using the wiggle-minus-no-wiggle difference $\Delta m = m_{\mathrm{w}} - m_{\mathrm{nw}}$ as the template, yielding a signed detection significance $S/N = (\Delta m^{\mathsf{T}} C^{-1} r)/\sqrt{\Delta m^{\mathsf{T}} C^{-1} \Delta m}$, where $r$ is the residual. We adopt this as our detection statistic, as it isolates the wiggle-specific information. Each redshift bin is treated independently, so their information combines at the covariance level: $C^{-1}_{\mathrm{comb}} = \sum_s C^{-1}_s$. Because the acoustic feature is a small fraction of the total signal and lies in the most covariant bins, we validate the analytic significance against empirical null distributions built by bootstrap resampling and sign-flipping the sky regions; the empirical and analytic significances agree.

\begin{figure}
\centering
\includegraphics[width=.48\textwidth]{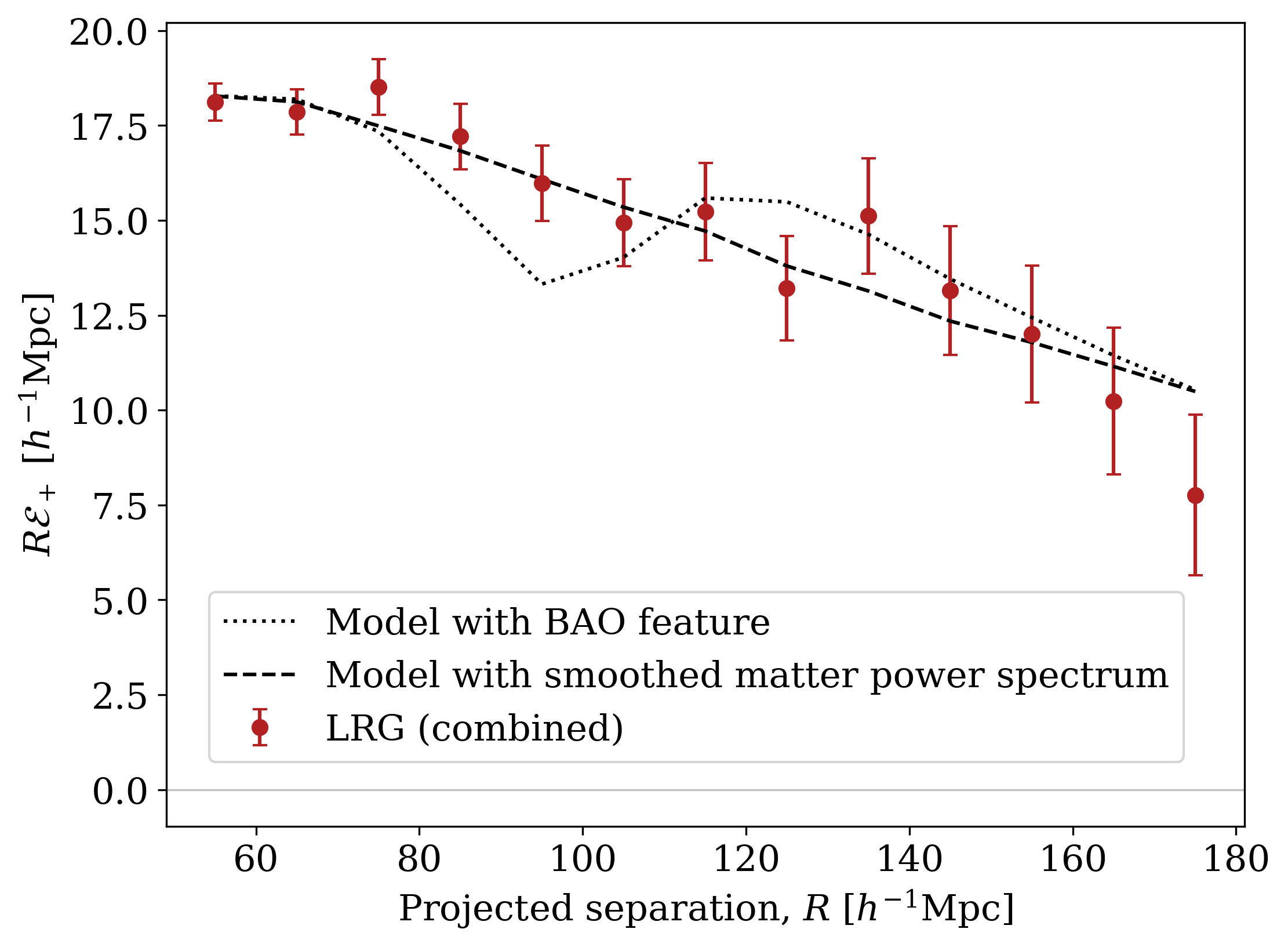}
\caption[]{Tidal alignment of LRG multiplets around the acoustic scale. Each of the three redshift bins was independently used to fit the model amplitudes before being combined at the covariance level. These measurements display no preference for the model with a BAO feature over the broadband.}
\label{fig:bao-lrg-measurement}
\vspace{.2in}
\end{figure}

\begin{figure}
\centering
\includegraphics[width=.45\textwidth]{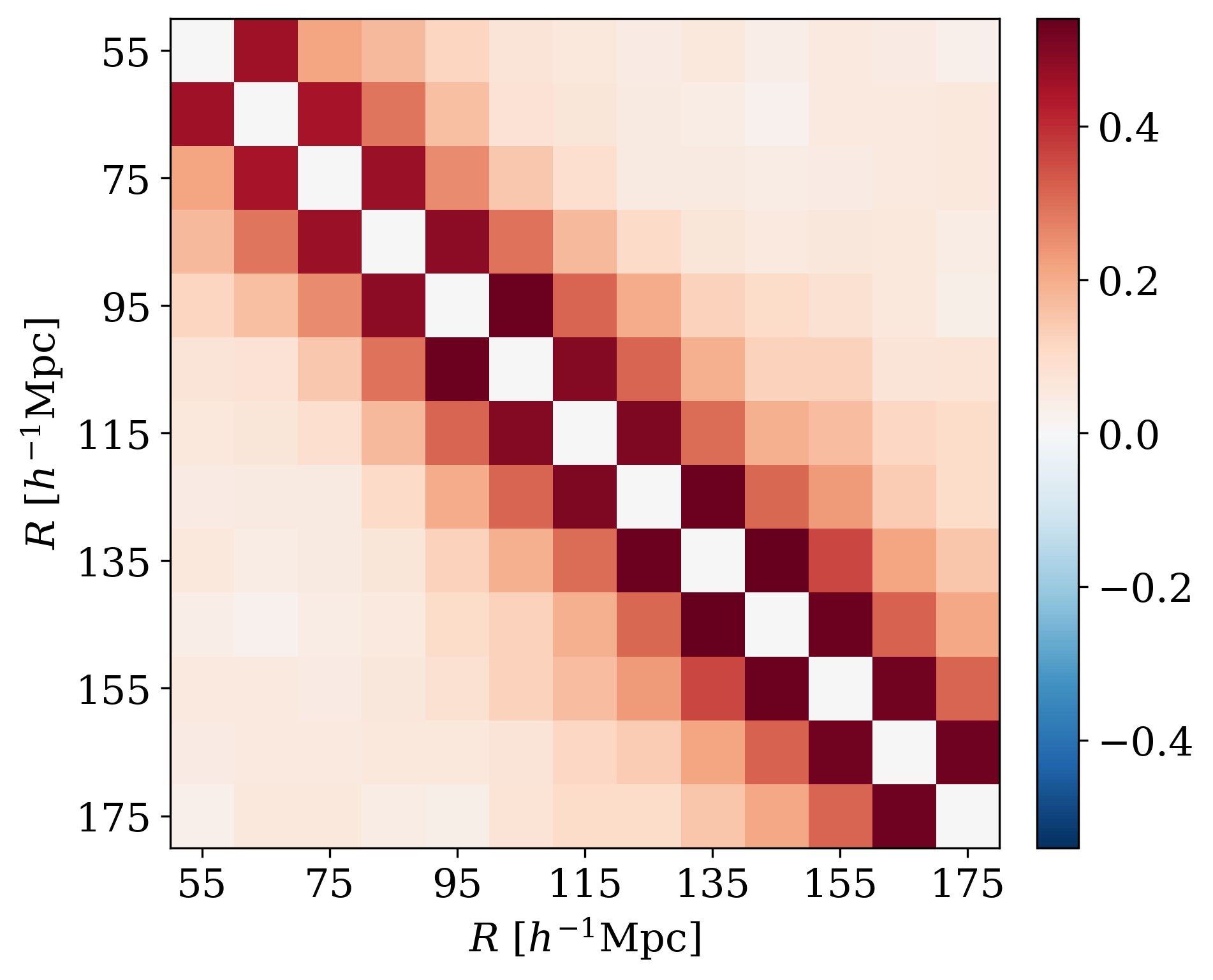}
\caption[]{The correlation matrix corresponding to the measurements shown in Figure \ref{fig:bao-lrg-measurement}, and accounted for in our BAO analysis. Here the identity matrix has been subtracted for clarity. Adjacent bins are positively correlated, although not as strongly as the larger-scale measurements shown in Figure \ref{fig:cov-LRG}.}
\label{fig:bao-lrg-cov}
\vspace{.2in}
\end{figure}

Figure~\ref{fig:bao-lrg-measurement} shows the inverse-variance-combined LRG signal against the two model templates, and Figure~\ref{fig:bao-lrg-cov} the corresponding combined correlation matrix, with the identity subtracted to highlight the off-diagonal structure. We find no evidence for a BAO feature in the multiplet alignment signal. In the three redshift bins individually, the matched-filter significances are $S/N = -0.56$, $-0.64$, and $-0.09$ for the low, middle, and high-redshift LRG samples respectively; combined, $S/N = -0.78$. The data prefer the smooth template, with a combined $\Delta\chi^2$ of -9.2, although the magnitude is well within the noise. Repeating the analysis with the merged regions of Section~\ref{sec:significance} reduces the sensitivity in the acoustic window by roughly $10\%$ and shifts the combined matched-filter significance by less than the scatter between merging-grid choices, leaving the conclusion unchanged.

To see if this non-detection reflects the true absence of a feature rather than an insensitive measurement, we performed an injection-recovery test. We injected a synthetic BAO signal of amplitude $A$, in units of the model-predicted wiggle amplitude ($A = 1$ corresponding to the feature at the strength predicted for each sample's fitted $\tau$), coherently into the per-region measurement vectors, and re-ran the analysis. The results indicate that our measurements would be sensitive to a BAO feature at the model-predicted amplitude, recovering it at $\sim3\sigma$. Its absence could indicate that the linear-alignment model is over-predicting the response of multiplet alignment to the BAO feature, although could simply be variance. 

\begin{figure}
\centering
\includegraphics[width=.48\textwidth]{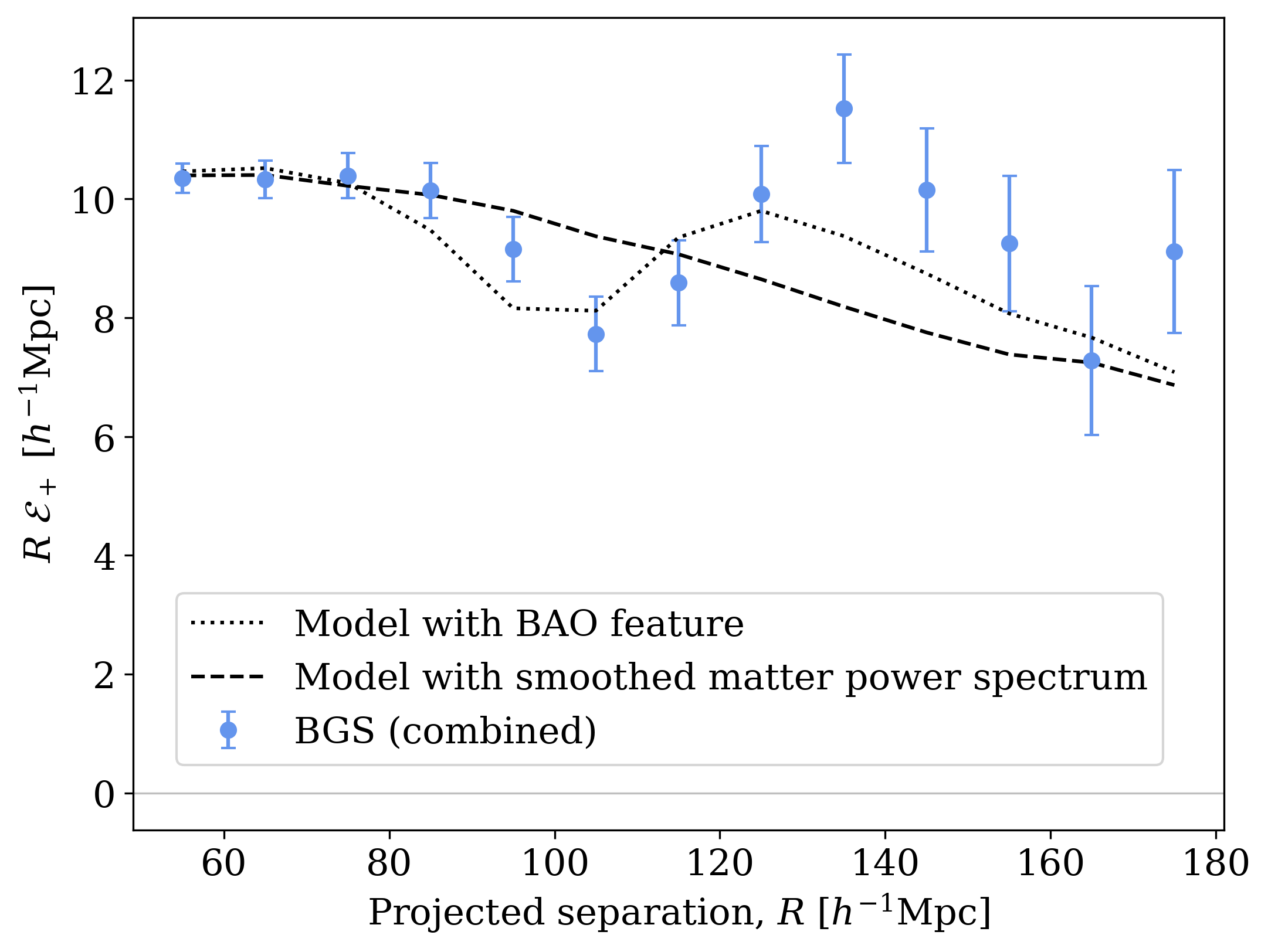}
\caption[]{Figure \ref{fig:bao-lrg-measurement} but averaged over the BGS redshift bins. Although the measurements appear to have a feature similar to the BAO expectation, they systematically rise with $R$ relative to both models. After fitting for an $R$-based correction, we do not find a consistent $R$ dependence of the correction between redshift bins, indicating that this excess is not due to a single large angular systematic.}
\label{fig:bao-bgs-measurement}
\end{figure}

Figure~\ref{fig:bao-bgs-measurement} similarly shows the combined signal of all BGS redshift bins. In contrast to the LRG sample, the BGS measurements are not well described by either model template. A feature appears to be present here, but the signal rises above both the wiggle and no-wiggle predictions at large separations, and the broadband fit is only moderately good ($\chi^2/{\rm dof} = 1.6$ for the wiggle model, $1.8$ for no-wiggle, combined over 60 ${\rm dof}$), with the weakest agreement concentrated in BGS3. This scale-dependent excess is strongest in the lowest-redshift bins. The combined measurements yield $S/N=-3.28$ and $\Delta\chi^2=13.1$, corresponding to $3.6\sigma$. Although $\Delta\chi^2$ indicates a preference for the BAO feature, fitting amplitudes separately to each template makes the test biased by excess at large $R$, where the BAO feature appears as a peak over the smooth model.

This large-scale excess in the BGS signal could be a sign of unaccounted angular systematics, so we test for one by fitting an empirical, per-redshift-bin correction to absorb $R$-dependent systematics. For each redshift bin, we fit the wiggle and no-wiggle amplitudes as above, average the two $\tau$-scaled model curves into a single broadband reference (so the small acoustic wiggle has negligible leverage), and fit a linear correction function to the data's residual against that reference. This is fit by generalized least squares under the same region-block covariance used throughout, and subtracted from the data on top of the existing per-region randoms subtraction. For this test we extend the $R$ bins to $55 < R < 210\,h^{-1}\mathrm{Mpc}$ to capture the excess where it is largest.

The correction improves the fit for both models, but neither becomes an acceptable description of the data. Notably, the fitted corrections are not consistent across redshifts. BGS1-3 require a growing positive term at large $R$ while BGS4-5 require a negative one. This argues against a single large angular systematic. Some residual geometric contamination cannot be ruled out, but the excess is equally consistent with noise. 

We conclude that a robust acoustic measurement in the alignment signal for both LRG and BGS is better pursued with 3D estimators and the greater statistical power of a future DESI data release.

\vspace{.1in}\section{Conclusion}

We present the largest-scale signal of intrinsic tidal shear yet detected. Multiplets of galaxies in DESI's DR2 LRG samples display significant alignment with the surrounding matter field at separations over 200 $h^{-1}$Mpc, and we find alignment of BGS and ELG multiplets over 100 $h^{-1}$Mpc. These represent an order of magnitude improvement over previous alignment measurements, reflecting advancements in data, measurement methods, and new multiplet definitions. 

We also present a sample-dependent method for identifying multiplets, which increases the maximum separation between multiplet members as galaxy density decreases. This results in larger multiplets relative to previous work and a higher signal-to-noise ratio of alignment measurements. We find no improvement from quantifying multiplets with a full ellipticity over a single orientation.

Although a weaker signal, the ELG measurements are especially notable as they present the opportunity to make alignment measurements in samples previously excluded from IA applications. They are also the sample with the most improved measurements over DR1, mostly due to the new multiplet definition, and the sample with the most potential for improvement as ELG density increases with future surveys.

We additionally measure, for the first time, the multiplet autocorrelation. Results are not consistent with model expectations calibrated to the cross-correlations, likely due to an underestimate of the galaxy bias. However, the autocorrelation provides the opportunity to estimate the linear galaxy bias directly from our samples, which we do for BGS.

Taking into account the uncertainty in galaxy bias and variations from a changing multiplet definition, we find the alignment strength of BGS multiplets to be positively correlated with redshift. Less significant redshift correlations are found in LRGs and ELGs, which may be related to these samples having a more consistent comoving density. In their redshift overlap, we compare the alignment strength of LRG and ELG multiplets by measuring them both relative to the full LRG sample. We find that ELGs have half the alignment amplitude of LRGs, demonstrating that the small-scale clustering of spiral galaxies is less correlated with their large-scale environment than that of large elliptical galaxies. It would be interesting to further examine these trends with hydrodynamic simulations.

Finally, we examine the potential presence of a BAO feature in LRG and BGS. The acoustic dip is present in mock data and an injection-recovery test indicates that our measurements should be sensitive to it. However we find no detection, even after a first-order accounting of an unknown angular systematic. This could still be a result of angular effects unaccounted for in the survey mask, or merely sample variance. We leave a more detailed analysis to future samples and a more suitable (3D) estimator.

Multiplets have now enabled the largest-scale detections of intrinsic tidal shear, and their orientations are robust to the imaging systematics that plague direct detections of IA with individual galaxy shapes. This unlocks the potential of intrinsic alignments as a practical cosmological probe. Especially promising applications involve exploring primordial features that have since inflated to the very large scales that multiplet alignment is sensitive to.

\vspace{.1in}
\section*{Acknowledgments}

CL acknowledges useful discussions with Christopher Hirata and the CCAPP Cosmo Group. CL is supported by an NSF Postdoctoral Fellowship under award 2502789.

This material is based upon work supported by the U.S. Department of Energy (DOE), Office of Science, Office of High-Energy Physics, under Contract No. DE–AC02–05CH11231, and by the National Energy Research Scientific Computing Center, a DOE Office of Science User Facility under the same contract. Additional support for DESI was provided by the U.S. National Science Foundation (NSF), Division of Astronomical Sciences under Contract No. AST-0950945 to the NSF’s National Optical-Infrared Astronomy Research Laboratory; the Science and Technology Facilities Council of the United Kingdom; the Gordon and Betty Moore Foundation; the Heising-Simons Foundation; the French Alternative Energies and Atomic Energy Commission (CEA); the Secretariat of Science, Humanities, Technology and Innovation (SECIHTI) of Mexico; the Ministry of Science, Innovation and Universities of Spain (MICIU/AEI/10.13039/501100011033), and by the DESI Member Institutions: \url{https://www.desi.lbl.gov/collaborating-institutions}. Any opinions, findings, and conclusions or recommendations expressed in this material are those of the author(s) and do not necessarily reflect the views of the U. S. National Science Foundation, the U. S. Department of Energy, or any of the listed funding agencies.

The authors are honored to be permitted to conduct scientific research on I'oligam Du'ag (Kitt Peak), a mountain with particular significance to the Tohono O’odham Nation.

Generative AI (Anthropic's Claude) was used in this work to assist with software development: optimizing existing functions, generating batch scripts, editing plotting scripts, writing unit tests, debugging, and testing pipelines for modeling and significance estimation. Final measurement code that included AI-generated changes was validated against the original pipeline and an independent TreeCorr-based implementation. The modeling and significance code was tested for consistency with previously validated human-made code and analytic computations.
\section*{Data Availability}

Data plotted in this paper can be downloaded from \href{https://zenodo.org/records/22697934}{zenodo.org/records/22697934}.

Code used in this work is publicly available in the repository \href{https://github.com/cmlamman/spec-IA}{github.com/cmlamman/spec-IA}.

\vspace{2in}
\bibliographystyle{mnras}
\bibliography{references}

@article{desicollaborationDataRelease12026,
	title = {Data {Release} 1 of the {Dark} {Energy} {Spectroscopic} {Instrument}},
	volume = {171},
	issn = {0004-6256},
	url = {https://ui.adsabs.harvard.edu/abs/2026AJ....171..285D},
	doi = {10.3847/1538-3881/ae4c43},
	urldate = {2026-09-18},
	journal = {The Astronomical Journal},
	publisher = {IOP},
	author = {{DESI Collaboration} and Abdul Karim, M. and Adame, A. G. and Aguado, D. and Aguilar, J. and Ahlen, S. and Alam, S. and Aldering, G. and Alexander, D. M. and Alfarsy, R. and Allen, L. and Allende Prieto, C. and Alves, O. and Anand, A. and Andrade, U. and Armengaud, E. and Avila, S. and Aviles, A. and Awan, H. and Bailey, S. and Baleato Lizancos, A. and Ballester, O. and Bault, A. and Bautista, J. and Bean, R. and Behera, J. and BenZvi, S. and Beraldo e Silva, L. and Bermejo-Climent, J. R. and Beutler, F. and Bianchi, D. and Blake, C. and Blum, R. and Bolton, A. S. and Bonici, M. and Brieden, S. and Brodzeller, A. and Brooks, D. and Buckley-Geer, E. and Burtin, E. and Byström, A. and Canning, R. and Carnero Rosell, A. and Carr, A. and Carrilho, P. and Casas, L. and Castander, F. J. and Cereskaite, R. and Cervantes-Cota, J. L. and Chaussidon, E. and Chaves-Montero, J. and Chen, S. and Chen, X. and Circosta, C. and Claybaugh, T. and Cole, S. and Cooper, A. P. and Cousinou, M.-C. and Cuceu, A. and Davis, T. M. and Dawson, K. S. and de Belsunce, R. and de la Cruz, R. and de la Macorra, A. and de Mattia, A. and Deiosso, N. and Della Costa, J. and Demina, R. and Demirbozan, U. and DeRose, J. and Dey, A. and Dey, B. and Ding, J. and Ding, Z. and Doel, P. and Douglass, K. and Dowicz, M. and Ebina, H. and Edelstein, J. and Eisenstein, D. J. and Elbers, W. and Emas, N. and Escoffier, S. and Fagrelius, P. and Fan, X. and Fanning, K. and Favole, G. and Fawcett, V. A. and Fernández-García, E. and Ferraro, S. and Findlay, N. and Font-Ribera, A. and Forero-Romero, J. E. and Forero-Sánchez, D. and Frenk, C. S. and Gänsicke, B. T. and Galbany, L. and García-Bellido, J. and Garcia-Quintero, C. and Garrison, L. H. and Gaztañaga, E. and Gil-Marín, H. and Gloudemans, A. and Gnedin, O. Y. and Gontcho A Gontcho, S. and Gonzalez, D. and Gonzalez-Morales, A. X. and Gonzalez-Perez, V. and Gordon, C. and Graur, O. and Green, D. and Gruen, D. and Gsponer, R. and Guandalin, C. and Gutierrez, G. and Guy, J. and Hahn, C. and Han, J. J. and Han, J. and He, S. and Herrera-Alcantar, H. K. and Heydenreich, S. and Honscheid, K. and Hou, J. and Howlett, C. and Huterer, D. and Iršič, V. and Ishak, M. and Jacques, A. and Jiang, L. and Jimenez, J. and Jing, Y. P. and Joachimi, B. and Joudaki, S. and Joyce, R. and Jullo, E. and Juneau, S. and Karaçaylı, N. G. and Karim, T. and Kehoe, R. and Kent, S. and Khederlarian, A. and Kirkby, D. and Kisner, T. and Kitaura, F.-S. and Kizhuprakkat, N. and Kong, H. and Koposov, S. E. and Kremin, A. and Krolewski, A. and Lahav, O. and Lai, Y. and Lamman, C. and Lan, T.-W. and Landriau, M. and Lang, D. and Lange, J. U. and Lasker, J. and Le Goff, J. M. and Le Guillou, L. and Leauthaud, A. and Levi, M. E. and Li, S. and Li, T. S. and Liu, W. and Lodha, K. and Lokken, M. and Luo, Y. and Magneville, C. and Manera, M. and Manser, C. J. and Margala, D. and Martini, P. and Maus, M. and McCullough, J. and McDonald, P. and Medina, G. E. and Medina-Varela, L. and Meisner, A. and Mena-Fernández, J. and Menegas, A. and Meneses-Rizo, J. and Mezcua, M. and Miquel, R. and Montero-Camacho, P. and Moon, J. and Moustakas, J. and Muñoz-Gutiérrez, A. and Mu noz-Santos, D. and Myers, A. D. and Myles, J. and Nadathur, S. and Najita, J. and Napolitano, L. and Newman, J. A. and Nikakhtar, F. and Nikutta, R. and Niz, G. and Noriega, H. E. and Nugent, P.},
	month = may,
	year = {2026},
	note = {ADS Bibcode: 2026AJ....171..285D},
	pages = {285},
}

@article{silberRoboticMultiobjectFocal2023,
	title = {The {Robotic} {Multiobject} {Focal} {Plane} {System} of the {Dark} {Energy} {Spectroscopic} {Instrument} ({DESI})},
	volume = {165},
	issn = {0004-6256},
	url = {https://ui.adsabs.harvard.edu/abs/2023AJ....165....9S},
	doi = {10.3847/1538-3881/ac9ab1},
	urldate = {2026-09-03},
	journal = {The Astronomical Journal},
	publisher = {IOP},
	author = {Silber, Joseph Harry and Fagrelius, Parker and Fanning, Kevin and Schubnell, Michael and Aguilar, Jessica Nicole and Ahlen, Steven and Ameel, Jon and Ballester, Otger and Baltay, Charles and Bebek, Chris and Benton Beard, Dominic and Besuner, Robert and Cardiel-Sas, Laia and Casas, Ricard and Castander, Francisco Javier and Claybaugh, Todd and Dobson, Carl and Duan, Yutong and Dunlop, Patrick and Edelstein, Jerry and Emmet, William T. and Elliott, Ann and Evatt, Matthew and Gershkovich, Irena and Guy, Julien and Harris, Stu and Heetderks, Henry and Heetderks, Ian and Honscheid, Klaus and Illa, Jose Maria and Jelinsky, Patrick and Jelinsky, Sharon R. and Jimenez, Jorge and Karcher, Armin and Kent, Stephen and Kirkby, David and Kneib, Jean-Paul and Lambert, Andrew and Lampton, Mike and Leitner, Daniela and Levi, Michael and McCauley, Jeremy and Meisner, Aaron and Miller, Timothy N. and Miquel, Ramon and Mundet, Juliá and Poppett, Claire and Rabinowitz, David and Reil, Kevin and Roman, David and Schlegel, David and Serrano, Santiago and Van Shourt, William and Sprayberry, David and Tarlé, Gregory and Tie, Suk Sien and Weaverdyck, Curtis and Zhang, Kai and Azzaro, Marco and Bailey, Stephen and Becerril, Santiago and Blackwell, Tami and Bouri, Mohamed and Brooks, David and Buckley-Geer, Elizabeth and Castro, Jose Peñate and Derwent, Mark and Dey, Arjun and Dhungana, Govinda and Doel, Peter and Eisenstein, Daniel J. and Fahim, Nasib and Garcia-Bellido, Juan and Gaztañaga, Enrique and A Gontcho, Satya Gontcho and Gutierrez, Gaston and Hörler, Philipp and Kehoe, Robert and Kisner, Theodore and Kremin, Anthony and Kronig, Luzius and Landriau, Martin and Le Guillou, Laurent and Martini, Paul and Moustakas, John and Palanque-Delabrouille, Nathalie and Peng, Xiyan and Percival, Will and Prada, Francisco and Allende Prieto, Carlos and de Rivera, Guillermo Gonzalez and Sanchez, Eusebio and Sanchez, Justo and Sharples, Ray and Soares-Santos, Marcelle and Schlafly, Edward and Weaver, Benjamin Alan and Zhou, Zhimin and Zhu, Yaling and Zou, Hu and {DESI Collaboration}},
	month = jan,
	year = {2023},
	note = {ADS Bibcode: 2023AJ....165....9S},
	pages = {9},
}

@article{mena-fernandezHODdependentSystematicsLuminous2025,
	title = {{HOD}-dependent systematics for luminous red galaxies in the {DESI} 2024 {BAO} analysis},
	volume = {2025},
	issn = {1475-7516},
	url = {https://ui.adsabs.harvard.edu/abs/2025JCAP...01..133M},
	doi = {10.1088/1475-7516/2025/01/133},
	urldate = {2026-09-03},
	journal = {Journal of Cosmology and Astroparticle Physics},
	publisher = {IOP},
	author = {Mena-Fernández, J. and Garcia-Quintero, C. and Yuan, S. and Hadzhiyska, B. and Alves, O. and Rashkovetskyi, M. and Seo, H. and Padmanabhan, N. and Nadathur, S. and Howlett, C. and Alam, S. and Rocher, A. and Ross, A. J. and Sanchez, E. and Ishak, M. and Aguilar, J. and Ahlen, S. and Andrade, U. and BenZvi, S. and Brooks, D. and Burtin, E. and Chen, S. and Chen, X. and Claybaugh, T. and Cole, S. and de la Macorra, A. and de Mattia, A. and Dey, Arjun and Dey, B. and Ding, Z. and Doel, P. and Fanning, K. and Forero-Romero, J. E. and Gaztañaga, E. and Gil-Marín, H. and Gontcho, S. Gontcho A. and Gutierrez, G. and Guy, J. and Hahn, C. and Honscheid, K. and Juneau, S. and Kremin, A. and Landriau, M. and Le Guillou, L. and Levi, M. E. and Manera, M. and Martini, P. and Medina-Varela, L. and Meisner, A. and Miquel, R. and Moustakas, J. and Mueller, E. and Muñoz-Gutiérrez, A. and Myers, A. D. and Newman, J. A. and Nie, J. and Niz, G. and Paillas, E. and Palanque-Delabrouille, N. and Percival, W. J. and Poppett, C. and Pérez-Fernández, A. and Rosado-Marin, A. and Rossi, G. and Ruggeri, R. and Saulder, C. and Schlegel, D. and Schubnell, M. and Sprayberry, D. and Tarlé, G. and Vargas-Magaña, M. and Weaver, B. A. and Yu, J. and Zhang, H. and Zou, H.},
	month = jan,
	year = {2025},
	note = {ADS Bibcode: 2025JCAP...01..133M},
	pages = {133},
}

@article{rossConstructionLargescaleStructure2025a,
	title = {The construction of large-scale structure catalogs for the {Dark} {Energy} {Spectroscopic} {Instrument}},
	volume = {2025},
	issn = {1475-7516},
	url = {https://ui.adsabs.harvard.edu/abs/2025JCAP...01..125R},
	doi = {10.1088/1475-7516/2025/01/125},
	urldate = {2026-09-03},
	journal = {Journal of Cosmology and Astroparticle Physics},
	publisher = {IOP},
	author = {Ross, A. J. and Aguilar, J. and Ahlen, S. and Alam, S. and Anand, A. and Bailey, S. and Bianchi, D. and Brieden, S. and Brooks, D. and Burtin, E. and Carnero Rosell, A. and Chaussidon, E. and Claybaugh, T. and Cole, S. and Dawson, K. and de la Macorra, A. and de Mattia, A. and Dey, A. and Dey, B. and Doel, P. and Fanning, K. and Ferraro, S. and Ereza, J. and Font-Ribera, A. and Forero-Romero, J. E. and Gaztañaga, E. and Gil-Marín, H. and Gontcho A Gontcho, S. and Gonzalez-Morales, A. X. and Guy, J. and Hahn, C. and Heydenreich, S. and Honscheid, K. and Howlett, C. and Ishak, M. and Karim, T. and Kirkby, D. and Kisner, T. and Kong, H. and Kremin, A. and Krolewski, A. and Lambert, A. and Landriau, M. and Lasker, J. and Guillou, L. L. and Levi, M. E. and Manera, M. and Martini, P. and McDonald, P. and Meisner, A. and Miquel, R. and Moon, J. and Moustakas, J. and Muñoz-Gutiérrez, A. and Myers, A. D. and Nadathur, S. and Napolitano, L. and Newman, J. A. and Nie, J. and Niz, G. and Palanque-Delabrouille, N. and Percival, W. J. and Poppett, C. and Prada, F. and Raichoor, A. and Ravoux, C. and Rezaie, M. and Rosado-Marin, A. and Rossi, G. and Samushia, L. and Sanchez, E. and Schlafly, E. F. and Schlegel, D. and Seo, H. and Smith, A. and Sprayberry, D. and Tarlé, G. and Valcin, D. and Vargas-Magaña, M. and Weaver, B. A. and Wilson, M. J. and Yu, J. and Zarrouk, P. and Zhao, C. and Zhou, R. and Zou, H.},
	month = jan,
	year = {2025},
	note = {ADS Bibcode: 2025JCAP...01..125R},
	pages = {125},
}

@article{hartlapWhyYourModel2007,
	title = {Why your model parameter confidences might be too optimistic. {Unbiased} estimation of the inverse covariance matrix},
	volume = {464},
	issn = {0004-6361},
	url = {https://ui.adsabs.harvard.edu/abs/2007A&A...464..399H},
	doi = {10.1051/0004-6361:20066170},
	urldate = {2026-08-31},
	journal = {Astronomy and Astrophysics},
	publisher = {EDP},
	author = {Hartlap, J. and Simon, P. and Schneider, P.},
	month = mar,
	year = {2007},
	note = {ADS Bibcode: 2007A\&A...464..399H},
	pages = {399--404},
}

@article{siegelIntrinsicAlignmentDemographics2025,
	title = {Intrinsic alignment demographics for next-generation lensing: {Revealing} galaxy property trends with {DESI} {Y1} direct measurements},
	shorttitle = {Intrinsic alignment demographics for next-generation lensing},
	url = {https://ui.adsabs.harvard.edu/abs/2025arXiv250711530S/abstract},
	doi = {10.48550/arXiv.2507.11530},
	language = {en},
	urldate = {2026-08-31},
	journal = {arXiv e-prints},
	author = {Siegel, J. and McCullough, J. and Amon, A. and Lamman, C. and Jeffrey, N. and Joachimi, B. and Hoekstra, H. and Heydenreich, S. and Ross, A. J. and Aguilar, J. and Ahlen, S. and Bianchi, D. and Blake, C. and Brooks, D. and Castander, F. J. and Claybaugh, T. and de la Macorra, A. and DeRose, J. and Doel, P. and Emas, N. and Ferraro, S. and Font-Ribera, A. and Forero-Romero, J. E. and Gaztañaga, E. and Gontcho, S. Gontcho A. and Gutierrez, G. and Honscheid, K. and Ishak, M. and Joudaki, S. and Kehoe, R. and Kirkby, D. and Kisner, T. and Krolewski, A. and Lahav, O. and Lambert, A. and Landriau, M. and Le Guillou, L. and Levi, M. E. and Manera, M. and Meisner, A. and Miquel, R. and Moustakas, J. and Nadathur, S. and Newman, J. A. and Niz, G. and Palanque-Delabrouille, N. and Percival, W. J. and Porredon, A. and Prada, F. and Pérez-Ràfols, I. and Rossi, G. and Sanchez, E. and Saulder, C. and Schlegel, D. and Schubnell, M. and Semenaite, A. and Silber, J. and Sprayberry, D. and Sun, Z. and Tarlé, G. and Weaver, B. A. and Zhou, R. and Zou, H.},
	month = jul,
	year = {2025},
	pages = {arXiv:2507.11530},
}

@misc{vaisakhDESIDR2Reference2026,
	title = {{DESI} {DR2} {Reference} {Mocks}: {Clustering} results from {UCHUU} {ELGs} and {QSOs}},
	shorttitle = {{DESI} {DR2} {Reference} {Mocks}},
	url = {https://ui.adsabs.harvard.edu/abs/2026arXiv260628559V},
	doi = {10.48550/arXiv.2606.28559},
	urldate = {2026-08-08},
	publisher = {arXiv},
	author = {Vaisakh, R. and Lasker, J. and Kehoe, R. and Amalbert, A. and Khan, N. and Fernandez-Garcia, E. and Prada, F. and Wang, M. S. and DeRose, J. and Bailey, S. and Ross, A. J. and Aguilar, J. and Ahlen, S. and Bianchi, D. and Brooks, D. and Castander, F. J. and Claybaugh, T. and Dawson, K. S. and de la Macorra, A. and Ferraro, S. and Forero-Romero, J. E. and Gaztanaga, E. and Gontcho, Satya Gontcho A and Gutierrez, G. and Hahn, C. and Ishak, M. and Joyce, R. and Juneau, S. and Kisner, T. and Kremin, A. and Lamman, C. and Landriau, M. and Levi, M. E. and Manera, M. and Miquel, R. and Myers, A. D. and Nadathur, S. and Percival, W. J. and Perez-Rafols, I. and Rossi, G. and Sanchez, E. and Schlegel, D. and Seo, H. and Tarle, G. and Weaver, B. A. and Zhou, R. and Zou, H.},
	month = jun,
	year = {2026},
	note = {ADS Bibcode: 2026arXiv260628559V},
}

@article{jarvisSkewnessApertureMass2004,
	title = {The skewness of the aperture mass statistic},
	volume = {352},
	issn = {0035-8711},
	url = {https://doi.org/10.1111/j.1365-2966.2004.07926.x},
	doi = {10.1111/j.1365-2966.2004.07926.x},
	number = {1},
	urldate = {2026-07-31},
	journal = {Monthly Notices of the Royal Astronomical Society},
	author = {Jarvis, M. and Bernstein, G. and Jain, B.},
	month = jul,
	year = {2004},
	pages = {338--352},
}

@article{moreOverdensityMassesFriendsoffriends2011,
	title = {The {Overdensity} and {Masses} of the {Friends}-of-friends {Halos} and {Universality} of {Halo} {Mass} {Function}},
	volume = {195},
	issn = {0067-0049},
	url = {https://ui.adsabs.harvard.edu/abs/2011ApJS..195....4M},
	doi = {10.1088/0067-0049/195/1/4},
	urldate = {2026-07-28},
	journal = {The Astrophysical Journal Supplement Series},
	publisher = {IOP},
	author = {More, Surhud and Kravtsov, Andrey V. and Dalal, Neal and Gottlöber, Stefan},
	month = jul,
	year = {2011},
	note = {ADS Bibcode: 2011ApJS..195....4M},
	pages = {4},
}

@article{paulApparentParityViolation2024,
	title = {Apparent {Parity} {Violation} in the {Observed} {Galaxy} {Trispectrum}},
	volume = {133},
	issn = {0031-9007},
	url = {https://ui.adsabs.harvard.edu/abs/2024PhRvL.133l1001P},
	doi = {10.1103/PhysRevLett.133.121001},
	urldate = {2026-07-28},
	journal = {Physical Review Letters},
	publisher = {APS},
	author = {Paul, Pritha and Clarkson, Chris and Maartens, Roy},
	month = sep,
	year = {2024},
	note = {ADS Bibcode: 2024PhRvL.133l1001P},
	pages = {121001},
}

@misc{achucarroInflationTheoryObservations2022,
	title = {Inflation: {Theory} and {Observations}},
	shorttitle = {Inflation},
	url = {https://ui.adsabs.harvard.edu/abs/2022arXiv220308128A},
	doi = {10.48550/arXiv.2203.08128},
	urldate = {2026-07-28},
	publisher = {arXiv},
	author = {Achúcarro, Ana and Biagetti, Matteo and Braglia, Matteo and Cabass, Giovanni and Caldwell, Robert and Castorina, Emanuele and Chen, Xingang and Coulton, William and Flauger, Raphael and Fumagalli, Jacopo and Ivanov, Mikhail M. and Lee, Hayden and Maleknejad, Azadeh and Meerburg, P. Daniel and Moradinezhad Dizgah, Azadeh and Palma, Gonzalo A. and Pimentel, Guilherme L. and Renaux-Petel, Sébastien and Wallisch, Benjamin and Wandelt, Benjamin D. and Witkowski, Lukas T. and Kimmy Wu, W. L.},
	month = mar,
	year = {2022},
	note = {ADS Bibcode: 2022arXiv220308128A},
}

@article{okumuraTestingTidalAlignment2020,
	title = {Testing tidal alignment models for anisotropic correlations of halo ellipticities with {N}-body simulations},
	volume = {494},
	issn = {0035-8711},
	url = {https://doi.org/10.1093/mnras/staa718},
	doi = {10.1093/mnras/staa718},
	number = {1},
	urldate = {2026-07-14},
	journal = {Monthly Notices of the Royal Astronomical Society},
	author = {Okumura, Teppei and Taruya, Atsushi and Nishimichi, Takahiro},
	month = may,
	year = {2020},
	pages = {694--702},
}

@article{ichikawaEnvironmentalDependenceHalo2026,
	title = {The {Environmental} {Dependence} of {Halo} {Intrinsic} {Alignments}: {Stronger} {Signals} in {Underdense} {Regions}},
	shorttitle = {The {Environmental} {Dependence} of {Halo} {Intrinsic} {Alignments}},
	url = {https://ui.adsabs.harvard.edu/abs/2026arXiv260711114I/abstract},
	language = {en},
	urldate = {2026-07-14},
	journal = {arXiv e-prints},
	author = {Ichikawa, Masaya and Shi, Jingjing and Kurita, Toshiki and Takada, Masahiro and Nishimichi, Takahiro and Blot, Linda},
	month = jul,
	year = {2026},
	pages = {arXiv:2607.11114},
}

@article{hervaspetersUNIONSDirectMeasurement2025,
	title = {{UNIONS}: {A} direct measurement of intrinsic alignment with {BOSS}/{eBOSS} spectroscopy},
	volume = {699},
	issn = {0004-6361},
	shorttitle = {{UNIONS}},
	url = {https://ui.adsabs.harvard.edu/abs/2025A&A...699A.201H},
	doi = {10.1051/0004-6361/202453442},
	urldate = {2026-07-12},
	journal = {Astronomy and Astrophysics},
	publisher = {EDP},
	author = {Hervas Peters, Fabian and Kilbinger, Martin and Paviot, Romain and Baumont, Lucie and Russier, Elisa and Zhang, Ziwen and Murray, Calum and Pettorino, Valeria and de Boer, Thomas and Fabbro, Sébastien and Guerrini, Sacha and Hildebrandt, Hendrik and Hudson, Michael J. and Van Waerbeke, Ludovic and Wittje, Anna},
	month = jul,
	year = {2025},
	note = {ADS Bibcode: 2025A\&A...699A.201H},
	pages = {A201},
}

@article{briedenModelagnosticInterpretation102022,
	title = {Model-agnostic interpretation of 10 billion years of cosmic evolution traced by {BOSS} and {eBOSS} data},
	volume = {2022},
	issn = {1475-7516},
	url = {https://ui.adsabs.harvard.edu/abs/2022JCAP...08..024B},
	doi = {10.1088/1475-7516/2022/08/024},
	urldate = {2026-07-07},
	journal = {Journal of Cosmology and Astroparticle Physics},
	publisher = {IOP},
	author = {Brieden, Samuel and Gil-Marín, Héctor and Verde, Licia},
	month = aug,
	year = {2022},
	note = {ADS Bibcode: 2022JCAP...08..024B},
	pages = {024},
}

@article{andradeValidationDESIDR22025,
	title = {Validation of the {DESI} {DR2} measurements of baryon acoustic oscillations from galaxies and quasars},
	volume = {112},
	issn = {1550-79980556-2821},
	url = {https://ui.adsabs.harvard.edu/abs/2025PhRvD.112h3512A},
	doi = {10.1103/kdys-w8vl},
	urldate = {2026-07-07},
	journal = {Physical Review D},
	publisher = {APS},
	author = {Andrade, U. and Paillas, E. and Mena-Fernández, J. and Li, Q. and Ross, A. J. and Nadathur, S. and Rashkovetskyi, M. and Pérez-Fernández, A. and Seo, H. and Sanders, N. and Alves, O. and Chen, X. and Deiosso, N. and de Mattia, A. and White, M. and Karim, M. Abdul and Ahlen, S. and Armengaud, E. and Aviles, A. and Bansal, P. and Behera, J. and Bianchi, D. and Brieden, S. and Brodzeller, A. and Brooks, D. and Burtin, E. and Calderon, R. and Canning, R. and Rosell, A. Carnero and Casas, L. and Castander, F. J. and Charles, M. and Chaussidon, E. and Chaves-Montero, J. and Claybaugh, T. and Cole, S. and Cooper, A. P. and Cuceu, A. and Dawson, K. S. and de la Macorra, A. and Della Costa, J. and Dey, A. and Dey, B. and Ding, Z. and Doel, P. and Eisenstein, D. J. and Elbers, W. and Fernández-García, E. and Ferraro, S. and Font-Ribera, A. and Forero-Romero, J. E. and Garcia-Quintero, C. and Garrison, L. H. and Gaztañaga, E. and Gil-Marín, H. and Gontcho, S. Gontcho A. and Gonzalez-Morales, A. X. and Gordon, C. and Gutierrez, G. and Guy, J. and Hahn, C. and He, S. and Herrera-Alcantar, H. K. and Honscheid, K. and Howlett, C. and Huterer, D. and Ishak, M. and Juneau, S. and Kehoe, R. and Kirkby, D. and Kisner, T. and Kremin, A. and Lahav, O. and Lamman, C. and Landriau, M. and Le Guillou, L. and Leauthaud, A. and Levi, M. E. and Magneville, C. and Manera, M. and Martini, P. and Matthewson, W. L. and Meisner, A. and Miquel, R. and Moustakas, J. and Muñoz-Gutiérrez, A. and Muñoz-Santos, D. and Myers, A. D. and Napolitano, L. and Newman, J. A. and Noriega, H. E. and Palanque-Delabrouille, N. and Pan, J. and Percival, W. J. and Pérez-Ràfols, I. and Poppett, C. and Prada, F. and Raichoor, A. and Ramírez-Pérez, C. and Ravoux, C. and Rossi, G. and Ruggeri, R. and Samushia, L. and Sanchez, E. and Schlegel, D. and Schubnell, M. and Sinigaglia, F. and Sprayberry, D. and Tan, T. and Tarlé, G. and Taylor, P. and Turner, W. and Vaisakh, R. and Vargas-Magaña, M. and Walther, M. and Weaver, B. A. and Wolfson, M. and Yèche, C. and Yu, J. and Zarrouk, P. and Zhou, R. and Zou, H. and {DESI Collaboration}},
	month = oct,
	year = {2025},
	note = {ADS Bibcode: 2025PhRvD.112h3512A},
	pages = {083512},
}

@misc{herleAssemblyBiasRedshift2026,
	title = {Assembly bias and the redshift evolution of intrinsic alignments for {LRGs}},
	url = {https://ui.adsabs.harvard.edu/abs/2026arXiv260700785H},
	doi = {10.48550/arXiv.2607.00785},
	urldate = {2026-07-07},
	publisher = {arXiv},
	author = {Herle, A. and Chisari, N. E. and Hoekstra, H. and Neumann, D.},
	month = jul,
	year = {2026},
	note = {ADS Bibcode: 2026arXiv260700785H},
}

@article{fernandez-garciaDESIDR2Reference2026,
	title = {{DESI} {DR2} reference mocks: clustering results from {UCHUU}-{BGS} and {LRG}},
	volume = {2026},
	issn = {1475-7516},
	shorttitle = {{DESI} {DR2} reference mocks},
	url = {https://ui.adsabs.harvard.edu/abs/2026JCAP...05..002F},
	doi = {10.1088/1475-7516/2026/05/002},
	urldate = {2026-07-02},
	journal = {Journal of Cosmology and Astroparticle Physics},
	publisher = {IOP},
	author = {Fernández-García, E. and Prada, F. and Smith, A. and DeRose, J. and Ross, A. J. and Bailey, S. and Wang, M. S. and Ding, Z. and Guandalin, C. and Lamman, C. and Vaisakh, R. and Kehoe, R. and Lasker, J. and Ishiyama, T. and Moore, S. M. and Cole, S. and Siudek, M. and Amalbert, A. and Salcedo, A. and Hearin, A. and Joachimi, B. and Rocher, A. and Saito, S. and Krolewski, A. and Slepian, Z. and Li, Q. and Dawson, K. S. and Jullo, E. and Aguilar, J. and Ahlen, S. and Bianchi, D. and Brooks, D. and Claybaugh, T. and de la Macorra, A. and Doel, P. and Ferraro, S. and Font-Ribera, A. and Forero-Romero, J. E. and Gontcho A Gontcho, S. and Gutierrez, G. and Honscheid, K. and Ishak, M. and Joyce, R. and Juneau, S. and Kirkby, D. and Kisner, T. and Kremin, A. and Lahav, O. and Lambert, A. and Landriau, M. and Levi, M. E. and Manera, M. and Miquel, R. and Moustakas, J. and Nadathur, S. and Percival, W. J. and Pérez-Ràfols, I. and Rossi, G. and Sanchez, E. and Schlegel, D. and Seo, H. and Silber, J. and Sprayberry, D. and Tarlé, G. and Weaver, B. A. and Zarrouk, P. and Zhou, R.},
	month = may,
	year = {2026},
	note = {ADS Bibcode: 2026JCAP...05..002F},
	pages = {002},
}

@article{2025RNAAS...9..305K,
	title = {Enhancing multiplet alignment measurements with imaging},
	volume = {9},
	doi = {10.3847/2515-5172/ae1c19},
	number = {11},
	journal = {Research Notes of the American Astronomical Society},
	author = {Kumwembe, Alexus Annika and Lamman, Claire and Eisenstein, Daniel and Aguilar, Jessica Nicole and Ahlen, Steven and Bianchi, Davide and Brooks, David and Claybaugh, Todd and Cuceu, Andrei and de la Macorra, Axel and Dey, Biprateep and Doel, Peter and Font-Ribera, Andreu and Forero-Romero, Jaime E. and Gaztañaga, Enrique and Gontcho A Gontcho, Satya and Gutierrez, Gaston and Ishak, Mustapha and Jimenez, Jorge and Joyce, Dick and Kehoe, Robert and Kisner, Theodore and Lahav, Ofer and Landriau, Martin and Manera, Marc and Miquel, Ramon and Nadathur, Seshadri and Palanque-Delabrouille, Nathalie and Pérez-Ràfols, Ignasi and Prada, Francisco and Rossi, Graziano and Sanchez, Eusebio and Schlegel, David and Seo, Hee-Jong and Silber, Joseph Harry and Sprayberry, David and Tarlé, Gregory and Weaver, Benjamin Alan},
	month = nov,
	year = {2025},
	note = {arXiv: 2510.08353 [astro-ph.CO]
Number: 305
tex.adsnote: Provided by the SAO/NASA Astrophysics Data System},
	pages = {305},
}

@article{abdulkarimDESIDR2Results2025,
	title = {{DESI} {DR2} results. {II}. {Measurements} of baryon acoustic oscillations and cosmological constraints},
	volume = {112},
	issn = {1550-79980556-2821},
	url = {https://ui.adsabs.harvard.edu/abs/2025PhRvD.112h3515A},
	doi = {10.1103/tr6y-kpc6},
	urldate = {2026-03-27},
	journal = {Physical Review D},
	publisher = {APS},
	author = {DESI Collaboration, . and Abdul Karim, M. and Aguilar, J. and Ahlen, S. and Alam, S. and Allen, L. and Allende Prieto, C. and Alves, O. and Anand, A. and Andrade, U. and Armengaud, E. and Aviles, A. and Bailey, S. and Baltay, C. and Bansal, P. and Bault, A. and Behera, J. and BenZvi, S. and Bianchi, D. and Blake, C. and Brieden, S. and Brodzeller, A. and Brooks, D. and Buckley-Geer, E. and Burtin, E. and Calderon, R. and Canning, R. and Rosell, A. Carnero and Carrilho, P. and Casas, L. and Castander, F. J. and Charles, M. and Chaussidon, E. and Chaves-Montero, J. and Chebat, D. and Chen, X. and Claybaugh, T. and Cole, S. and Cooper, A. P. and Cuceu, A. and Dawson, K. S. and de la Macorra, A. and de Mattia, A. and Deiosso, N. and Della Costa, J. and Demina, R. and Dey, A. and Dey, B. and Ding, Z. and Doel, P. and Edelstein, J. and Eisenstein, D. J. and Elbers, W. and Fagrelius, P. and Fanning, K. and Fernández-García, E. and Ferraro, S. and Font-Ribera, A. and Forero-Romero, J. E. and Frenk, C. S. and Garcia-Quintero, C. and Garrison, L. H. and Gaztañaga, E. and Gil-Marín, H. and Gontcho A Gontcho, S. and Gonzalez, D. and Gonzalez-Morales, A. X. and Gordon, C. and Green, D. and Gutierrez, G. and Guy, J. and Hadzhiyska, B. and Hahn, C. and He, S. and Herbold, M. and Herrera-Alcantar, H. K. and Ho, M.-F. and Honscheid, K. and Howlett, C. and Huterer, D. and Ishak, M. and Juneau, S. and Kamble, N. V. and Karaçaylı, N. G. and Kehoe, R. and Kent, S. and Kim, A. G. and Kirkby, D. and Kisner, T. and Koposov, S. E. and Kremin, A. and Krolewski, A. and Lahav, O. and Lamman, C. and Landriau, M. and Lang, D. and Lasker, J. and Le Goff, J. M. and Le Guillou, L. and Leauthaud, A. and Levi, M. E. and Li, Q. and Li, T. S. and Lodha, K. and Lokken, M. and Lozano-Rodríguez, F. and Magneville, C. and Manera, M. and Martini, P. and Matthewson, W. L. and Meisner, A. and Mena-Fernández, J. and Menegas, A. and Mergulhão, T. and Miquel, R. and Moustakas, J. and Muñoz-Gutiérrez, A. and Muñoz-Santos, D. and Myers, A. D. and Nadathur, S. and Naidoo, K. and Napolitano, L. and Newman, J. A. and Niz, G. and Noriega, H. E. and Paillas, E. and Palanque-Delabrouille, N. and Pan, J. and Peacock, J. A. and Pellejero Ibanez, M. and Percival, W. J. and Pérez-Fernández, A. and Pérez-Ràfols, I. and Pieri, M. M. and Poppett, C. and Prada, F. and Rabinowitz, D. and Raichoor, A. and Ramírez-Pérez, C. and Rashkovetskyi, M. and Ravoux, C. and Rich, J. and Rocher, A. and Rockosi, C. and Rohlf, J. and Román-Herrera, J. O. and Ross, A. J. and Rossi, G. and Ruggeri, R. and Ruhlmann-Kleider, V. and Samushia, L. and Sanchez, E. and Sanders, N. and Schlegel, D. and Schubnell, M. and Seo, H. and Shafieloo, A. and Sharples, R. and Silber, J. and Sinigaglia, F. and Sprayberry, D. and Tan, T. and Tarlé, G. and Taylor, P. and Turner, W. and Ureña-López, L. A. and Vaisakh, R. and Valdes, F. and Valogiannis, G. and Vargas-Magaña, M. and Verde, L. and Walther, M. and Weaver, B. A. and Weinberg, D. H. and White, M. and Wolfson, M. and Yèche, C. and Yu, J. and Zaborowski, E. A. and Zarrouk, P. and Zhai, Z. and Zhang, H. and Zhao, C. and Zhao, G. B. and Zhou, R. and Zou, H. and {DESI Collaboration}},
	month = oct,
	year = {2025},
	note = {ADS Bibcode: 2025PhRvD.112h3515A},
	pages = {083515},
}

@article{poppettOverviewFiberSystem2024,
	title = {Overview of the {Fiber} {System} for the {Dark} {Energy} {Spectroscopic} {Instrument}},
	volume = {168},
	issn = {0004-6256},
	url = {https://ui.adsabs.harvard.edu/abs/2024AJ....168..245P},
	doi = {10.3847/1538-3881/ad76a4},
	urldate = {2026-03-23},
	journal = {The Astronomical Journal},
	publisher = {IOP},
	author = {Poppett, Claire and Tyas, Luke and Aguilar, J. and Bebek, Christopher and Bramall, D. and Claybaugh, T. and Edelstein, J. and Fagrelius, P. and Heetderks, H. and Jelinsky, P. and Jelinsky, S. and Lafever, Robin and Lambert, A. and Lampton, M. and Levi, Michael E. and Martini, P. and Rockosi, C. and Schmoll, J. and Sharples, Ray M. and Sirk, Martin and Wishnow, Edward and Yu, Jiaxi and Ahlen, S. and Bault, A. and BenZvi, S. and Brooks, D. and Cole, S. and de la Macorra, A. and Dey, Arjun and Doel, P. and Fanning, K. and Font-Ribera, A. and Forero-Romero, J. E. and Gaztañaga, E. and Gontcho A Gontcho, S. and Gonzalez-Morales, A. X. and Hahn, C. and Honscheid, K. and Jimenez, J. and Juneau, S. and Kirkby, D. and Kremin, A. and Landriau, M. and Le Guillou, L. and Manera, M. and Meisner, A. and Miquel, R. and Moustakas, J. and Mueller, E. and Muñoz-Gutiérrez, A. and Myers, A. D. and Nie, J. and Niz, G. and Palanque-Delabrouille, N. and Percival, W. J. and Prada, F. and Rabinowitz, D. and Rezaie, M. and Rossi, G. and Sanchez, E. and Schlafly, Edward F. and Schlegel, D. and Schubnell, M. and Seo, H. and Sprayberry, D. and Tarlé, G. and Vargas-Magaña, M. and Weaver, B. A. and Zhou, R.},
	month = dec,
	year = {2024},
	note = {ADS Bibcode: 2024AJ....168..245P},
	pages = {245},
}

@article{chisariRisingTideIntrinsic2025,
	title = {A rising tide: intrinsic alignments since the turn of the millennium},
	volume = {33},
	issn = {0935-4956},
	shorttitle = {A rising tide},
	url = {https://ui.adsabs.harvard.edu/abs/2025A&ARv..33....5C},
	doi = {10.1007/s00159-025-00161-8},
	urldate = {2025-10-29},
	journal = {Astronomy and Astrophysics Review},
	publisher = {Springer},
	author = {Chisari, Nora Elisa},
	month = oct,
	year = {2025},
	note = {ADS Bibcode: 2025A\&ARv..33....5C},
	pages = {5},
}

@misc{kuritaParityViolationGalaxy2025,
	title = {Parity {Violation} in {Galaxy} {Shapes}: {Primordial} {Non}-{Gaussianity}},
	shorttitle = {Parity {Violation} in {Galaxy} {Shapes}},
	url = {https://ui.adsabs.harvard.edu/abs/2025arXiv250908787K},
	doi = {10.48550/arXiv.2509.08787},
	urldate = {2025-10-02},
	publisher = {arXiv},
	author = {Kurita, Toshiki and Jamieson, Drew and Komatsu, Eiichiro and Schmidt, Fabian},
	month = sep,
	year = {2025},
	note = {ADS Bibcode: 2025arXiv250908787K},
}

@misc{lammanOptimalIntrinsicAlignment2025,
	title = {Optimal intrinsic alignment estimators in the presence of redshift-space distortions},
	url = {https://ui.adsabs.harvard.edu/abs/2025arXiv250416076L},
	doi = {10.48550/arXiv.2504.16076},
	urldate = {2025-05-05},
	publisher = {arXiv},
	author = {Lamman, Claire and Blazek, Jonathan and Eisenstein, Daniel J.},
	month = apr,
	year = {2025},
	note = {ADS Bibcode: 2025arXiv250416076L},
}

@article{DESI2016b.Instr,
	title = {The {DESI} experiment part {II}: {Instrument} design},
	journal = {arXiv e-prints},
	author = {{DESI Collaboration} and Aghamousa, Amir and Aguilar, Jessica and Ahlen, Steve and Alam, Shadab and Allen, Lori E. and Allende Prieto, Carlos and Annis, James and Bailey, Stephen and Balland, Christophe and Ballester, Otger and Baltay, Charles and Beaufore, Lucas and Bebek, Chris and Beers, Timothy C. and Bell, Eric F. and Bernal, José Luis and Besuner, Robert and Beutler, Florian and Blake, Chris and Bleuler, Hannes and Blomqvist, Michael and Blum, Robert and Bolton, Adam S. and Briceno, Cesar and Brooks, David and Brownstein, Joel R. and Buckley-Geer, Elizabeth and Burden, Angela and Burtin, Etienne and Busca, Nicolas G. and Cahn, Robert N. and Cai, Yan-Chuan and Cardiel-Sas, Laia and Carlberg, Raymond G. and Carton, Pierre-Henri and Casas, Ricard and Castander, Francisco J. and Cervantes-Cota, Jorge L. and Claybaugh, Todd M. and Close, Madeline and Coker, Carl T. and Cole, Shaun and Comparat, Johan and Cooper, Andrew P. and Cousinou, M. -C. and Crocce, Martin and Cuby, Jean-Gabriel and Cunningham, Daniel P. and Davis, Tamara M. and Dawson, Kyle S. and de la Macorra, Axel and De Vicente, Juan and Delubac, Timothée and Derwent, Mark and Dey, Arjun and Dhungana, Govinda and Ding, Zhejie and Doel, Peter and Duan, Yutong T. and Ealet, Anne and Edelstein, Jerry and Eftekharzadeh, Sarah and Eisenstein, Daniel J. and Elliott, Ann and Escoffier, Stéphanie and Evatt, Matthew and Fagrelius, Parker and Fan, Xiaohui and Fanning, Kevin and Farahi, Arya and Farihi, Jay and Favole, Ginevra and Feng, Yu and Fernandez, Enrique and Findlay, Joseph R. and Finkbeiner, Douglas P. and Fitzpatrick, Michael J. and Flaugher, Brenna and Flender, Samuel and Font-Ribera, Andreu and Forero-Romero, Jaime E. and Fosalba, Pablo and Frenk, Carlos S. and Fumagalli, Michele and Gaensicke, Boris T. and Gallo, Giuseppe and Garcia-Bellido, Juan and Gaztanaga, Enrique and Pietro Gentile Fusillo, Nicola and Gerard, Terry and Gershkovich, Irena and Giannantonio, Tommaso and Gillet, Denis and Gonzalez-de-Rivera, Guillermo and Gonzalez-Perez, Violeta and Gott, Shelby and Graur, Or and Gutierrez, Gaston and Guy, Julien and Habib, Salman and Heetderks, Henry and Heetderks, Ian and Heitmann, Katrin and Hellwing, Wojciech A. and Herrera, David A. and Ho, Shirley and Holland, Stephen and Honscheid, Klaus and Huff, Eric and Hutchinson, Timothy A. and Huterer, Dragan and Hwang, Ho Seong and Illa Laguna, Joseph Maria and Ishikawa, Yuzo and Jacobs, Dianna and Jeffrey, Niall and Jelinsky, Patrick and Jennings, Elise and Jiang, Linhua and Jimenez, Jorge and Johnson, Jennifer and Joyce, Richard and Jullo, Eric and Juneau, Stéphanie and Kama, Sami and Karcher, Armin and Karkar, Sonia and Kehoe, Robert and Kennamer, Noble and Kent, Stephen and Kilbinger, Martin and Kim, Alex G. and Kirkby, David and Kisner, Theodore and Kitanidis, Ellie and Kneib, Jean-Paul and Koposov, Sergey and Kovacs, Eve and Koyama, Kazuya and Kremin, Anthony and Kron, Richard and Kronig, Luzius and Kueter-Young, Andrea and Lacey, Cedric G. and Lafever, Robin and Lahav, Ofer and Lambert, Andrew and Lampton, Michael and Landriau, Martin and Lang, Dustin and Lauer, Tod R. and Le Goff, Jean-Marc and Le Guillou, Laurent and Le Van Suu, Auguste and Lee, Jae Hyeon and Lee, Su-Jeong and Leitner, Daniela and Lesser, Michael and Levi, Michael E. and L'Huillier, Benjamin and Li, Baojiu and Liang, Ming and Lin, Huan and Linder, Eric and Loebman, Sarah R. and Lukić, Zarija and Ma, Jun and MacCrann, Niall and Magneville, Christophe and Makarem, Laleh and Manera, Marc and Manser, Christopher J. and Marshall, Robert and Martini, Paul and Massey, Richard and Matheson, Thomas and McCauley, Jeremy and McDonald, Patrick and McGreer, Ian D. and Meisner, Aaron and Metcalfe, Nigel and Miller, Timothy N. and Miquel, Ramon and Moustakas, John and Myers, Adam and Naik, Milind and Newman, Jeffrey A. and Nichol, Robert C. and Nicola, Andrina and Nicolati da Costa, Luiz and Nie, Jundan and Niz, Gustavo and Norberg, Peder and Nord, Brian and Norman, Dara and Nugent, Peter and O'Brien, Thomas and Oh, Minji and Olsen, Knut A. G. and Padilla, Cristobal and Padmanabhan, Hamsa and Padmanabhan, Nikhil and Palanque-Delabrouille, Nathalie and Palmese, Antonella and Pappalardo, Daniel and Pâris, Isabelle and Park, Changbom and Patej, Anna and Peacock, John A. and Peiris, Hiranya V. and Peng, Xiyan and Percival, Will J. and Perruchot, Sandrine and Pieri, Matthew M. and Pogge, Richard and Pollack, Jennifer E. and Poppett, Claire and Prada, Francisco and Prakash, Abhishek and Probst, Ronald G. and Rabinowitz, David and Raichoor, Anand and Ree, Chang Hee and Refregier, Alexandre and Regal, Xavier and Reid, Beth and Reil, Kevin and Rezaie, Mehdi and Rockosi, Constance M. and Roe, Natalie and Ronayette, Samuel and Roodman, Aaron and Ross, Ashley J. and Ross, Nicholas P. and Rossi, Graziano and Rozo, Eduardo and Ruhlmann-Kleider, Vanina and Rykoff, Eli S. and Sabiu, Cristiano and Samushia, Lado and Sanchez, Eusebio and Sanchez, Javier and Schlegel, David J. and Schneider, Michael and Schubnell, Michael and Secroun, Aurélia and Seljak, Uros and Seo, Hee-Jong and Serrano, Santiago and Shafieloo, Arman and Shan, Huanyuan and Sharples, Ray and Sholl, Michael J. and Shourt, William V. and Silber, Joseph H. and Silva, David R. and Sirk, Martin M. and Slosar, Anze and Smith, Alex and Smoot, George F. and Som, Debopam and Song, Yong-Seon and Sprayberry, David and Staten, Ryan and Stefanik, Andy and Tarle, Gregory and Sien Tie, Suk and Tinker, Jeremy L. and Tojeiro, Rita and Valdes, Francisco and Valenzuela, Octavio and Valluri, Monica and Vargas-Magana, Mariana and Verde, Licia and Walker, Alistair R. and Wang, Jiali and Wang, Yuting and Weaver, Benjamin A. and Weaverdyck, Curtis and Wechsler, Risa H. and Weinberg, David H. and White, Martin and Yang, Qian and Yeche, Christophe and Zhang, Tianmeng and Zhao, Gong-Bo and Zheng, Yi and Zhou, Xu and Zhou, Zhimin and Zhu, Yaling and Zou, Hu and Zu, Ying},
	month = oct,
	year = {2016},
	note = {arXiv: 1611.00037 [astro-ph.IM]
Number: arXiv:1611.00037
tex.adsnote: Provided by the SAO/NASA Astrophysics Data System},
	pages = {arXiv:1611.00037},
}

@article{millerOpticalCorrectorDark2024,
	title = {The {Optical} {Corrector} for the {Dark} {Energy} {Spectroscopic} {Instrument}},
	volume = {168},
	issn = {0004-6256},
	url = {https://ui.adsabs.harvard.edu/abs/2024AJ....168...95M},
	doi = {10.3847/1538-3881/ad45fe},
	urldate = {2025-03-28},
	journal = {The Astronomical Journal},
	publisher = {IOP},
	author = {Miller, Timothy N. and Doel, Peter and Gutierrez, Gaston and Besuner, Robert and Brooks, David and Gallo, Giuseppe and Heetderks, Henry and Jelinsky, Patrick and Kent, Stephen M. and Lampton, Michael and Levi, Michael E. and Liang, Ming and Meisner, Aaron and Sholl, Michael J. and Silber, Joseph Harry and Sprayberry, David and Aguilar, Jessica Nicole and de la Macorra, Axel and Eisenstein, Daniel and Fanning, Kevin and Font-Ribera, Andreu and Gaztañaga, Enrique and Gontcho A Gontcho, Satya and Honscheid, Klaus and Jimenez, Jorge and Joyce, Dick and Kehoe, Robert and Kisner, Theodore and Kremin, Anthony and Landriau, Martin and Le Guillou, Laurent and Magneville, Christophe and Martini, Paul and Miquel, Ramon and Moustakas, John and Nie, Jundan and Percival, Will and Poppett, Claire and Prada, Francisco and Rossi, Graziano and Schlegel, David and Schubnell, Michael and Seo, Hee-Jong and Sharples, Ray and Tarlé, Gregory and Vargas-Magaña, Mariana and Zhou, Zhimin and {the DESI Collaboration}},
	month = aug,
	year = {2024},
	note = {ADS Bibcode: 2024AJ....168...95M},
	pages = {95},
}

@article{kuritaConstraintsAnisotropicPrimordial2023,
	title = {Constraints on anisotropic primordial non-{Gaussianity} from intrinsic alignments of {SDSS}-{III} {BOSS} galaxies},
	volume = {108},
	issn = {1550-79980556-2821},
	url = {https://ui.adsabs.harvard.edu/abs/2023PhRvD.108h3533K},
	doi = {10.1103/PhysRevD.108.083533},
	urldate = {2025-03-28},
	journal = {Physical Review D},
	publisher = {APS},
	author = {Kurita, Toshiki and Takada, Masahiro},
	month = oct,
	year = {2023},
	note = {ADS Bibcode: 2023PhRvD.108h3533K},
	pages = {083533},
}

@article{samuroffDarkEnergySurvey2023,
	title = {The {Dark} {Energy} {Survey} {Year} 3 and {eBOSS}: constraining galaxy intrinsic alignments across luminosity and colour space},
	volume = {524},
	issn = {0035-8711},
	shorttitle = {The {Dark} {Energy} {Survey} {Year} 3 and {eBOSS}},
	url = {https://ui.adsabs.harvard.edu/abs/2023MNRAS.524.2195S},
	doi = {10.1093/mnras/stad2013},
	urldate = {2025-03-28},
	journal = {Monthly Notices of the Royal Astronomical Society},
	publisher = {OUP},
	author = {Samuroff, S. and Mandelbaum, R. and Blazek, J. and Campos, A. and MacCrann, N. and Zacharegkas, G. and Amon, A. and Prat, J. and Singh, S. and Elvin-Poole, J. and Ross, A. J. and Alarcon, A. and Baxter, E. and Bechtol, K. and Becker, M. R. and Bernstein, G. M. and Rosell, A. Carnero and Kind, M. Carrasco and Cawthon, R. and Chang, C. and Chen, R. and Choi, A. and Crocce, M. and Davis, C. and DeRose, J. and Dodelson, S. and Doux, C. and Drlica-Wagner, A. and Eckert, K. and Everett, S. and Ferté, A. and Gatti, M. and Giannini, G. and Gruen, D. and Gruendl, R. A. and Harrison, I. and Herner, K. and Huff, E. M. and Jarvis, M. and Kuropatkin, N. and Leget, P. -F. and Lemos, P. and McCullough, J. and Myles, J. and Navarro-Alsina, A. and Pandey, S. and Porredon, A. and Raveri, M. and Rodriguez-Monroy, M. and Rollins, R. P. and Roodman, A. and Rossi, G. and Rykoff, E. S. and Sánchez, C. and Secco, L. F. and Sevilla-Noarbe, I. and Sheldon, E. and Shin, T. and Troxel, M. A. and Tutusaus, I. and Weaverdyck, N. and Yanny, B. and Yin, B. and Zhang, Y. and Zuntz, J. and Aguena, M. and Alves, O. and Annis, J. and Bacon, D. and Bertin, E. and Bocquet, S. and Brooks, D. and Burke, D. L. and Carretero, J. and Costanzi, M. and da Costa, L. N. and Pereira, M. E. S. and De Vicente, J. and Desai, S. and Diehl, H. T. and Dietrich, J. P. and Doel, P. and Ferrero, I. and Flaugher, B. and Frieman, J. and García-Bellido, J. and Hinton, S. R. and Hollowood, D. L. and Honscheid, K. and James, D. J. and Kuehn, K. and Lahav, O. and Marshall, J. L. and Melchior, P. and Mena-Fernández, J. and Menanteau, F. and Miquel, R. and Newman, J. and Palmese, A. and Pieres, A. and Malagón, A. A. Plazas and Sanchez, E. and Scarpine, V. and Smith, M. and Suchyta, E. and Swanson, M. E. C. and Tarle, G. and To, C. and {DES Collaboration}},
	month = sep,
	year = {2023},
	note = {ADS Bibcode: 2023MNRAS.524.2195S},
	pages = {2195--2223},
}

@article{lammanDetectionLargescaleTidal2024,
	title = {Detection of the large-scale tidal field with galaxy multiplet alignment in the {DESI} {Y1} spectroscopic survey},
	volume = {534},
	issn = {0035-8711},
	url = {https://ui.adsabs.harvard.edu/abs/2024MNRAS.534.3540L},
	doi = {10.1093/mnras/stae2290},
	urldate = {2025-03-28},
	journal = {Monthly Notices of the Royal Astronomical Society},
	publisher = {OUP},
	author = {Lamman, Claire and Eisenstein, Daniel and Forero-Romero, Jaime E. and Aguilar, Jessica Nicole and Ahlen, Steven and Bailey, Stephen and Bianchi, Davide and Brooks, David and Claybaugh, Todd and de la Macorra, Axel and Doel, Peter and Ferraro, Simone and Font-Ribera, Andreu and Gaztañaga, Enrique and Gontcho A Gontcho, Satya and Gutierrez, Gaston and Honscheid, Klaus and Howlett, Cullan and Kremin, Anthony and Lambert, Andrew and Landriau, Martin and Le Guillou, Laurent and Levi, Michael E. and Meisner, Aaron and Miquel, Ramon and Moustakas, John and Newman, Jeffrey A. and Niz, Gustavo and Prada, Francisco and Pérez-Ràfols, Ignasi and Ross, Ashley J. and Rossi, Graziano and Sanchez, Eusebio and Schubnell, Michael and Sprayberry, David and Tarlé, Gregory and Vargas-Magaña, Mariana and Weaver, Benjamin Alan and Zou, Hu},
	month = nov,
	year = {2024},
	note = {ADS Bibcode: 2024MNRAS.534.3540L},
	pages = {3540--3551},
}

@article{deyOverviewDESILegacy2019,
	title = {Overview of the {DESI} {Legacy} {Imaging} {Surveys}},
	volume = {157},
	issn = {0004-6256},
	url = {https://ui.adsabs.harvard.edu/abs/2019AJ....157..168D},
	doi = {10.3847/1538-3881/ab089d},
	urldate = {2022-11-17},
	journal = {The Astronomical Journal},
	author = {Dey, Arjun and Schlegel, David J. and Lang, Dustin and Blum, Robert and Burleigh, Kaylan and Fan, Xiaohui and Findlay, Joseph R. and Finkbeiner, Doug and Herrera, David and Juneau, Stéphanie and Landriau, Martin and Levi, Michael and McGreer, Ian and Meisner, Aaron and Myers, Adam D. and Moustakas, John and Nugent, Peter and Patej, Anna and Schlafly, Edward F. and Walker, Alistair R. and Valdes, Francisco and Weaver, Benjamin A. and Yèche, Christophe and Zou, Hu and Zhou, Xu and Abareshi, Behzad and Abbott, T. M. C. and Abolfathi, Bela and Aguilera, C. and Alam, Shadab and Allen, Lori and Alvarez, A. and Annis, James and Ansarinejad, Behzad and Aubert, Marie and Beechert, Jacqueline and Bell, Eric F. and BenZvi, Segev Y. and Beutler, Florian and Bielby, Richard M. and Bolton, Adam S. and Briceño, César and Buckley-Geer, Elizabeth J. and Butler, Karen and Calamida, Annalisa and Carlberg, Raymond G. and Carter, Paul and Casas, Ricard and Castander, Francisco J. and Choi, Yumi and Comparat, Johan and Cukanovaite, Elena and Delubac, Timothée and DeVries, Kaitlin and Dey, Sharmila and Dhungana, Govinda and Dickinson, Mark and Ding, Zhejie and Donaldson, John B. and Duan, Yutong and Duckworth, Christopher J. and Eftekharzadeh, Sarah and Eisenstein, Daniel J. and Etourneau, Thomas and Fagrelius, Parker A. and Farihi, Jay and Fitzpatrick, Mike and Font-Ribera, Andreu and Fulmer, Leah and Gänsicke, Boris T. and Gaztanaga, Enrique and George, Koshy and Gerdes, David W. and Gontcho, Satya Gontcho A. and Gorgoni, Claudio and Green, Gregory and Guy, Julien and Harmer, Diane and Hernandez, M. and Honscheid, Klaus and Huang, Lijuan Wendy and James, David J. and Jannuzi, Buell T. and Jiang, Linhua and Joyce, Richard and Karcher, Armin and Karkar, Sonia and Kehoe, Robert and Kneib, Jean-Paul and Kueter-Young, Andrea and Lan, Ting-Wen and Lauer, Tod R. and Le Guillou, Laurent and Le Van Suu, Auguste and Lee, Jae Hyeon and Lesser, Michael and Perreault Levasseur, Laurence and Li, Ting S. and Mann, Justin L. and Marshall, Robert and Martínez-Vázquez, C. E. and Martini, Paul and du Mas des Bourboux, Hélion and McManus, Sean and Meier, Tobias Gabriel and Ménard, Brice and Metcalfe, Nigel and Muñoz-Gutiérrez, Andrea and Najita, Joan and Napier, Kevin and Narayan, Gautham and Newman, Jeffrey A. and Nie, Jundan and Nord, Brian and Norman, Dara J. and Olsen, Knut A. G. and Paat, Anthony and Palanque-Delabrouille, Nathalie and Peng, Xiyan and Poppett, Claire L. and Poremba, Megan R. and Prakash, Abhishek and Rabinowitz, David and Raichoor, Anand and Rezaie, Mehdi and Robertson, A. N. and Roe, Natalie A. and Ross, Ashley J. and Ross, Nicholas P. and Rudnick, Gregory and Safonova, Sasha and Saha, Abhijit and Sánchez, F. Javier and Savary, Elodie and Schweiker, Heidi and Scott, Adam and Seo, Hee-Jong and Shan, Huanyuan and Silva, David R. and Slepian, Zachary and Soto, Christian and Sprayberry, David and Staten, Ryan and Stillman, Coley M. and Stupak, Robert J. and Summers, David L. and Sien Tie, Suk and Tirado, H. and Vargas-Magaña, Mariana and Vivas, A. Katherina and Wechsler, Risa H. and Williams, Doug and Yang, Jinyi and Yang, Qian and Yapici, Tolga and Zaritsky, Dennis and Zenteno, A. and Zhang, Kai and Zhang, Tianmeng and Zhou, Rongpu and Zhou, Zhimin},
	month = may,
	year = {2019},
	note = {ADS Bibcode: 2019AJ....157..168D},
	pages = {168},
}

@article{samuroffAdvancesConstrainingIntrinsic2021,
	title = {Advances in constraining intrinsic alignment models with hydrodynamic simulations},
	volume = {508},
	issn = {0035-8711},
	url = {https://ui.adsabs.harvard.edu/abs/2021MNRAS.508..637S},
	doi = {10.1093/mnras/stab2520},
	urldate = {2022-03-23},
	journal = {Monthly Notices of the Royal Astronomical Society},
	author = {Samuroff, S. and Mandelbaum, R. and Blazek, J.},
	month = nov,
	year = {2021},
	note = {ADS Bibcode: 2021MNRAS.508..637S},
	pages = {637--664},
}

@article{desicollaborationOverviewInstrumentationDark2022,
	title = {Overview of the {Instrumentation} for the {Dark} {Energy} {Spectroscopic} {Instrument}},
	volume = {164},
	issn = {0004-6256},
	url = {https://ui.adsabs.harvard.edu/abs/2022AJ....164..207D},
	doi = {10.3847/1538-3881/ac882b},
	urldate = {2023-04-01},
	journal = {The Astronomical Journal},
	author = {{DESI Collaboration} and Abareshi, B. and Aguilar, J. and Ahlen, S. and Alam, Shadab and Alexander, David M. and Alfarsy, R. and Allen, L. and Allende Prieto, C. and Alves, O. and Ameel, J. and Armengaud, E. and Asorey, J. and Aviles, Alejandro and Bailey, S. and Balaguera-Antolínez, A. and Ballester, O. and Baltay, C. and Bault, A. and Beltran, S. F. and Benavides, B. and BenZvi, S. and Berti, A. and Besuner, R. and Beutler, Florian and Bianchi, D. and Blake, C. and Blanc, P. and Blum, R. and Bolton, A. and Bose, S. and Bramall, D. and Brieden, S. and Brodzeller, A. and Brooks, D. and Brownewell, C. and Buckley-Geer, E. and Cahn, R. N. and Cai, Z. and Canning, R. and Capasso, R. and Carnero Rosell, A. and Carton, P. and Casas, R. and Castander, F. J. and Cervantes-Cota, J. L. and Chabanier, S. and Chaussidon, E. and Chuang, C. and Circosta, C. and Cole, S. and Cooper, A. P. and da Costa, L. and Cousinou, M. -C. and Cuceu, A. and Davis, T. M. and Dawson, K. and de la Cruz-Noriega, R. and de la Macorra, A. and de Mattia, A. and Della Costa, J. and Demmer, P. and Derwent, M. and Dey, A. and Dey, B. and Dhungana, G. and Ding, Z. and Dobson, C. and Doel, P. and Donald-McCann, J. and Donaldson, J. and Douglass, K. and Duan, Y. and Dunlop, P. and Edelstein, J. and Eftekharzadeh, S. and Eisenstein, D. J. and Enriquez-Vargas, M. and Escoffier, S. and Evatt, M. and Fagrelius, P. and Fan, X. and Fanning, K. and Fawcett, V. A. and Ferraro, S. and Ereza, J. and Flaugher, B. and Font-Ribera, A. and Forero-Romero, J. E. and Frenk, C. S. and Fromenteau, S. and Gänsicke, B. T. and Garcia-Quintero, C. and Garrison, L. and Gaztañaga, E. and Gerardi, F. and Gil-Marín, H. and Gontcho a Gontcho, S. and Gonzalez-Morales, Alma X. and Gonzalez-de-Rivera, G. and Gonzalez-Perez, V. and Gordon, C. and Graur, O. and Green, D. and Grove, C. and Gruen, D. and Gutierrez, G. and Guy, J. and Hahn, C. and Harris, S. and Herrera, D. and Herrera-Alcantar, Hiram K. and Honscheid, K. and Howlett, C. and Huterer, D. and Iršič, V. and Ishak, M. and Jelinsky, P. and Jiang, L. and Jimenez, J. and Jing, Y. P. and Joyce, R. and Jullo, E. and Juneau, S. and Karaçaylı, N. G. and Karamanis, M. and Karcher, A. and Karim, T. and Kehoe, R. and Kent, S. and Kirkby, D. and Kisner, T. and Kitaura, F. and Koposov, S. E. and Kovács, A. and Kremin, A. and Krolewski, Alex and L'Huillier, B. and Lahav, O. and Lambert, A. and Lamman, C. and Lan, Ting-Wen and Landriau, M. and Lane, S. and Lang, D. and Lange, J. U. and Lasker, J. and Le Guillou, L. and Leauthaud, A. and Le Van Suu, A. and Levi, Michael E. and Li, T. S. and Magneville, C. and Manera, M. and Manser, Christopher J. and Marshall, B. and Martini, Paul and McCollam, W. and McDonald, P. and Meisner, Aaron M. and Mena-Fernández, J. and Meneses-Rizo, J. and Mezcua, M. and Miller, T. and Miquel, R. and Montero-Camacho, P. and Moon, J. and Moustakas, J. and Mueller, E. and Muñoz-Gutiérrez, Andrea and Myers, Adam D. and Nadathur, S. and Najita, J. and Napolitano, L. and Neilsen, E. and Newman, Jeffrey A. and Nie, J. D. and Ning, Y. and Niz, G. and Norberg, P. and Noriega, Hernán E. and O'Brien, T. and Obuljen, A. and Palanque-Delabrouille, N. and Palmese, A. and Zhiwei, P. and Pappalardo, D. and PENG, X. and Percival, W. J. and Perruchot, S. and Pogge, R. and Poppett, C. and Porredon, A. and Prada, F. and Prochaska, J. and Pucha, R. and Pérez-Fernández, A. and Pérez-Ràfols, I. and Rabinowitz, D. and Raichoor, A. and Ramirez-Solano, S. and Ramírez-Pérez, César and Ravoux, C. and Reil, K. and Rezaie, M. and Rocher, A. and Rockosi, C. and Roe, N. A. and Roodman, A. and Ross, A. J. and Rossi, G. and Ruggeri, R. and Ruhlmann-Kleider, V. and Sabiu, C. G. and Safonova, S. and Said, K. and Saintonge, A. and Salas Catonga, Javier and Samushia, L. and Sanchez, E. and Saulder, C. and Schaan, E. and Schlafly, E. and Schlegel, D. and Schmoll, J. and Scholte, D. and Schubnell, M. and Secroun, A. and Seo, H. and Serrano, S. and Sharples, Ray M. and Sholl, Michael J. and Silber, Joseph Harry and Silva, D. R. and Sirk, M. and Siudek, M. and Smith, A. and Sprayberry, D. and Staten, R. and Stupak, B. and Tan, T. and Tarlé, Gregory and Tie, Suk Sien and Tojeiro, R. and Ureña-López, L. A. and Valdes, F. and Valenzuela, O. and Valluri, M. and Vargas-Magaña, M. and Verde, L. and Walther, M. and Wang, B. and Wang, M. S. and Weaver, B. A. and Weaverdyck, C. and Wechsler, R. and Wilson, Michael J. and Yang, J. and Yu, Y. and Yuan, S. and Yèche, Christophe and Zhang, H. and Zhang, K. and Zhao, Cheng and Zhou, Rongpu and Zhou, Zhimin and Zou, H. and Zou, J. and Zou, S. and Zu, Y. and {DESI Collaboration}},
	month = nov,
	year = {2022},
	note = {ADS Bibcode: 2022AJ....164..207D},
	pages = {207},
}

@article{myersTargetselectionPipelineDark2023,
	title = {The {Target}-selection {Pipeline} for the {Dark} {Energy} {Spectroscopic} {Instrument}},
	volume = {165},
	issn = {0004-6256},
	url = {https://ui.adsabs.harvard.edu/abs/2023AJ....165...50M},
	doi = {10.3847/1538-3881/aca5f9},
	urldate = {2023-11-27},
	journal = {The Astronomical Journal},
	author = {Myers, Adam D. and Moustakas, John and Bailey, Stephen and Weaver, Benjamin A. and Cooper, Andrew P. and Forero-Romero, Jaime E. and Abolfathi, Bela and Alexander, David M. and Brooks, David and Chaussidon, Edmond and Chuang, Chia-Hsun and Dawson, Kyle and Dey, Arjun and Dey, Biprateep and Dhungana, Govinda and Doel, Peter and Fanning, Kevin and Gaztañaga, Enrique and Gontcho A Gontcho, Satya and Gonzalez-Morales, Alma X. and Hahn, ChangHoon and Herrera-Alcantar, Hiram K. and Honscheid, Klaus and Ishak, Mustapha and Karim, Tanveer and Kirkby, David and Kisner, Theodore and Koposov, Sergey E. and Kremin, Anthony and Lan, Ting-Wen and Landriau, Martin and Lang, Dustin and Levi, Michael E. and Magneville, Christophe and Napolitano, Lucas and Martini, Paul and Meisner, Aaron and Newman, Jeffrey A. and Palanque-Delabrouille, Nathalie and Percival, Will and Poppett, Claire and Prada, Francisco and Raichoor, Anand and Ross, Ashley J. and Schlafly, Edward F. and Schlegel, David and Schubnell, Michael and Tan, Ting and Tarle, Gregory and Wilson, Michael J. and Yèche, Christophe and Zhou, Rongpu and Zhou, Zhimin and Zou, Hu},
	month = feb,
	year = {2023},
	note = {ADS Bibcode: 2023AJ....165...50M},
	pages = {50},
}

@article{zhouTargetSelectionValidation2023,
	title = {Target {Selection} and {Validation} of {DESI} {Luminous} {Red} {Galaxies}},
	volume = {165},
	issn = {0004-6256, 1538-3881},
	url = {http://arxiv.org/abs/2208.08515},
	doi = {10.3847/1538-3881/aca5fb},
	number = {2},
	urldate = {2024-08-03},
	journal = {The Astronomical Journal},
	author = {Zhou, Rongpu and Dey, Biprateep and Newman, Jeffrey A. and Eisenstein, Daniel J. and Dawson, K. and Bailey, S. and Berti, A. and Guy, J. and Lan, Ting-Wen and Zou, H. and Aguilar, J. and Ahlen, S. and Alam, Shadab and Brooks, D. and de la Macorra, A. and Dey, A. and Dhungana, G. and Fanning, K. and Font-Ribera, A. and Gontcho, S. Gontcho A. and Honscheid, K. and Ishak, Mustapha and Kisner, T. and Kovács, A. and Kremin, A. and Landriau, M. and Levi, Michael E. and Magneville, C. and Manera, Marc and Martini, P. and Meisner, Aaron M. and Miquel, R. and Moustakas, J. and Myers, Adam D. and Nie, Jundan and Palanque-Delabrouille, N. and Percival, W. J. and Poppett, C. and Prada, F. and Raichoor, A. and Ross, A. J. and Schlafly, E. and Schlegel, D. and Schubnell, M. and Tarlé, Gregory and Weaver, B. A. and Wechsler, R. H. and Yèche, Christophe and Zhou, Zhimin},
	month = feb,
	year = {2023},
	note = {arXiv:2208.08515 [astro-ph]},
	pages = {58},
}

@article{guySpectroscopicDataProcessing2023,
	title = {The {Spectroscopic} {Data} {Processing} {Pipeline} for the {Dark} {Energy} {Spectroscopic} {Instrument}},
	volume = {165},
	issn = {0004-6256},
	url = {https://ui.adsabs.harvard.edu/abs/2023AJ....165..144G},
	doi = {10.3847/1538-3881/acb212},
	urldate = {2023-11-27},
	journal = {The Astronomical Journal},
	author = {Guy, J. and Bailey, S. and Kremin, A. and Alam, Shadab and Alexander, D. M. and Allende Prieto, C. and BenZvi, S. and Bolton, A. S. and Brooks, D. and Chaussidon, E. and Cooper, A. P. and Dawson, K. and de la Macorra, A. and Dey, A. and Dey, Biprateep and Dhungana, G. and Eisenstein, D. J. and Font-Ribera, A. and Forero-Romero, J. E. and Gaztañaga, E. and Gontcho A Gontcho, S. and Green, D. and Honscheid, K. and Ishak, M. and Kehoe, R. and Kirkby, D. and Kisner, T. and Koposov, Sergey E. and Lan, Ting-Wen and Landriau, M. and Le Guillou, L. and Levi, Michael E. and Magneville, C. and Manser, Christopher J. and Martini, P. and Meisner, Aaron M. and Miquel, R. and Moustakas, J. and Myers, Adam D. and Newman, Jeffrey A. and Nie, Jundan and Palanque-Delabrouille, N. and Percival, W. J. and Poppett, C. and Prada, F. and Raichoor, A. and Ravoux, C. and Ross, A. J. and Schlafly, E. F. and Schlegel, D. and Schubnell, M. and Sharples, Ray M. and Tarlé, Gregory and Weaver, B. A. and Yéche, Christophe and Zhou, Rongpu and Zhou, Zhimin and Zou, H.},
	month = apr,
	year = {2023},
	note = {ADS Bibcode: 2023AJ....165..144G},
	pages = {144},
}

@article{hahnDESIBrightGalaxy2023,
	title = {{DESI} {Bright} {Galaxy} {Survey}: {Final} {Target} {Selection}, {Design}, and {Validation}},
	volume = {165},
	issn = {0004-6256, 1538-3881},
	shorttitle = {{DESI} {Bright} {Galaxy} {Survey}},
	url = {http://arxiv.org/abs/2208.08512},
	doi = {10.3847/1538-3881/accff8},
	number = {6},
	urldate = {2024-08-03},
	journal = {The Astronomical Journal},
	author = {Hahn, ChangHoon and Wilson, Michael J. and Ruiz-Macias, Omar and Cole, Shaun and Weinberg, David H. and Moustakas, John and Kremin, Anthony and Tinker, Jeremy L. and Smith, Alex and Wechsler, Risa H. and Ahlen, Steven and Alam, Shadab and Bailey, Stephen and Brooks, David and Cooper, Andrew P. and Davis, Tamara M. and Dawson, Kyle and Dey, Arjun and Dey, Biprateep and Eftekharzadeh, Sarah and Eisenstein, Daniel J. and Fanning, Kevin and Forero-Romero, Jaime E. and Frenk, Carlos S. and Gaztañaga, Enrique and Gontcho, Satya Gontcho A. and Guy, Julien and Honscheid, Klaus and Ishak, Mustapha and Juneau, Stéphanie and Kehoe, Robert and Kisner, Theodore and Lan, Ting-Wen and Landriau, Martin and Guillou, Laurent Le and Levi, Michael E. and Magneville, Christophe and Martini, Paul and Meisner, Aaron and Myers, Adam D. and Nie, Jundan and Norberg, Peder and Palanque-Delabrouille, Nathalie and Percival, Will J. and Poppett, Claire and Prada, Francisco and Raichoor, Anand and Ross, Ashley J. and Safonova, Sasha and Saulder, Christoph and Schlafly, Eddie and Schlegel, David and Sierra-Porta, David and Tarle, Gregory and Weaver, Benjamin A. and Yèche, Christophe and Zarrouk, Pauline and Zhou, Rongpu and Zhou, Zhimin and Zou, Hu},
	month = jun,
	year = {2023},
	note = {arXiv:2208.08512 [astro-ph]},
	pages = {253},
}

@article{schlaflySurveyOperationsDark2023,
	title = {Survey {Operations} for the {Dark} {Energy} {Spectroscopic} {Instrument}},
	volume = {166},
	issn = {0004-6256},
	url = {https://ui.adsabs.harvard.edu/abs/2023AJ....166..259S},
	doi = {10.3847/1538-3881/ad0832},
	urldate = {2024-08-03},
	journal = {The Astronomical Journal},
	publisher = {IOP},
	author = {Schlafly, Edward F. and Kirkby, David and Schlegel, David J. and Myers, Adam D. and Raichoor, Anand and Dawson, Kyle and Aguilar, Jessica and Allende Prieto, Carlos and Bailey, Stephen and BenZvi, Segev and Bermejo-Climent, Jose and Brooks, David and de la Macorra, Axel and Dey, Arjun and Doel, Peter and Fanning, Kevin and Font-Ribera, Andreu and Forero-Romero, Jaime E. and García-Bellido, Juan and Gontcho A Gontcho, Satya and Guy, Julien and Hahn, ChangHoon and Honscheid, Klaus and Ishak, Mustapha and Juneau, Stéphanie and Kehoe, Robert and Kisner, Theodore and Kremin, Anthony and Landriau, Martin and Lang, Dustin A. and Lasker, James and Levi, Michael E. and Magneville, Christophe and Manser, Christopher J. and Martini, Paul and Meisner, Aaron M. and Miquel, Ramon and Moustakas, John and Newman, Jeffrey A. and Nie, Jundan and Palanque-Delabrouille, Nathalie. and Percival, Will J. and Poppett, Claire and Rockosi, Constance and Ross, Ashley J. and Rossi, Graziano and Tarlé, Gregory and Weaver, Benjamin A. and Yèche, Christophe and Zhou, Rongpu and {DESI Collaboration}},
	month = dec,
	year = {2023},
	note = {ADS Bibcode: 2023AJ....166..259S},
	pages = {259},
}

@article{raichoorTargetSelectionValidation2023,
	title = {Target {Selection} and {Validation} of {DESI} {Emission} {Line} {Galaxies}},
	volume = {165},
	issn = {0004-6256},
	url = {https://ui.adsabs.harvard.edu/abs/2023AJ....165..126R},
	doi = {10.3847/1538-3881/acb213},
	urldate = {2023-11-27},
	journal = {The Astronomical Journal},
	author = {Raichoor, A. and Moustakas, J. and Newman, Jeffrey A. and Karim, T. and Ahlen, S. and Alam, Shadab and Bailey, S. and Brooks, D. and Dawson, K. and de la Macorra, A. and de Mattia, A. and Dey, A. and Dey, Biprateep and Dhungana, G. and Eftekharzadeh, S. and Eisenstein, D. J. and Fanning, K. and Font-Ribera, A. and García-Bellido, J. and Gaztañaga, E. and A Gontcho, S. Gontcho and Guy, J. and Honscheid, K. and Ishak, M. and Kehoe, R. and Kisner, T. and Kremin, Anthony and Lan, Ting-Wen and Landriau, M. and Le Guillou, L. and Levi, Michael E. and Magneville, C. and Manera, M. and Martini, P. and Meisner, Aaron M. and Myers, Adam D. and Nie, Jundan and Palanque-Delabrouille, N. and Percival, W. J. and Poppett, C. and Prada, F. and Ross, A. J. and Ruhlmann-Kleider, V. and Sabiu, C. G. and Schlafly, E. F. and Schlegel, D. and Tarlé, Gregory and Weaver, B. A. and Yèche, Christophe and Zhou, Rongpu and Zhou, Zhimin and Zou, H.},
	month = mar,
	year = {2023},
	note = {ADS Bibcode: 2023AJ....165..126R},
	pages = {126},
}

@article{lammanIAGuideBreakdown2024,
	title = {The {IA} {Guide}: {A} {Breakdown} of {Intrinsic} {Alignment} {Formalisms}},
	volume = {7},
	shorttitle = {The {IA} {Guide}},
	url = {https://astro.theoj.org/article/94228-the-ia-guide-a-breakdown-of-intrinsic-alignment-formalisms},
	doi = {10.21105/astro.2309.08605},
	language = {en},
	urldate = {2025-03-25},
	journal = {The Open Journal of Astrophysics},
	publisher = {Maynooth Academic Publishing},
	author = {Lamman, Claire and Tsaprazi, Eleni and Shi, Jingjing and Šarčević, Nikolina Niko and Pyne, Susan and Legnani, Elisa and Ferreira, Tassia},
	month = feb,
	year = {2024},
}

@article{bartelmannWeakGravitationalLensing2001,
	title = {Weak gravitational lensing},
	volume = {340},
	issn = {0370-1573},
	url = {https://www.sciencedirect.com/science/article/pii/S037015730000082X?via%3Dihub},
	doi = {10.1016/S0370-1573(00)00082-X},
	number = {4-5},
	urldate = {2023-04-28},
	journal = {Physics Reports},
	author = {Bartelmann, Matthias and Schneider, Peter},
	month = jan,
	year = {2001},
	note = {ADS Bibcode: 2001PhR...340..291B},
	pages = {291--472},
}

@misc{thelsstdarkenergysciencecollaborationLSSTDarkEnergy2018,
	title = {The {LSST} {Dark} {Energy} {Science} {Collaboration} ({DESC}) {Science} {Requirements} {Document}},
	url = {https://ui.adsabs.harvard.edu/abs/2018arXiv180901669T},
	doi = {10.48550/arXiv.1809.01669},
	urldate = {2024-09-27},
	author = {{The LSST Dark Energy Science Collaboration} and Mandelbaum, Rachel and Eifler, Tim and Hložek, Renée and Collett, Thomas and Gawiser, Eric and Scolnic, Daniel and Alonso, David and Awan, Humna and Biswas, Rahul and Blazek, Jonathan and Burchat, Patricia and Chisari, Nora Elisa and Dell'Antonio, Ian and Digel, Seth and Frieman, Josh and Goldstein, Daniel A. and Hook, Isobel and Ivezić, Željko and Kahn, Steven M. and Kamath, Sowmya and Kirkby, David and Kitching, Thomas and Krause, Elisabeth and Leget, Pierre-François and Marshall, Philip J. and Meyers, Joshua and Miyatake, Hironao and Newman, Jeffrey A. and Nichol, Robert and Rykoff, Eli and Sanchez, F. Javier and Slosar, Anže and Sullivan, Mark and Troxel, M. A.},
	month = sep,
	year = {2018},
	note = {Publication Title: arXiv e-prints
ADS Bibcode: 2018arXiv180901669T},
}

@article{vandompselerAlignmentGalaxiesBaryon2023b,
	title = {The alignment of galaxies at the {Baryon} {Acoustic} {Oscillation} scale},
	volume = {6},
	issn = {2565-6120},
	url = {https://ui.adsabs.harvard.edu/abs/2023OJAp....6E..19V},
	doi = {10.21105/astro.2301.04649},
	urldate = {2024-09-24},
	journal = {The Open Journal of Astrophysics},
	author = {van Dompseler, Dennis and Georgiou, Christos and Chisari, Nora Elisa},
	month = jun,
	year = {2023},
	note = {ADS Bibcode: 2023OJAp....6E..19V},
	pages = {19},
}

@article{bridleDarkEnergyConstraints2007,
	title = {Dark energy constraints from cosmic shear power spectra: impact of intrinsic alignments on photometric redshift requirements},
	volume = {9},
	issn = {1367-2630},
	shorttitle = {Dark energy constraints from cosmic shear power spectra},
	url = {https://ui.adsabs.harvard.edu/abs/2007NJPh....9..444B},
	doi = {10.1088/1367-2630/9/12/444},
	urldate = {2022-03-13},
	journal = {New Journal of Physics},
	author = {Bridle, Sarah and King, Lindsay},
	month = dec,
	year = {2007},
	note = {ADS Bibcode: 2007NJPh....9..444B},
	pages = {444},
}

@article{troxelIntrinsicAlignmentGalaxies2015,
	title = {The intrinsic alignment of galaxies and its impact on weak gravitational lensing in an era of precision cosmology},
	volume = {558},
	issn = {0370-1573},
	url = {https://ui.adsabs.harvard.edu/abs/2015PhR...558....1T},
	doi = {10.1016/j.physrep.2014.11.001},
	urldate = {2022-07-02},
	journal = {Physics Reports},
	author = {Troxel, M. A. and Ishak, Mustapha},
	month = feb,
	year = {2015},
	note = {ADS Bibcode: 2015PhR...558....1T},
	pages = {1--59},
}

@article{brownMeasurementIntrinsicAlignments2002,
	title = {Measurement of intrinsic alignments in galaxy ellipticities},
	volume = {333},
	issn = {0035-8711},
	url = {https://ui.adsabs.harvard.edu/abs/2002MNRAS.333..501B},
	doi = {10.1046/j.1365-8711.2002.05354.x},
	urldate = {2024-01-15},
	journal = {Monthly Notices of the Royal Astronomical Society},
	author = {Brown, M. L. and Taylor, A. N. and Hambly, N. C. and Dye, S.},
	month = jul,
	year = {2002},
	note = {ADS Bibcode: 2002MNRAS.333..501B},
	pages = {501--509},
}

@article{singhIntrinsicAlignmentsBOSS2016,
	title = {Intrinsic alignments of {BOSS} {LOWZ} galaxies {II}: {Impact} of shape measurement methods},
	volume = {457},
	issn = {0035-8711, 1365-2966},
	shorttitle = {Intrinsic alignments of {BOSS} {LOWZ} galaxies {II}},
	url = {http://arxiv.org/abs/1510.06752},
	doi = {10.1093/mnras/stw144},
	number = {3},
	urldate = {2023-03-14},
	journal = {Monthly Notices of the Royal Astronomical Society},
	author = {Singh, Sukhdeep and Mandelbaum, Rachel},
	month = apr,
	year = {2016},
	note = {arXiv:1510.06752 [astro-ph]},
	pages = {2301--2317},
}

@article{akitsuImprintAnisotropicPrimordial2021,
	title = {Imprint of anisotropic primordial non-{Gaussianity} on halo intrinsic alignments in simulations},
	volume = {103},
	issn = {1550-79980556-2821},
	url = {https://ui.adsabs.harvard.edu/abs/2021PhRvD.103h3508A},
	doi = {10.1103/PhysRevD.103.083508},
	urldate = {2023-05-15},
	journal = {Physical Review D},
	author = {Akitsu, Kazuyuki and Kurita, Toshiki and Nishimichi, Takahiro and Takada, Masahiro and Tanaka, Satoshi},
	month = apr,
	year = {2021},
	note = {ADS Bibcode: 2021PhRvD.103h3508A},
	pages = {083508},
}

@article{xuEvidenceBaryonAcoustic2023,
	title = {Evidence for baryon acoustic oscillations from galaxy-ellipticity correlations.},
	volume = {7},
	issn = {2397-3366},
	url = {https://ui.adsabs.harvard.edu/abs/2023NatAs...7.1259X},
	doi = {10.1038/s41550-023-02035-4},
	urldate = {2023-11-27},
	journal = {Nature Astronomy},
	author = {Xu, Kun and Jing, Y. P. and Zhao, Gong-Bo and Cuesta, Antonio J.},
	month = oct,
	year = {2023},
	note = {ADS Bibcode: 2023NatAs...7.1259X},
	pages = {1259--1264},
}

@article{chisariCosmologicalInformationIntrinsic2013,
	title = {Cosmological information in the intrinsic alignments of luminous red galaxies},
	volume = {12},
	issn = {1475-7516},
	url = {http://adsabs.harvard.edu/abs/2013JCAP...12..029C},
	doi = {10.1088/1475-7516/2013/12/029},
	urldate = {2020-01-25},
	journal = {Journal of Cosmology and Astroparticle Physics},
	author = {Chisari, Nora Elisa and Dvorkin, Cora},
	month = dec,
	year = {2013},
	pages = {029},
}

@article{joachimiGalaxyAlignmentsOverview2015,
	title = {Galaxy {Alignments}: {An} {Overview}},
	volume = {193},
	issn = {1572-9672},
	url = {https://doi.org/10.1007/s11214-015-0177-4},
	doi = {10.1007/s11214-015-0177-4},
	number = {1},
	urldate = {2023-04-24},
	journal = {Space Science Reviews},
	author = {Joachimi, Benjamin and Cacciato, Marcello and Kitching, Thomas D. and Leonard, Adrienne and Mandelbaum, Rachel and Schäfer, Björn Malte and Sifón, Cristóbal and Hoekstra, Henk and Kiessling, Alina and Kirk, Donnacha and Rassat, Anais},
	month = nov,
	year = {2015},
	pages = {1--65},
}

@article{singhIntrinsicAlignmentsSDSSIII2015,
	title = {Intrinsic alignments of {SDSS}-{III} {BOSS} {LOWZ} sample galaxies},
	volume = {450},
	issn = {0035-8711},
	url = {https://ui.adsabs.harvard.edu/abs/2015MNRAS.450.2195S},
	doi = {10.1093/mnras/stv778},
	urldate = {2022-10-31},
	journal = {Monthly Notices of the Royal Astronomical Society},
	author = {Singh, Sukhdeep and Mandelbaum, Rachel and More, Surhud},
	month = jun,
	year = {2015},
	note = {ADS Bibcode: 2015MNRAS.450.2195S},
	pages = {2195--2216},
}

@article{johnstonKiDSGAMAIntrinsic2019,
	title = {{KiDS}+{GAMA}: {Intrinsic} alignment model constraints for current and future weak lensing cosmology},
	volume = {624},
	issn = {0004-6361},
	shorttitle = {{KiDS}+{GAMA}},
	url = {https://ui.adsabs.harvard.edu/abs/2019A&A...624A..30J/abstract},
	doi = {10.1051/0004-6361/201834714},
	language = {en},
	urldate = {2022-10-31},
	journal = {Astronomy and Astrophysics},
	author = {Johnston, Harry and Georgiou, Christos and Joachimi, Benjamin and Hoekstra, Henk and Chisari, Nora Elisa and Farrow, Daniel and Fortuna, Maria Cristina and Heymans, Catherine and Joudaki, Shahab and Kuijken, Konrad and Wright, Angus},
	month = apr,
	year = {2019},
	pages = {A30},
}

@article{joachimiConstraintsIntrinsicAlignment2011,
	title = {Constraints on intrinsic alignment contamination of weak lensing surveys using the {MegaZ}-{LRG} sample},
	volume = {527},
	issn = {0004-6361},
	url = {https://ui.adsabs.harvard.edu/abs/2011A&A...527A..26J/abstract},
	doi = {10.1051/0004-6361/201015621},
	language = {en},
	urldate = {2022-10-31},
	journal = {Astronomy and Astrophysics},
	author = {Joachimi, B. and Mandelbaum, R. and Abdalla, F. B. and Bridle, S. L.},
	month = mar,
	year = {2011},
	pages = {A26},
}

@article{mandelbaumDetectionLargescaleIntrinsic2006,
	title = {Detection of large-scale intrinsic ellipticity-density correlation from the {Sloan} {Digital} {Sky} {Survey} and implications for weak lensing surveys},
	volume = {367},
	issn = {0035-8711},
	url = {https://ui.adsabs.harvard.edu/abs/2006MNRAS.367..611M},
	doi = {10.1111/j.1365-2966.2005.09946.x},
	urldate = {2022-10-31},
	journal = {Monthly Notices of the Royal Astronomical Society},
	author = {Mandelbaum, Rachel and Hirata, Christopher M. and Ishak, Mustapha and Seljak, Uroš and Brinkmann, Jonathan},
	month = apr,
	year = {2006},
	note = {ADS Bibcode: 2006MNRAS.367..611M},
	pages = {611--626},
}

@article{maksimovaAbacusSummitMassiveSet2021,
	title = {{AbacusSummit}: a massive set of high-accuracy, high-resolution {N}-body simulations},
	volume = {508},
	issn = {0035-8711},
	shorttitle = {{AbacusSummit}},
	url = {https://doi.org/10.1093/mnras/stab2484},
	doi = {10.1093/mnras/stab2484},
	number = {3},
	urldate = {2021-11-17},
	journal = {Monthly Notices of the Royal Astronomical Society},
	author = {Maksimova, Nina A and Garrison, Lehman H and Eisenstein, Daniel J and Hadzhiyska, Boryana and Bose, Sownak and Satterthwaite, Thomas P},
	month = dec,
	year = {2021},
	pages = {4017--4037},
}

\onecolumngrid
\vspace{.35in}
\noindent\makebox[\textwidth][c]{\normalsize\scshape 
Author Affiliations}
\par
\vspace{0.35in}          
\twocolumngrid
 
\noindent\footnotesize
\setlength{\parindent}{0pt}
 
$^{1}$Center for Cosmology and AstroParticle Physics (CCAPP), Ohio State University, Columbus, OH 43210\\
$^{2}$Department of Astronomy, The Ohio State University, Columbus, OH 43210, USA\\
$^{3}$Department of Physics, The Ohio State University, Columbus, OH 43210, USA\\
$^{4}$Lawrence Berkeley National Laboratory, 1 Cyclotron Road, Berkeley, CA 94720, USA\\
$^{5}$Department of Physics, Boston University, 590 Commonwealth Avenue, Boston, MA 02215 USA\\
$^{6}$Instituto Avanzado de Cosmolog\'{i}a A.~C., San Marcos 11 - Atenas 202. Magdalena Contreras. Ciudad de M\'{e}xico C.~P.~10720, M\'{e}xico\\
$^{7}$Instituto de Ciencias F\'{i}sicas, Universidad Nacional Aut\'onoma de M\'exico, Av. Universidad s/n, Cuernavaca, Morelos, C.~P.~62210, M\'exico\\
$^{8}$Dipartimento di Fisica ``Aldo Pontremoli'', Universit\`a degli Studi di Milano, Via Celoria 16, I-20133 Milano, Italy\\
$^{9}$INAF-Osservatorio Astronomico di Brera, Via Brera 28, 20122 Milano, Italy\\
$^{10}$Department of Physics \& Astronomy, University College London, Gower Street, London, WC1E 6BT, UK\\
$^{11}$Departamento de Astrof\'{i}sica, Universidad de La Laguna (ULL), E-38206, La Laguna, Tenerife, Spain\\
$^{12}$Instituto de Astrof\'{i}sica de Canarias, C/ V\'{i}a L\'{a}ctea, s/n, E-38205 La Laguna, Tenerife, Spain\\
$^{13}$Institut d'Estudis Espacials de Catalunya (IEEC), c/ Esteve Terradas 1, Edifici RDIT, Campus PMT-UPC, 08860 Castelldefels, Spain\\
$^{14}$Institute of Space Sciences, ICE-CSIC, Campus UAB, Carrer de Can Magrans s/n, 08913 Bellaterra, Barcelona, Spain\\
$^{15}$Instituto de F\'{i}sica, Universidad Nacional Aut\'{o}noma de M\'{e}xico, Circuito de la Investigaci\'{o}n Cient\'{i}fica, Ciudad Universitaria, Cd. de M\'{e}xico C.~P.~04510, M\'{e}xico\\
$^{16}$Center for Astrophysics $|$ Harvard \& Smithsonian, 60 Garden Street, Cambridge, MA 02138, USA\\
$^{17}$Instituci\'{o} Catalana de Recerca i Estudis Avan\c{c}ats, Passeig de Llu\'{i}s Companys, 23, 08010 Barcelona, Spain\\
$^{18}$Institut de F\'{i}sica d'Altes Energies (IFAE), The Barcelona Institute of Science and Technology, Edifici Cn, Campus UAB, 08193, Bellaterra (Barcelona), Spain\\
$^{19}$Departamento de F\'isica, Universidad de los Andes, Cra. 1 No. 18A-10, Edificio Ip, CP 111711, Bogot\'a, Colombia\\
$^{20}$Observatorio Astron\'omico, Universidad de los Andes, Cra. 1 No. 18A-10, Edificio H, CP 111711 Bogot\'a, Colombia\\
$^{21}$Institute of Cosmology and Gravitation, University of Portsmouth, Dennis Sciama Building, Portsmouth, PO1 3FX, UK\\
$^{22}$Fermi National Accelerator Laboratory, PO Box 500, Batavia, IL 60510, USA\\
$^{23}$The Ohio State University, Columbus, 43210 OH, USA\\
$^{24}$Department of Physics, The University of Texas at Dallas, 800 W. Campbell Rd., Richardson, TX 75080, USA\\
$^{25}$NSF NOIRLab, 950 N. Cherry Ave., Tucson, AZ 85719, USA\\
$^{26}$Department of Astronomy \& Astrophysics, University of Toronto, Toronto, ON M5S 3H4, Canada\\
$^{27}$Department of Physics and Astronomy, University of California, Irvine, 92697, USA\\
$^{28}$Departament de F\'{i}sica, Serra H\'{u}nter, Universitat Aut\`{o}noma de Barcelona, 08193 Bellaterra (Barcelona), Spain\\
$^{29}$IRFU, CEA, Universit\'{e} Paris-Saclay, F-91191 Gif-sur-Yvette, France\\
$^{30}$Department of Physics and Astronomy, University of Waterloo, 200 University Ave W, Waterloo, ON N2L 3G1, Canada\\
$^{31}$Perimeter Institute for Theoretical Physics, 31 Caroline St. North, Waterloo, ON N2L 2Y5, Canada\\
$^{32}$Waterloo Centre for Astrophysics, University of Waterloo, 200 University Ave W, Waterloo, ON N2L 3G1, Canada\\
$^{33}$Instituto de Astrof\'{i}sica de Andaluc\'{i}a (CSIC), Glorieta de la Astronom\'{i}a, s/n, E-18008 Granada, Spain\\
$^{34}$Departament de F\'isica, EEBE, Universitat Polit\`ecnica de Catalunya, c/Eduard Maristany 10, 08930 Barcelona, Spain\\
$^{35}$Universit\'{e} Clermont-Auvergne, CNRS, LPCA, 63000 Clermont-Ferrand, France\\
$^{36}$Department of Physics and Astronomy, Sejong University, 209 Neungdong-ro, Gwangjin-gu, Seoul 05006, Republic of Korea\\
$^{37}$Queensland University of Technology, School of Chemistry \& Physics, George St, Brisbane 4001, Australia\\
$^{38}$Abastumani Astrophysical Observatory, Tbilisi, GE-0179, Georgia\\
$^{39}$Department of Physics, Kansas State University, 116 Cardwell Hall, Manhattan, KS 66506, USA\\
$^{40}$CIEMAT, Avenida Complutense 40, E-28040 Madrid, Spain\\
$^{41}$Max Planck Institute for Extraterrestrial Physics, Gie\ss enbachstra\ss e 1, 85748 Garching, Germany\\
$^{42}$Department of Physics, University of Michigan, 450 Church Street, Ann Arbor, MI 48109, USA\\
$^{43}$University of Michigan, 500 S. State Street, Ann Arbor, MI 48109, USA\\
$^{44}$Department of Physics, Southern Methodist University, 3215 Daniel Avenue, Dallas, TX 75275, USA
 
\normalsize
\vspace{.1in}

\end{document}